\documentclass{pasj00}
\draft
\color{black}

\begin{document}
\SetRunningHead{K.~Tokoi et al.}
{Temperature and Abundance Profiles of HCG~62 with Suzaku}
\Received{----/--/--}
\Accepted{----/--/--}

\title{
Suzaku Observation of HCG 62: 
Temperature, Abundance, and Extended Hard X-ray Emission Profiles
}

\author{
Kazuyo {\sc Tokoi},$^1$ 
Kosuke {\sc Sato},$^2$
Yoshitaka {\sc Ishisaki},$^1$
Takaya {\sc Ohashi},$^1$
Noriko Y. {\sc Yamasaki},$^3$\\
Kazuhiro {\sc Nakazawa},$^4$
Kyoko {\sc Matsushita},$^2$
Yasushi {\sc Fukazawa},$^5$
Akio {\sc Hoshino},$^1$\\
Takayuki {\sc Tamura},$^3$
Chihiro {\sc Egawa},$^5$
Naomi {\sc Kawano},$^5$
Naomi {\sc Ota},$^3$
Naoki {\sc Isobe},$^6$\\
Madoka {\sc Kawaharada},$^6$
Hisamitsu {\sc Awaki},$^7$
and John P. {\sc Hughes}$^8$
}

\altaffiltext{1}
{Department of  Physics, Tokyo Metropolitan University, 1-1 Minami-Osawa,
Hachioji, Tokyo 192-0397}
\altaffiltext{2}
{Department of  Physics, Tokyo University of Science, 1-3 Kagurazaka, Sinjuku-ku, Tokyo 162-8601}
\altaffiltext{3}
{Institute of Space and Astronautical Science (ISAS), \\
Japan Aerospace Exploration Agency, 3-1-1 Yoshinodai, Sagamihara, 
Kanagawa 229-8510}
\altaffiltext{4}
{Department of  Physics, University of Tokyo, Hongo 7-3-1,
Bunkyo-ku, Tokyo 113-0033}
\altaffiltext{5}
{Department of  Physical Science, Hiroshima University, \\
1-3-1 Kagamiyama, Higashi-Hiroshima, Hiroshima 739-8526}
\altaffiltext{6}
{Cosmic Radiation Laboratory, RIKEN, 2-1 Hirosawa, Wako, Saitama 351-0198, 
Japan}
\altaffiltext{7}
{Department of  Physics, Ehime University, 2-5 Bunkyou-machi,
Matsuyama-shi, Ehime 790-8577}
\altaffiltext{8}
{Department of Physics and Astronomy, Rutgers University, \\
136 Frelinghuysen Road, Piscataway, NJ 08854-8019, USA}


\maketitle

\begin{abstract}
We present results of 120~ks observation of a compact group of
galaxies HCG~62 ($z=0.0145$) with Suzaku XIS and HXD-PIN\@. The XIS
spectra for four annular regions were fitted with two temperature {\it
vapec} model with variable abundance, combined with the foreground
Galactic component. The Galactic component was constrained to have a
common surface brightness among the four annuli, and two temperature
{\it apec} model was preferred to single temperature model.  We
confirmed the multi-temperature nature of the intra-group medium
reported with Chandra and XMM-Newton, with a doughnut-like
high temperature ring at radii 3.3--6.5$'$ in a hardness image. We
found Mg, Si, S, and Fe abundances to be fairly robust. We examined
the possible ``high-abundance arc'' at $\sim 2'$ southwest from the
center, however Suzaku data did not confirm it.
We suspect that it is a misidentification
of an excess hot component in this region as the Fe line.
Careful background study showed no positive detection of
the extended hard X-rays previously reported with ASCA, in 5--12~keV
with XIS and 12--40~keV with HXD-PIN, although our upper limit did not
exclude the ASCA result. There is an indication that the X-ray
intensity in $r<3.3'$ region is $70\pm 19$\% higher than the nominal
CXB level (5--12~keV), and Chandra and Suzaku data suggest that most
of this excess could be due to concentration of hard X-ray sources
with an average photon index of $\Gamma=1.38\pm 0.06$.  Cumulative
mass of O, Fe and Mg in the group gas and the metal mass-to-light
ratio were derived and compared with those in other groups. Possible
role of AGN or galaxy mergers in this group is also discussed.
\end{abstract}

\section{Introduction}

Groups of galaxies play a key role in the formation of the
universe. Most importantly, they act as building blocks in the
framework of hierarchical formation of structures (e.g.\
\cite{Navarro1995}). In this sense, properties of groups should be
critically compared with those in clusters of galaxies, to test the
general hypothesis that groups of galaxies indeed represent the
condition in clusters when they have not been evolved. For example,
baryon to dark matter ratio and metal abundance in groups of galaxies
can be a measure to be tested whether rich clusters are indeed simple
superposition of many groups or some other mechanism may be involved.

The groups are characterized by a short dynamical time scale, and we
expect galaxy encounters to take place frequently in such a high
density environment. The X-ray search for an evidence of dynamical
processes in the form of hard non-thermal emission would be an
important subject of observational studies. The relatively low gas
temperature compared with rich clusters greatly helps us find such
emission even with standard CCD instruments. Also it is well known
that group gas tends to contain significantly high entropy if one
extrapolates the scaling relation holding for clusters
\citep{Ponman1999}. A simulation study indicated that simple galactic
winds were unable to significantly raise the entropy in cluster size
hot gas \citep{Borgani2005}. This indicates that a certain heat input
or preheating occurred over a widespread region and groups of galaxies
offer us an opportunity to closely look into such an effect.

Another aspect concerning groups of galaxies is their role in the
chemical enrichment of cluster hot gas.
It has been shown that rich clusters maintain all the metals produced
by the cluster galaxies \citep{Fukazawa1998}, however early-type
galaxies are thought to have released most of the metals formed
through past supernova explosions \citep{Makishima2001}. We should
note that these galaxies almost always reside in groups and clusters
of galaxies and that their X-ray haloes may not be clearly separable
from the surrounding gas.
Groups of galaxies lie in the region where the
metal confinement can be marginally possible. This means that by
looking into the distribution of various metals in groups, we can
measure the efficiency of metal confinement and further estimate how
each metal has been injected into the intergalactic space. The low
temperature ($< 2$ keV) of the group gas also greatly helps us study
emission lines from various metals ranging from oxygen to iron.

HCG~62 is one of the brightest and well-studied groups of galaxies
among the Hickson compact groups of galaxies \citep{Hickson1989}
located at $z = 0.0145$ \citep{Mulchaey2003}.
\citet{Zabludoff2000} extensively surveyed a $1.5\times 1.5$ deg$^2$
region around the group and measured redshifts of 154 galaxies.
They identified 63 of them as members of the group within a
radius of $50'$ (900~kpc), and the measured velocity dispersion,
$\sigma_r = 390^{+37}_{-34}$ km~s$^{-1}$, is a typical value for
galaxy groups. The central region within $r<1.1'$ is dominated
by three galaxies, HCG~62a (NGC~4778), HCG~62b (NGC~4776),
and HCG~62c (NGC~4761), and they are all classified as S0 galaxies.
Kinematics of these galaxies suggests possible interaction
among them \citep{Spavone2006,Rampazzo1998}. Such a high galaxy
density and the low velocity dispersion are considered to result in a
galaxy merger, and \citet{Ponman1993} suggested that they should
culminate in a final merger within a few billion years,
consisting of a large elliptical galaxy embedded in an extended X-ray halo.

HCG~62 is also known as the first compact galaxy group that was
detected in X-ray to have an extended hot gas
(intra-cluster or group medium; ICM),
and the ICM properties have been extensively studied using ROSAT and ASCA
(e.g., \cite{Ponman1993,Pildis1995,Finoguenov1999,Mulchaey2003}).
Furthermore, spatially extended hard X-ray excess over the ICM emission
was detected with ASCA \citep{Fukazawa2001,Nakazawa2007},
and two symmetrical cavities in the central region
at a projected radius of $\sim 25''$ (7.4~kpc)
were discovered by Chandra \citep{Vrtilek2002}.
Detailed hot-gas properties including the cavities
and the gravitational mass have been studied
based on the combined data from Chandra and XMM-Newton \citep{Morita2006}.
Recently, \citet{Gu2007} reported that a region at $2'$ offset
from the central galaxy showed a twice higher metal abundance
than the surrounding gas (the ``high-abundance arc'').

In this paper, we report the results from Suzaku observation of HCG~62\@.
Owing to the low and stable background as well as
good sensitivity to emission lines below $\sim 1$~keV
of the X-ray Imaging Spectrometer (XIS; \cite{Koyama2007}),
accurate determination of O and Mg abundances to outer regions
have become feasible. XIS is also effective to constrain the
extended hard X-ray emission in combination with
the Hard X-ray Detector (HXD; \cite{Takahashi2007}).
This paper is organized as follows:
In sections~\ref{sec:obs}~and~\ref{sec:data},
we describe the Suzaku observation and the data reduction,
and show images of HCG~62 obtained with Suzaku in section~\ref{sec:image}.
In sections~\ref{sec:spec}--\ref{sec:abun},
we describe spectral analysis and the derived
temperature and abundance profiles,
and examine the existence of the ``high-abundance arc''
in section~\ref{sec:arc}.
In section~\ref{sec:hard-emission},
significance of the extended hard X-ray emission with Suzaku
is carefully investigated.
We discuss temperature and metallicity distributions,
metal mass-to-light ratio, and galaxy mergers in section~\ref{sec:discussion},
and a summary is given in section~\ref{sec:summary}.

We use $H_{0} = 0.7\; h_{100} = 70$ km~s$^{-1}$~Mpc$^{-1}$,
$\Omega_{\rm \Lambda}=1-\Omega_{\rm M} = 0.73$ throughout this paper.
At the redshift of $z = 0.0145$, 1$'$ corresponds to 17.8~kpc.
The virial radius, $r_{180}\equiv 1.95\; h_{100}^{-1}
\sqrt{k\langle T\rangle /10~\rm keV} = 1.08$~Mpc
\citep{Markevitch1998,Evrard1996}
for the average temperature of $k\langle T\rangle = 1.5$~keV\@.
We adopt a Galactic hydrogen column density of  
$N_{\rm H} = 3.03 \times 10^{20}$~cm$^{-2}$ in the direction of
HCG~62 \citep{Dickey1990}.
As for the definition of the solar abundance ratio, 
we followed \citet{Anders1989}.
Errors are 90\% confidence range for a single
interesting parameter.

\section{Observation}\label{sec:obs}

\begin{figure*}[tbgp]
\FigureFile(0.33\textwidth,1cm){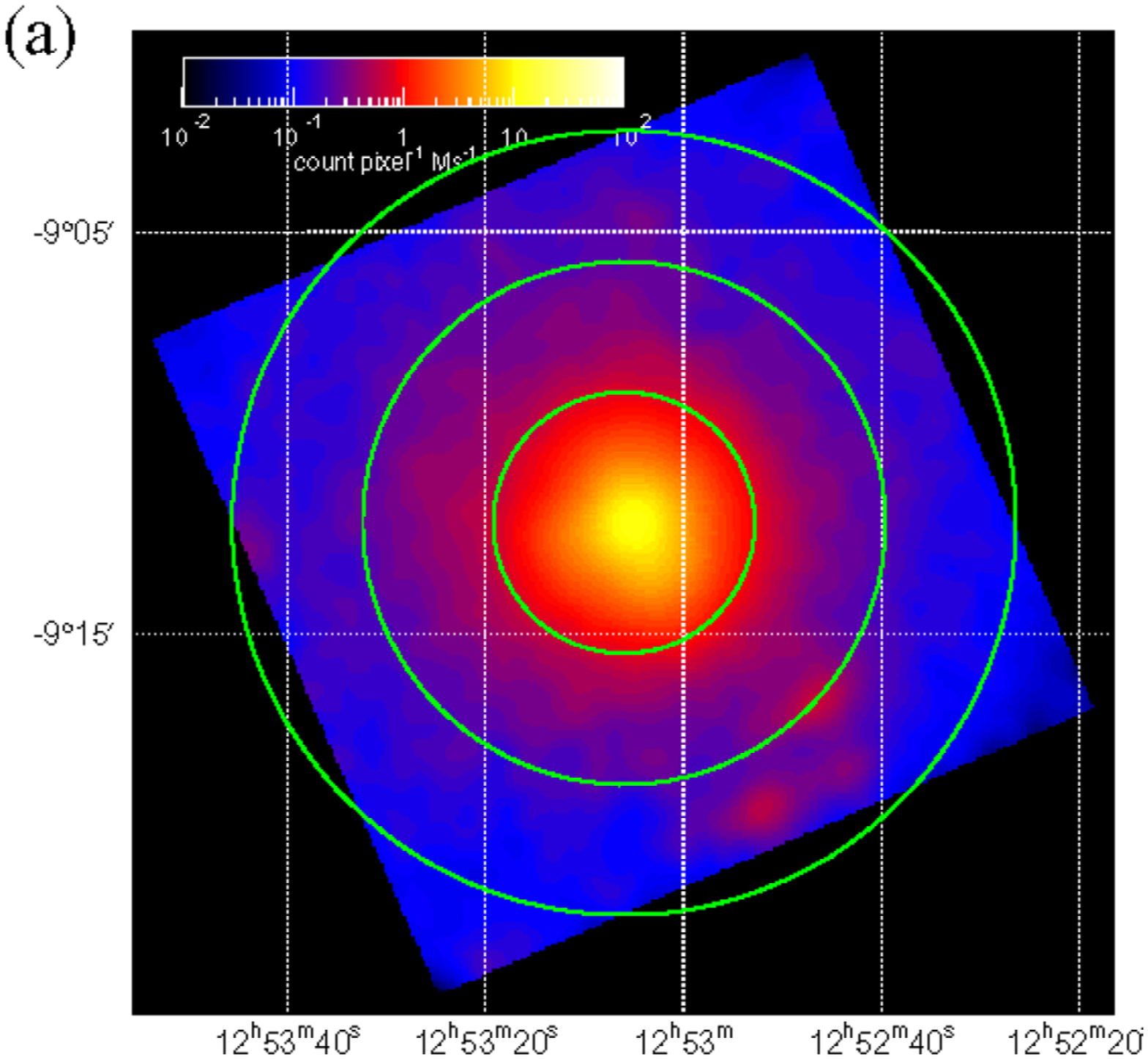}\hfill
\FigureFile(0.33\textwidth,1cm){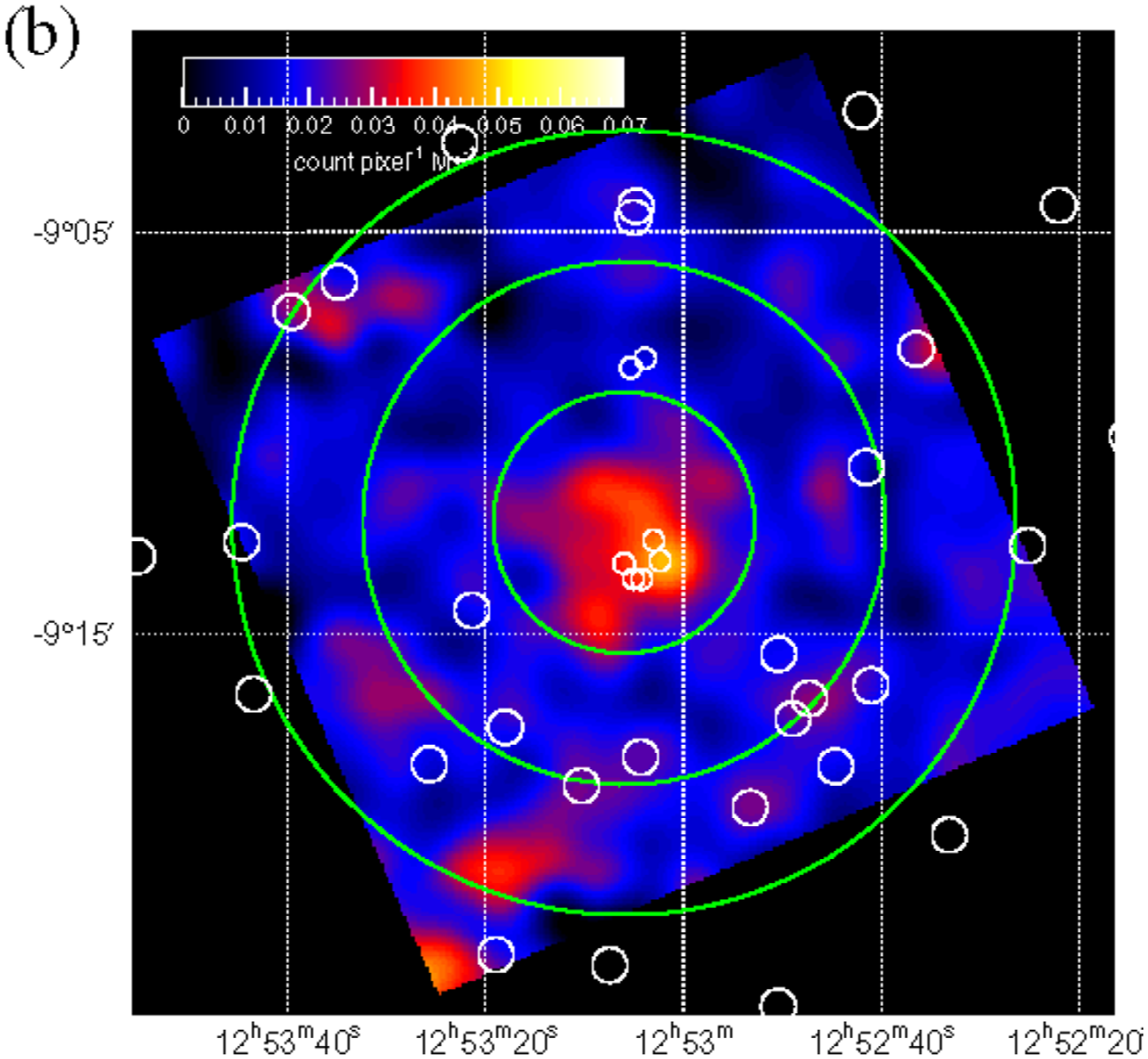}\hfill
\FigureFile(0.33\textwidth,1cm){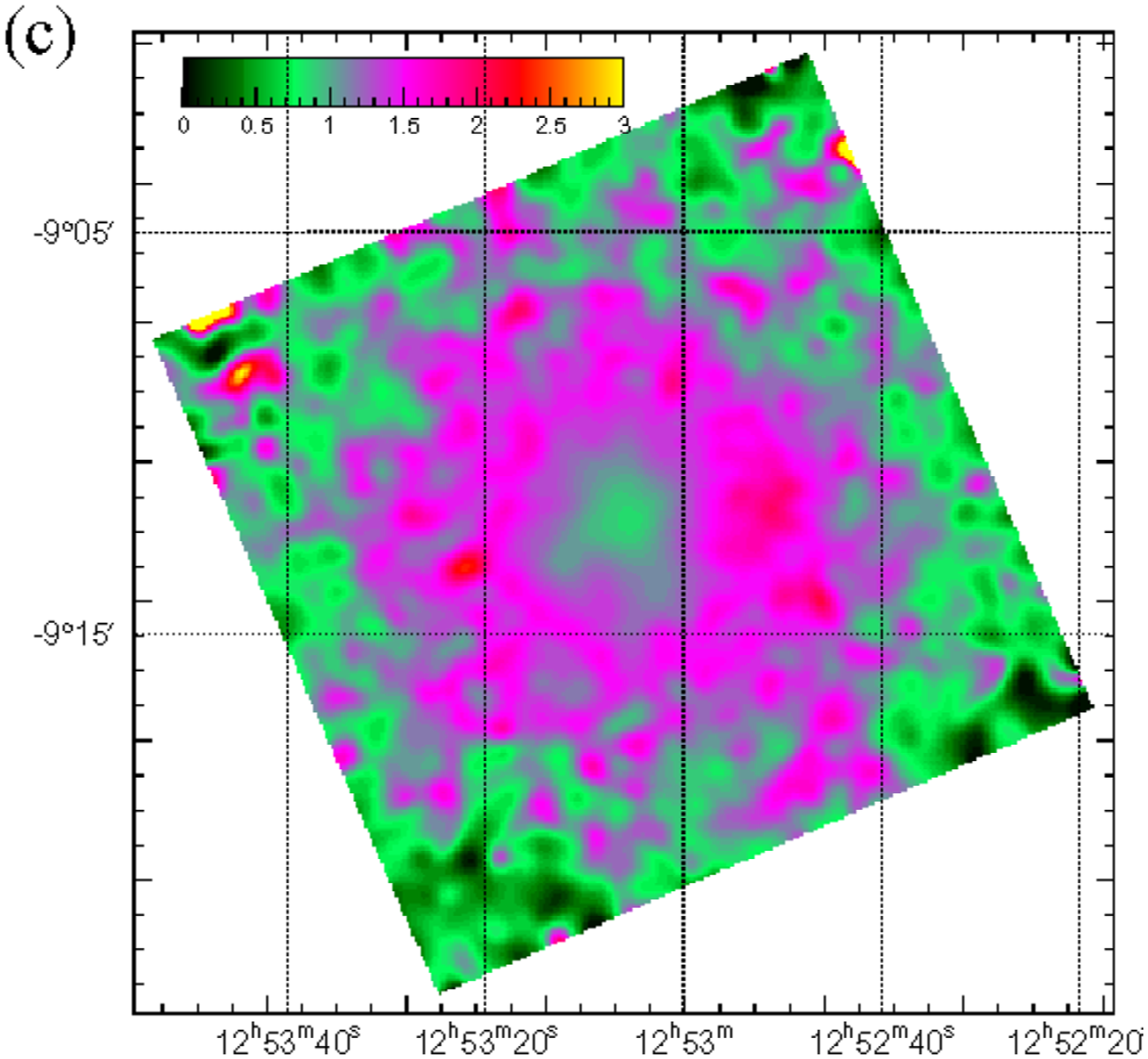}
\caption{
(a) Exposure corrected XIS image in 0.5--4.0~keV,
(b) in 6--10~keV energy range, and
(c) hardness ratio image of 1--2~keV to 0.5--1~keV bands.
Images of (a) and (c) are smoothed with gaussian of $\sigma = 17''$,
and (b) is with $\sigma = 33''$.
Estimated components of the NXB and CXB are subtracted,
but no vignetting correction is applied. 
Green circles corresponds to 3.3$'$, 6.5$'$, and 9.8$'$ radii,
from inner to outer, centered on HCG~62.
No {\it COR}\/ selection is applied, and both BI and FI sensors are co-added,
in order to maximize the photon statistics.
The ${}^{55}$Fe calibration source areas at corners are included
in (a) and (c), but excluded in (b).
White circles in (b) represent locations of point sources detected
by XMM-Newton. Sources shown in smaller circles are also detected
with Chandra.
}\label{fig:image}
\end{figure*}

Suzaku carried out observation of the central part of 
HCG~62 in January 2006 as a part of the Science Working Group (SWG)
time. The observation log is shown in table~\ref{tab:obslog}.
The XIS instrument was set to the normal clocking mode with data formats 
of $5\times 5$ and $3\times 3$ editing modes.
See \citet{Koyama2007} for details of the operation mode.
The XIS images are shown in figure~\ref{fig:image}.
We used the version 0.7 processing data, 
and the analysis was performed with HEAsoft 6.1.1 and XSPEC 11.3.2t.
After applying the standard data selection criteria: elevation from 
the sunlit Earth rim, ${\it DYE\_ELV} > 20^{\circ}$,
elevation from the Earth rim, ${\it ELV} > 5^{\circ}$,
time after the South Atlantic Anomaly (SAA) passage,
${\it T\_SAA\_HXD} > 256$~s, the exposure time of XIS was 119.4~ks.
Events of bad CCD event grades, bad columns, and hot/flickering pixels
were removed, by choosing ${\it GRADE} = 0$, 2, 3, 4, or 6,
${\it STATUS}< 262144$, and applying ``sisclean'' FTOOLS\@.

The HXD PIN and PMT were operated with nominal high-voltage supply 
and setups \citep{Kokubun2007}.
From the background modeling limitations as of the beginning of 2006, 
we selected time regions with the magnetic cut-off rigidity ({\it COR}\/)
larger than 8~GV, and ${\it T\_SAA\_HXD} > 1000$~s.
With these screenings, we obtained 75.1~ks of exposure for the PIN detector.
We did not analyze the GSO data of the HXD instrument in this paper.

\begin{table*}
\caption{Suzaku observation log of HCG 62.}
\label{tab:obslog}
\centering
\begin{tabular}{ll}
\hline\hline
Observation ID $\dotfill$
& 800013020 \\
Target Coordinates (J2000) \makebox[2em][l]{$\dotfill$}
& (RA\_NOM, DEC\_NOM) = (\timeform{12h53m06s}, \timeform{-9D12'14''}) \\
Date of Observation $\dotfill$
& 2006 January 20, 09:19 -- 23, 12:00\\
Exposure Time $\dotfill$
& XIS: \ \makebox[0in][r]{1}19.3~ks \hspace*{1em} HXD-PIN: --- \hspace*{1em} (No {\it COR} selection) \\
& XIS: \ 85.4~ks \hspace*{1em} HXD-PIN: 75.1~ks\hspace*{1em} (${\it COR} > 8$ GV) \\
\hline
\end{tabular}
\end{table*}

\begin{table*}
\caption{
Area, coverage of whole annulus, {\scriptsize SOURCE\_RATIO\_REG},
and observed/estimated counts for each annular region.
{\scriptsize SOURCE\_RATIO\_REG} represents the
flux ratio in the assumed spatial distribution on the sky
($3\beta$-model) inside the accumulation region
to the entire model, and written in the header keyword of
the calculated ARF response by ``xissimarfgen''.
}\label{tab:region}
\centerline{
\begin{tabular}{lrrrcrrrrcrrrr}
\hline\hline
\makebox[3em][l]{Region\,$^\ast$} & \multicolumn{1}{c}{Area\makebox[0in][l]{\,$^\dagger$}} & Coverage\makebox[0in][l]{\,$^\dagger$}\hspace*{-0.5em} & \makebox[4.2em][r]{\scriptsize SOURCE\_\makebox[0in][l]{\,$^\ddagger$}}\hspace*{-0.5em} & Energy & \multicolumn{4}{c}{BI counts\makebox[0in][l]{\,$^\S$}} &$\!\!\!\!$& \multicolumn{4}{c}{FI counts\makebox[0in][l]{\,$^\S$}} \\
\cline{6-9}\cline{11-14}
& \makebox[2em][c]{(arcmin$^2$)} &      & \makebox[4.2em][r]{\scriptsize RATIO\_REG}\hspace*{-0.5em} & (keV) & OBS & NXB & CXB & $f_{\rm BGD}$ &$\!\!\!\!$& OBS & NXB & CXB & $f_{\rm BGD}$ \\
\hline\\[-2ex]
0.0--3.3$'$& 33.4&100.0\%& 51.3\%& 0.4--7.5&$\!\!35,067$&$\!\!  906$&$\!\!1,020$&$\!\! 5.5\%$&$\!\!\!\!$&$\!\!65,887$&$\!\!1,445$&$\!\!2,353$&$\!\! 5.8\%$\\
3.3--6.5$'$&100.1&100.0\%& 16.8\%& 0.4--7.5&$\!\!20,095$&$\!\!2,691$&$\!\!2,620$&$\!\!26.4\%$&$\!\!\!\!$&$\!\!34,990$&$\!\!3,944$&$\!\!5,966$&$\!\!28.3\%$\\
6.5--9.8$'$&147.1& 88.2\%& 13.5\%& 0.4--7.5&$\!\!16,943$&$\!\!3,674$&$\!\!2,902$&$\!\!38.8\%$&$\!\!\!\!$&$\!\!27,659$&$\!\!5,256$&$\!\!6,251$&$\!\!41.6\%$\\
9.8--13$'$ & 36.8& 15.8\%&  2.1\%& 0.4--4.0&$\!\! 2,726$&$\!\!  629$&$\!\!  492$&$\!\!41.1\%$&$\!\!\!\!$&$\!\! 4,523$&$\!\!  980$&$\!\!1,051$&$\!\!44.9\%$\\
\hline\\[-2ex]
$r<1.1'$   &  3.8&100.0\%& 33.2\%& 0.4--7.5&$\!\!11,191$&$\!\!  105$&$\!\!  120$&$\!\! 2.0\%$&$\!\!\!\!$&$\!\!22,794$&$\!\!  166$&$\!\!  279$&$\!\! 2.0\%$\\
NE arc     & 14.8&100.0\%&  9.0\%& 0.4--7.5&$\!\!10,420$&$\!\!  405$&$\!\!  451$&$\!\! 8.2\%$&$\!\!\!\!$&$\!\!20,389$&$\!\!  650$&$\!\!1,045$&$\!\! 8.3\%$\\
SW arc     & 14.8&100.0\%&  9.0\%& 0.4--7.5&$\!\!13,358$&$\!\!  416$&$\!\!  433$&$\!\! 6.4\%$&$\!\!\!\!$&$\!\!22,526$&$\!\!  646$&$\!\!1,011$&$\!\! 7.4\%$\\
\hline
\end{tabular}
}

\medskip
\parbox{\textwidth}{\footnotesize
\footnotemark[$\ast$]
See figure~\ref{fig:image}(a) for the first four annuli,
figure~\ref{fig:arc}(a) for the latter three.

\footnotemark[$\dagger$]
The average values among four sensors are presented.

\footnotemark[$\ddagger$]
$\makebox{\scriptsize\rm SOURCE\_RATIO\_REG}\equiv
\makebox{\scriptsize\rm COVERAGE}\;\times
\int_{r_{\rm in}}^{r_{\rm out}} S(r)\; r\,dr / 
\int_{0}^{\infty} S(r)\; r\,dr$,
where $S(r)$ represents the assumed radial profile
of HCG~62, and we defined $S(r)$ in $26'\times 26'$ region on the sky.

\footnotemark[$\S$]
OBS denotes the observed counts including NXB and CXB in 0.4--7.5~keV
or 0.4--4~keV\@. NXB and CXB are the estimated counts.
}
\end{table*}

\section{Data Reduction}\label{sec:data}

In this section, we describe the spectral analysis of
the ICM with XIS\@.  First we extracted spectra
from four annular regions of 0.0--3.3$'$, 3.3--6.5$'$, 6.5--9.8$'$,
and 9.8--13$'$, centered on the target coordinates in table~\ref{tab:obslog}.
As for the background, we assumed a nominal cosmic X-ray background (CXB)
spectrum with a power-law photon index, $\Gamma=1.4$, and surface brightness of
$5.97\times 10^{-8}$ erg~cm$^{-2}$~s$^{-1}$~sr$^{-1}$
in 2--10~keV \citep{Kushino2002}.
The non X-ray background (NXB) spectra were estimated
from a database of the dark Earth observations with Suzaku
for the same detector area and with the same distribution of
{\it COR}\/ \citep{Tawa2007}. In order to increase the signal to noise ratio
by reducing the NXB count rate,
especially for the group outskirts and in high energy bands ($E\gtrsim 4$~keV),
we further required ${\it COR} > 8$~GV for the spectral analysis.
After this screening, the exposure time dropped to 85.4~ks,
however, the fit residuals in higher energy band were reduced
and the spectral fits became almost acceptable as described in a later section. 

\begin{figure*}[tbg]
\FigureFile(0.25\textwidth,1cm){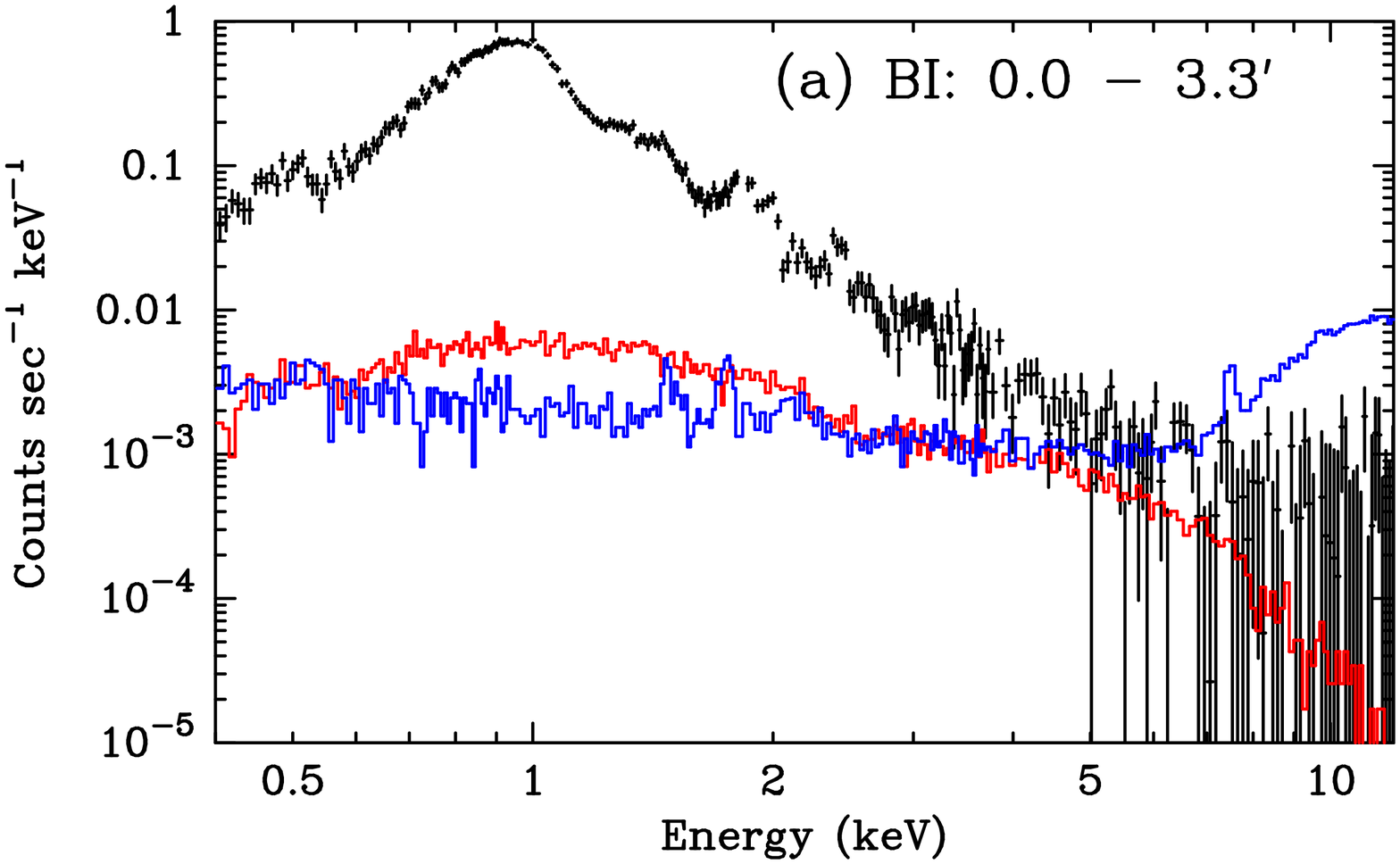}%
\FigureFile(0.25\textwidth,1cm){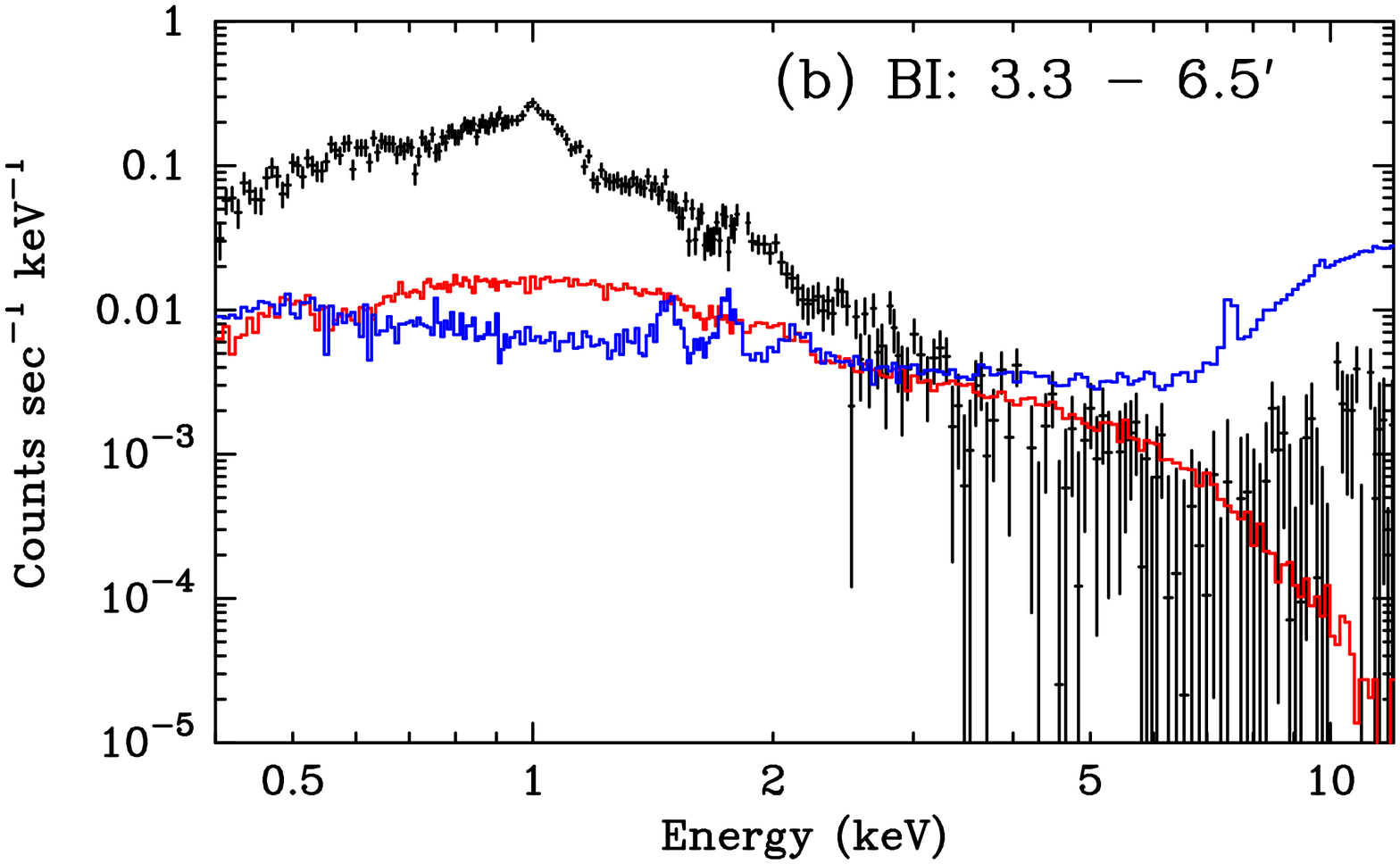}%
\FigureFile(0.25\textwidth,1cm){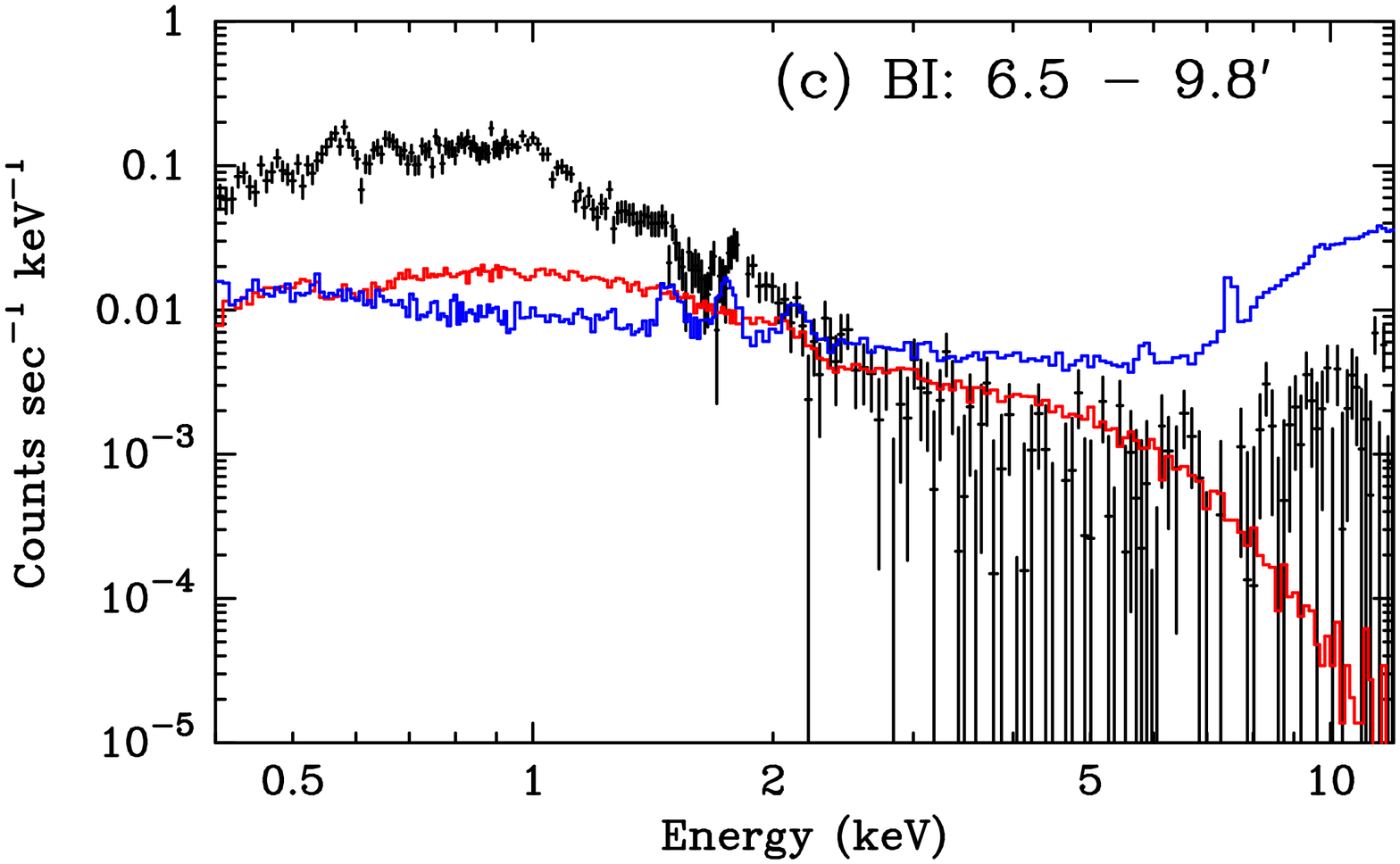}%
\FigureFile(0.25\textwidth,1cm){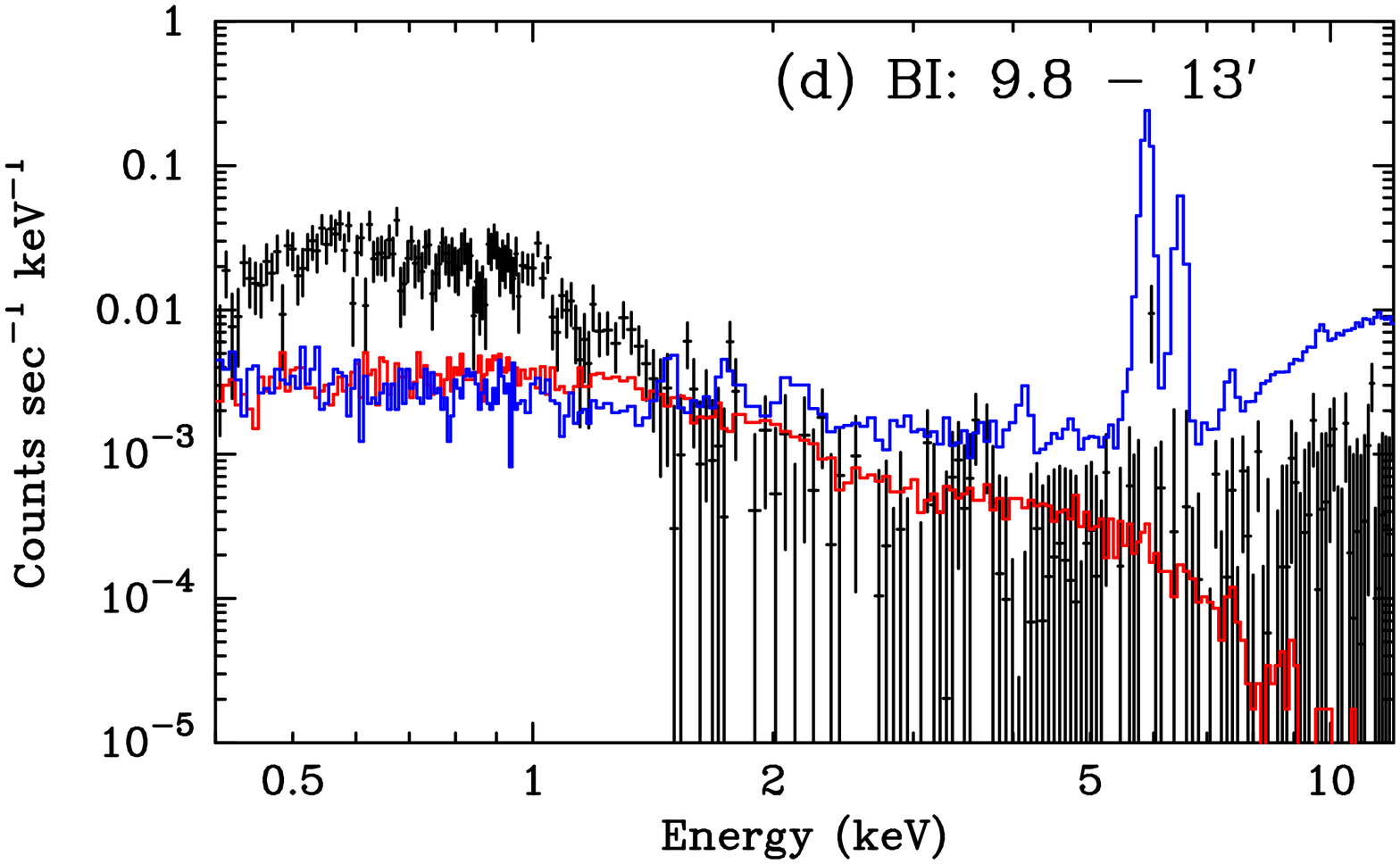}
\\[-1ex]
\FigureFile(0.25\textwidth,1cm){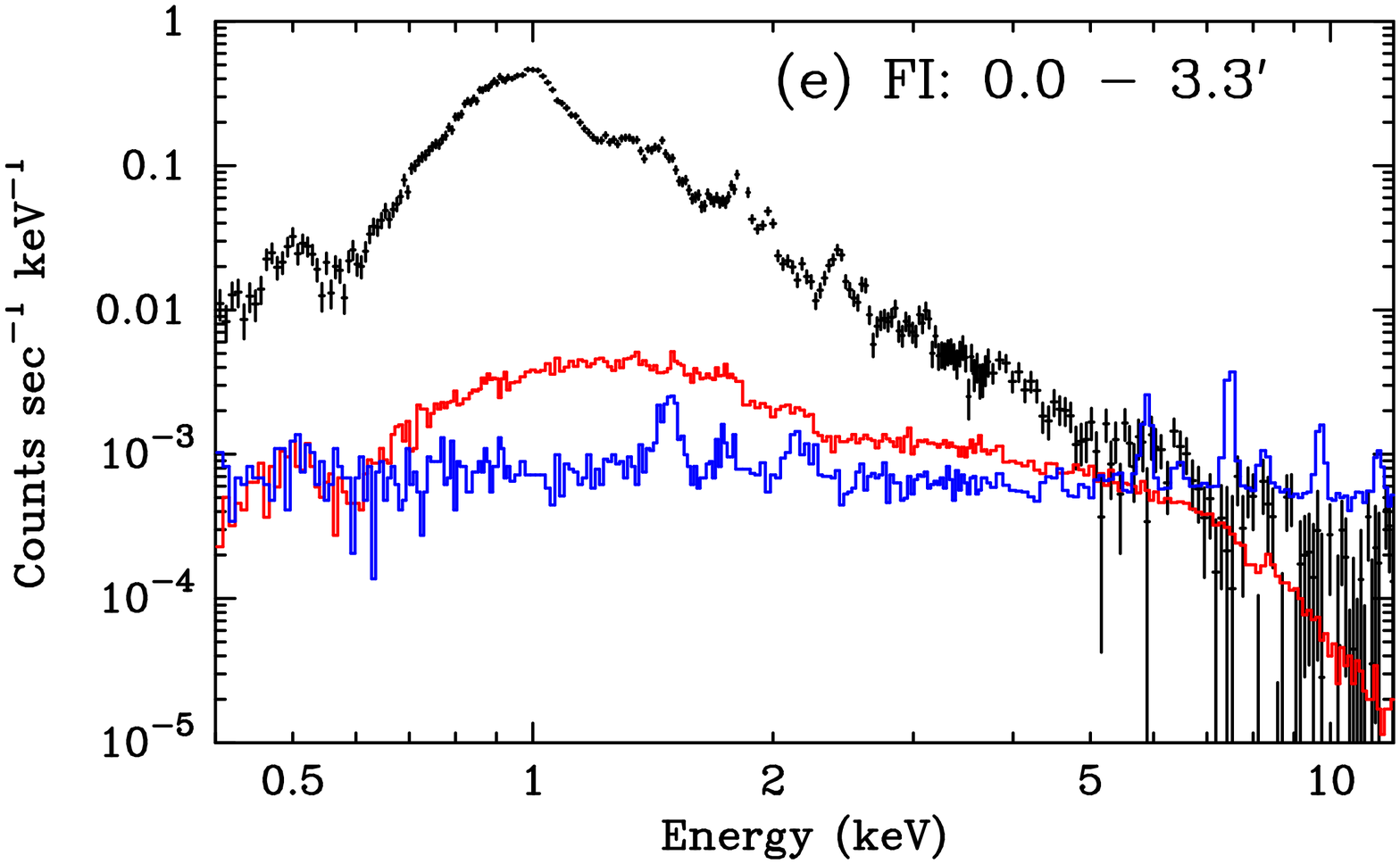}%
\FigureFile(0.25\textwidth,1cm){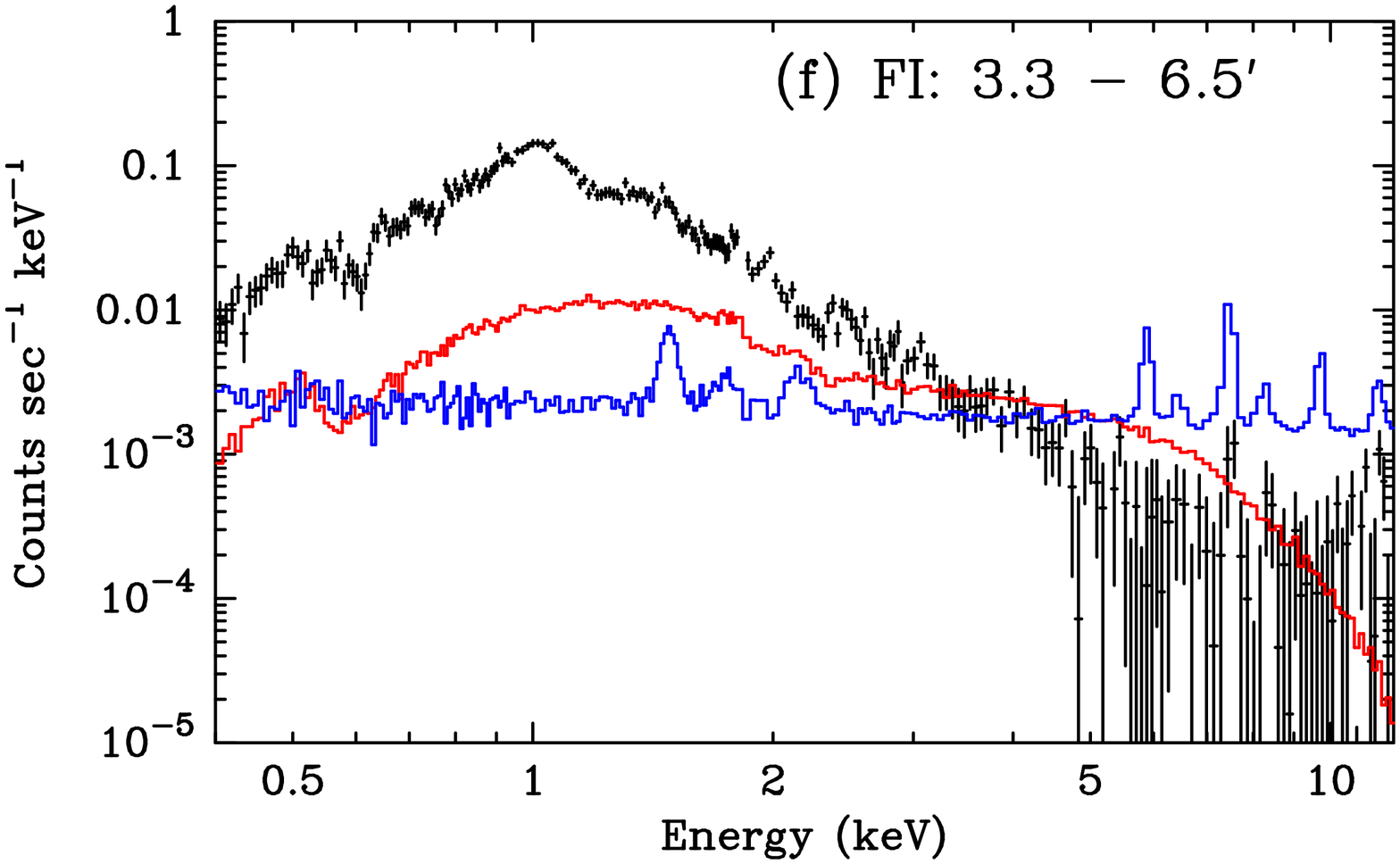}%
\FigureFile(0.25\textwidth,1cm){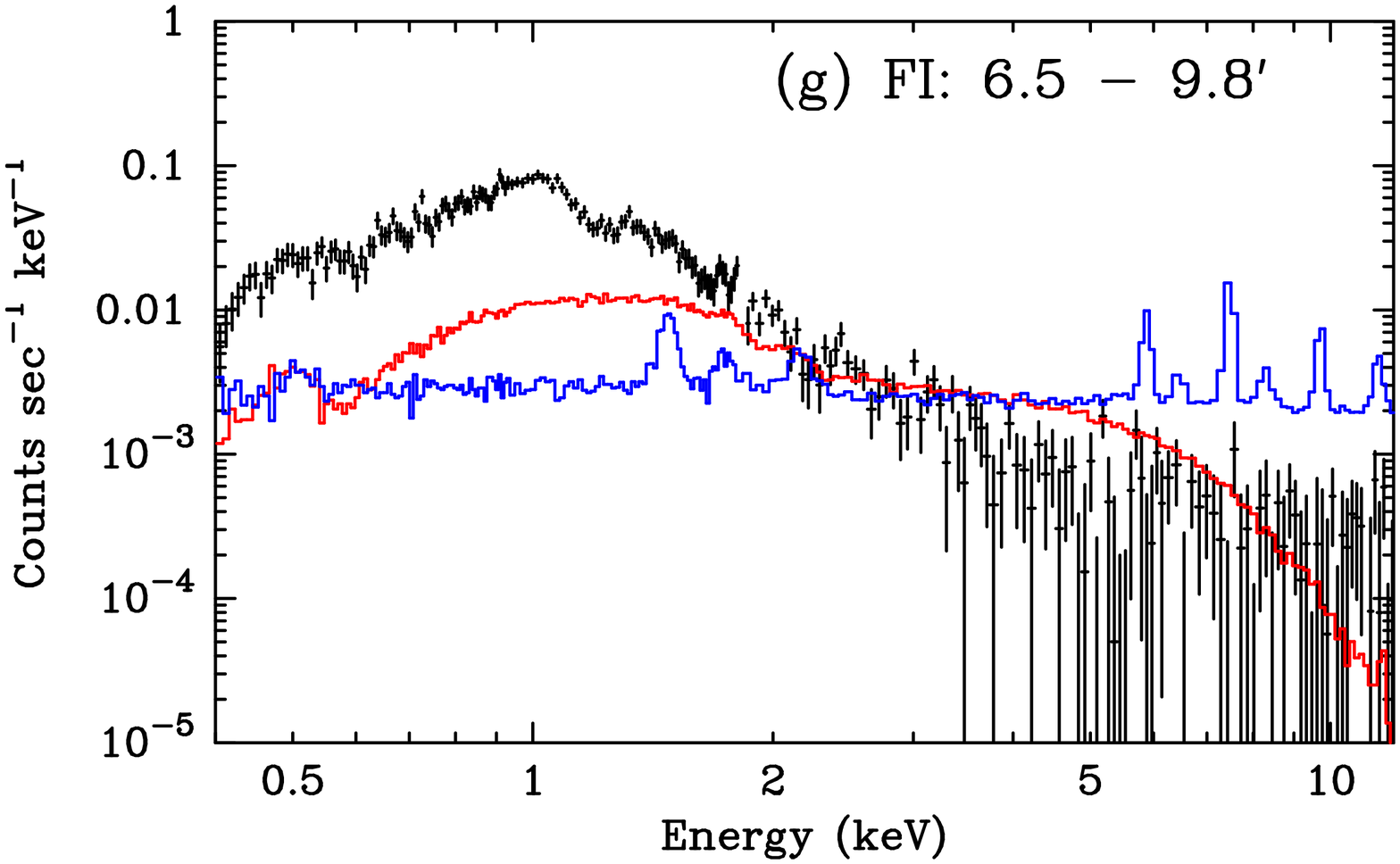}%
\FigureFile(0.25\textwidth,1cm){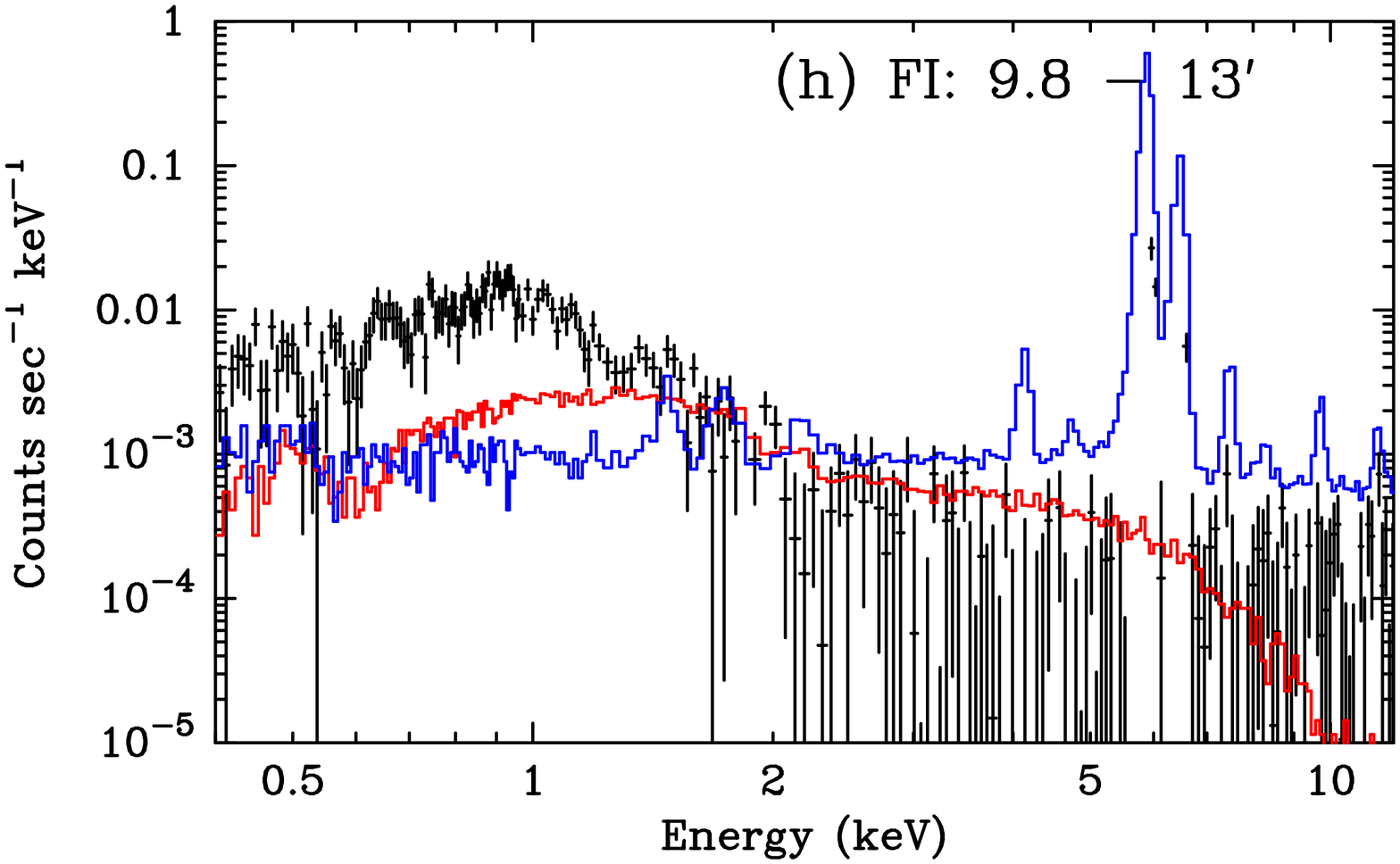}
\caption{
The observed spectra at the annular regions for (a)--(d) BI and
(e)--(h) FI sensors.
Estimated components of the NXB and CXB are subtracted,
which are indicated by blue and red histograms.
The ${\it COR} > 8$~GV screening is applied.
The ${}^{55}$Fe calibration source areas are included
for the accumulation regions of (d) and (h), but excluded for others.
}\label{fig:nxb_cxb}
\FigureFile(0.25\textwidth,1cm){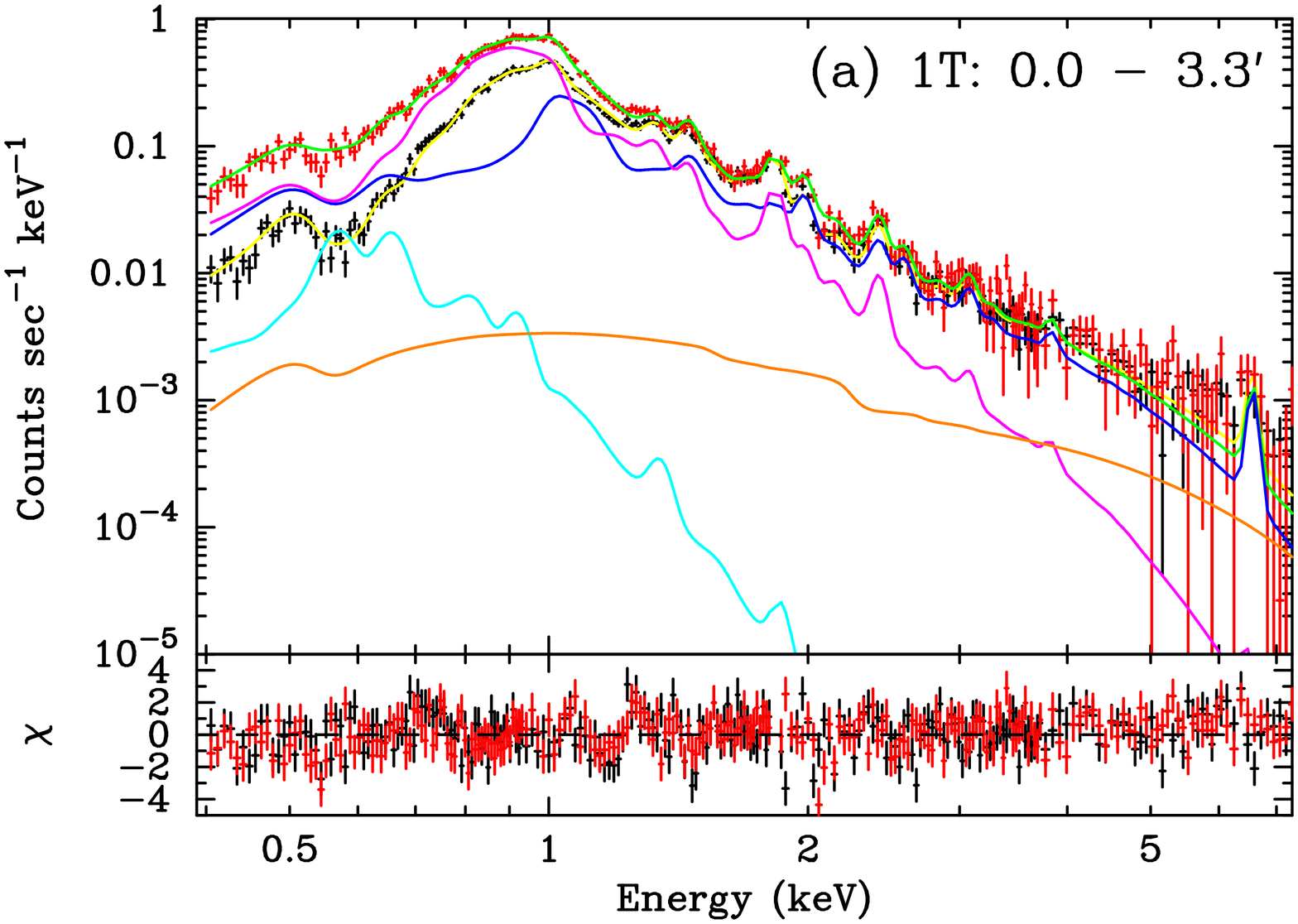}%
\FigureFile(0.25\textwidth,1cm){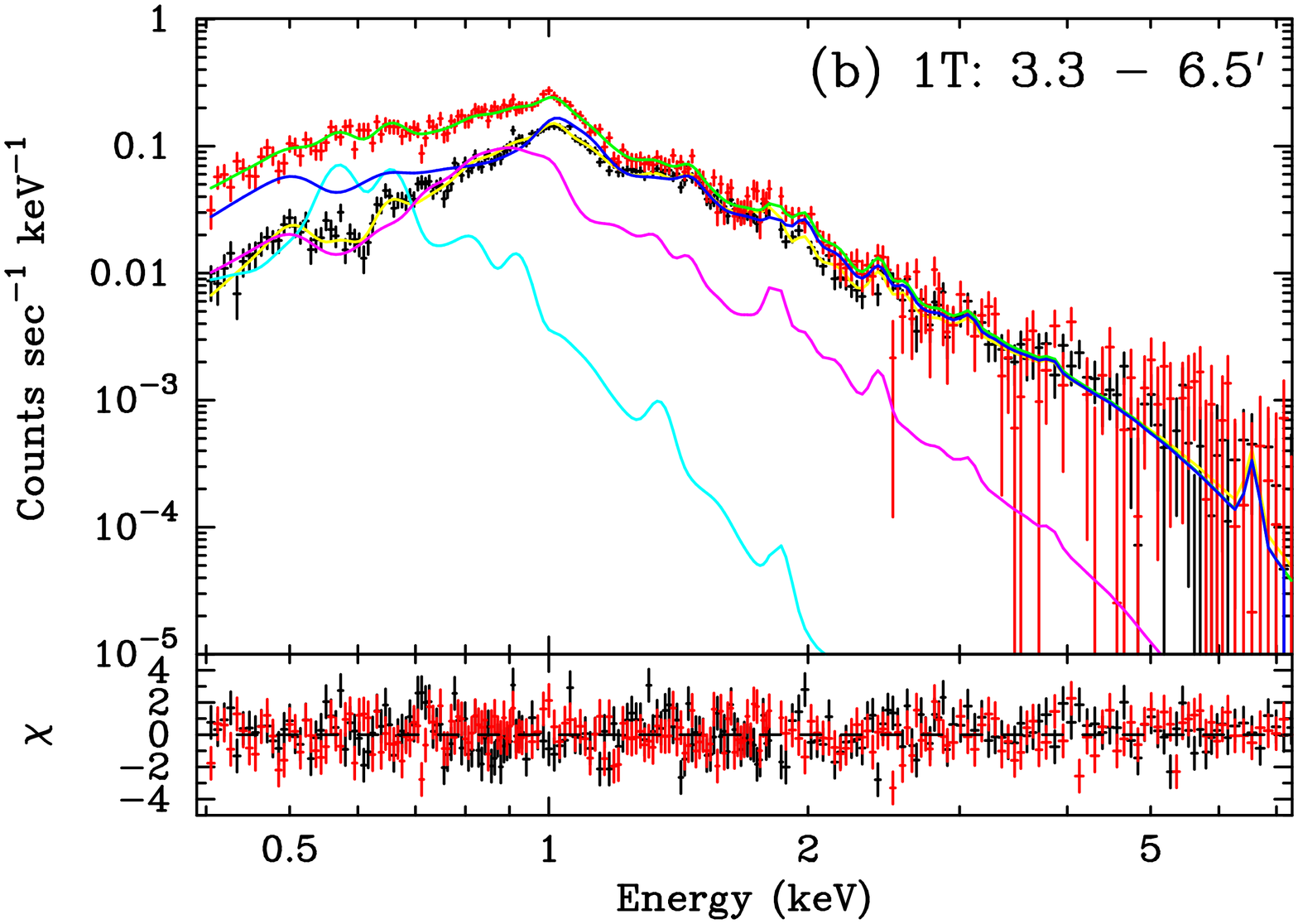}%
\FigureFile(0.25\textwidth,1cm){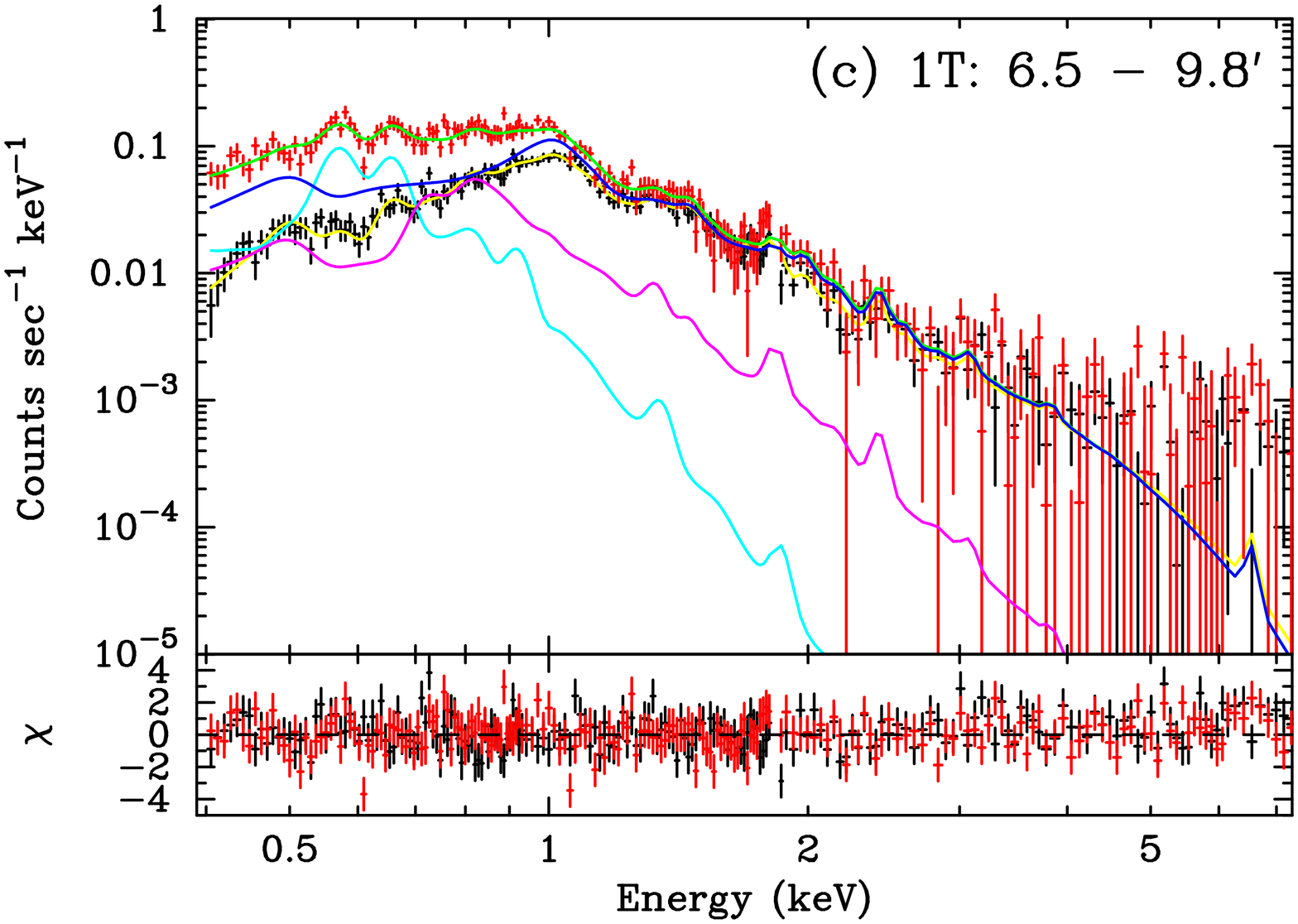}%
\FigureFile(0.25\textwidth,1cm){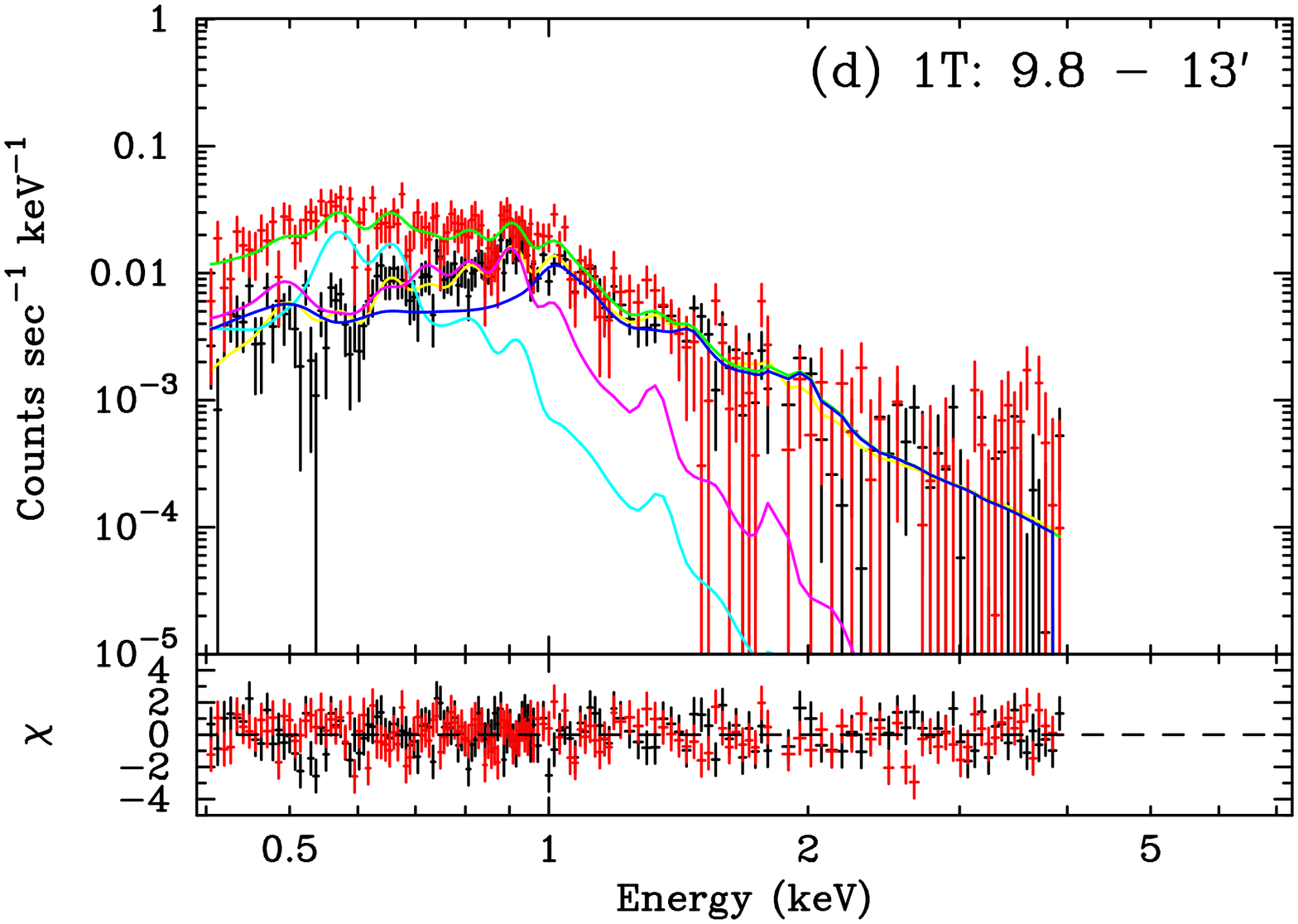}
\\[-1ex]
\FigureFile(0.25\textwidth,1cm){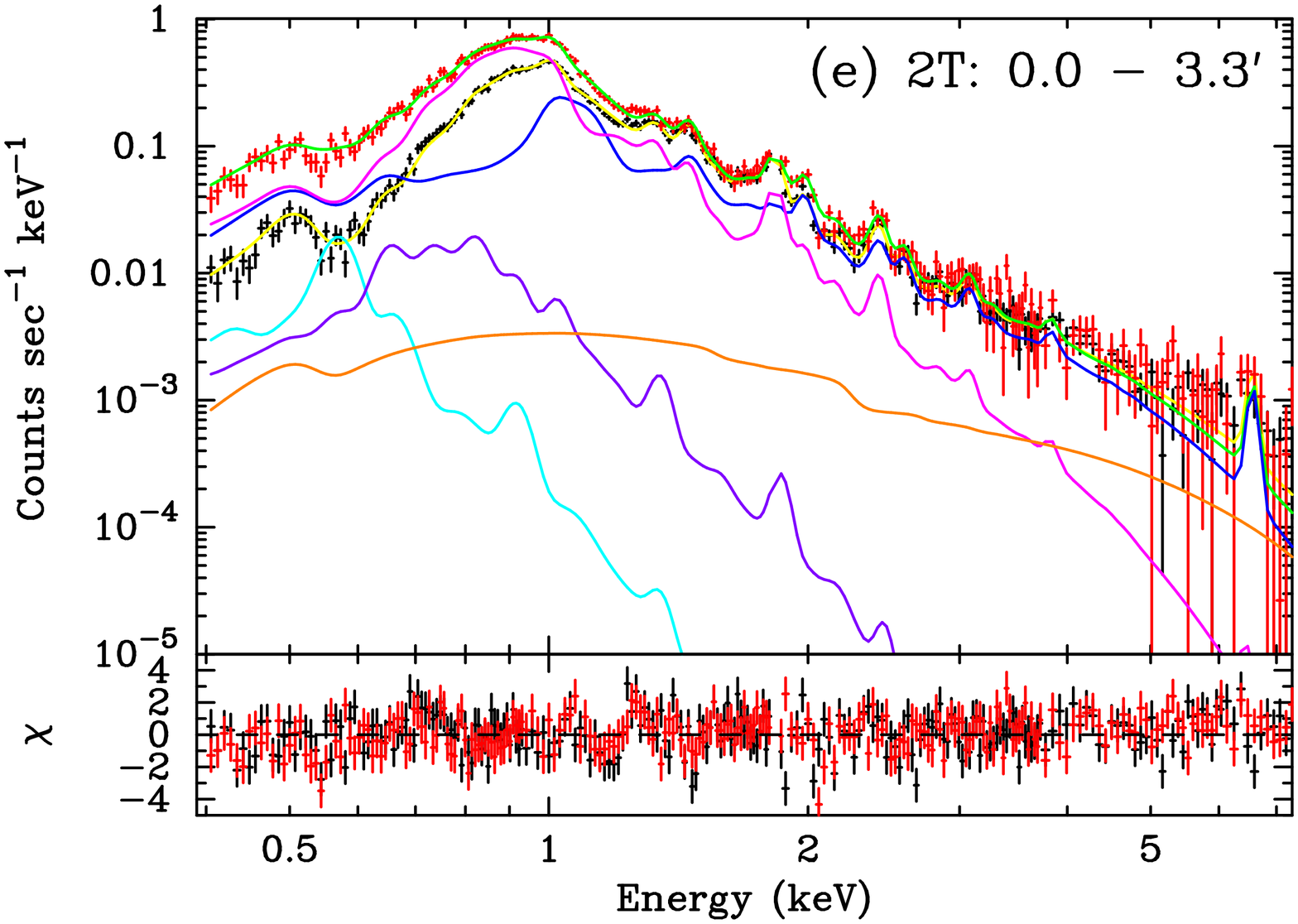}%
\FigureFile(0.25\textwidth,1cm){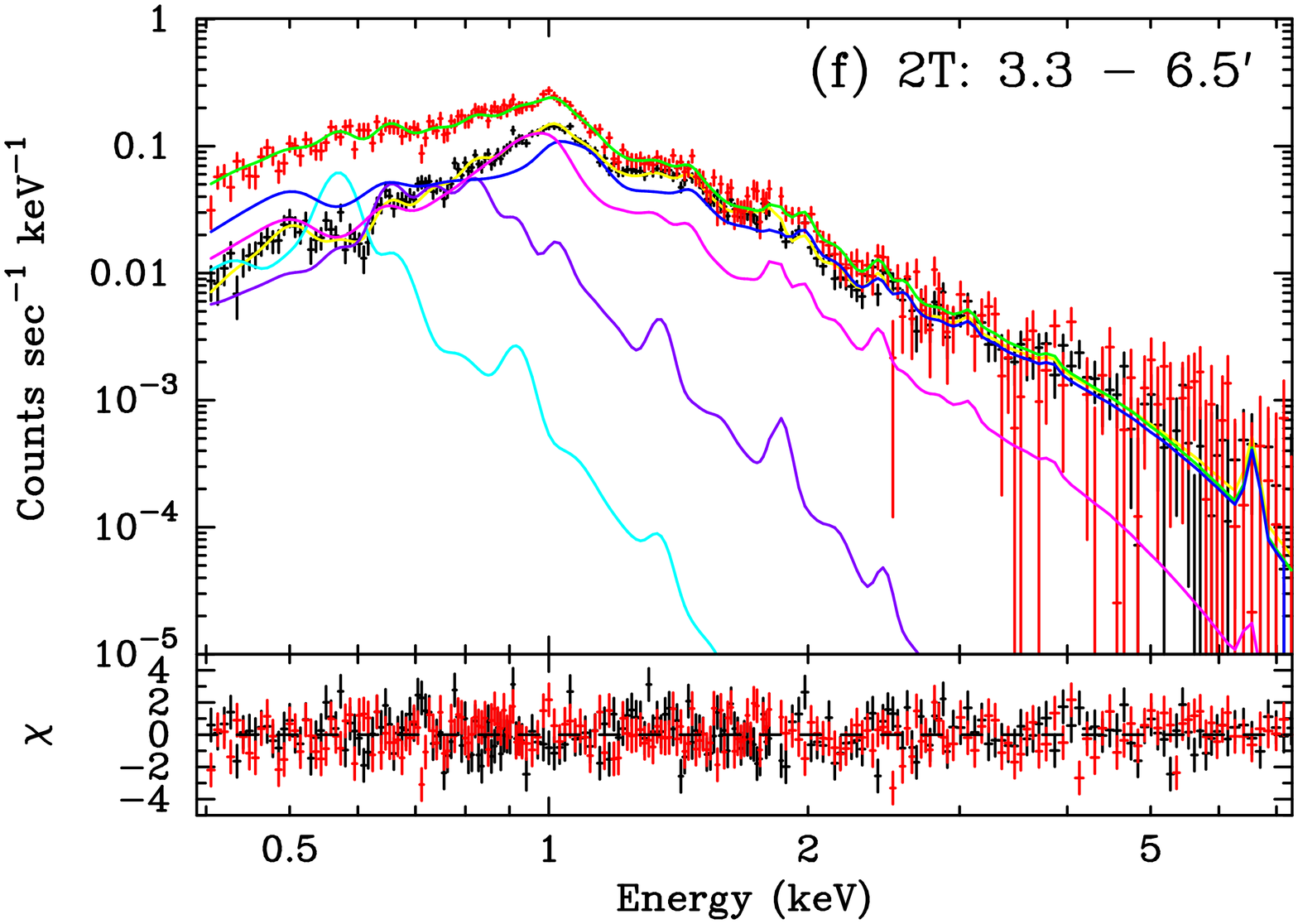}%
\FigureFile(0.25\textwidth,1cm){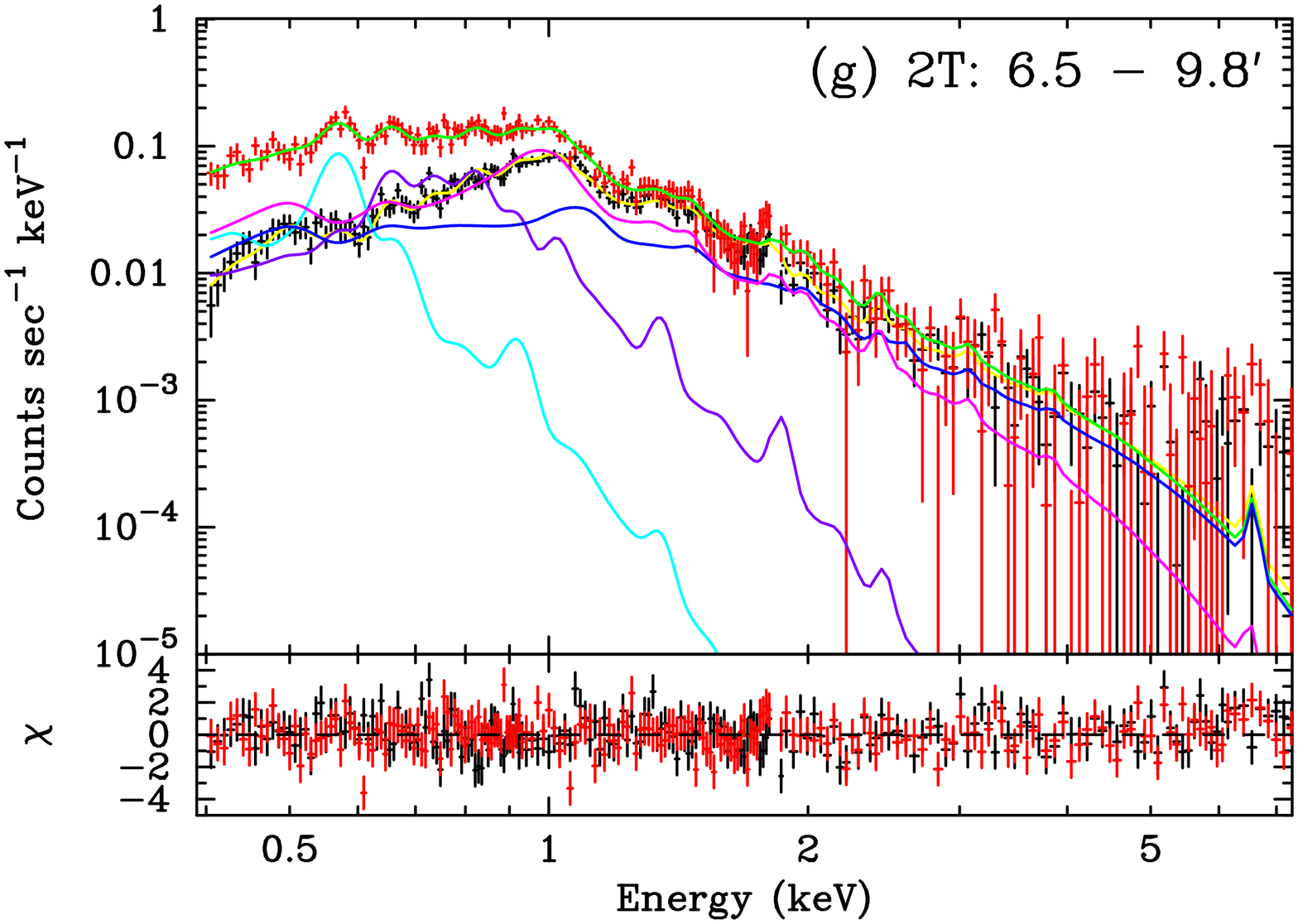}%
\FigureFile(0.25\textwidth,1cm){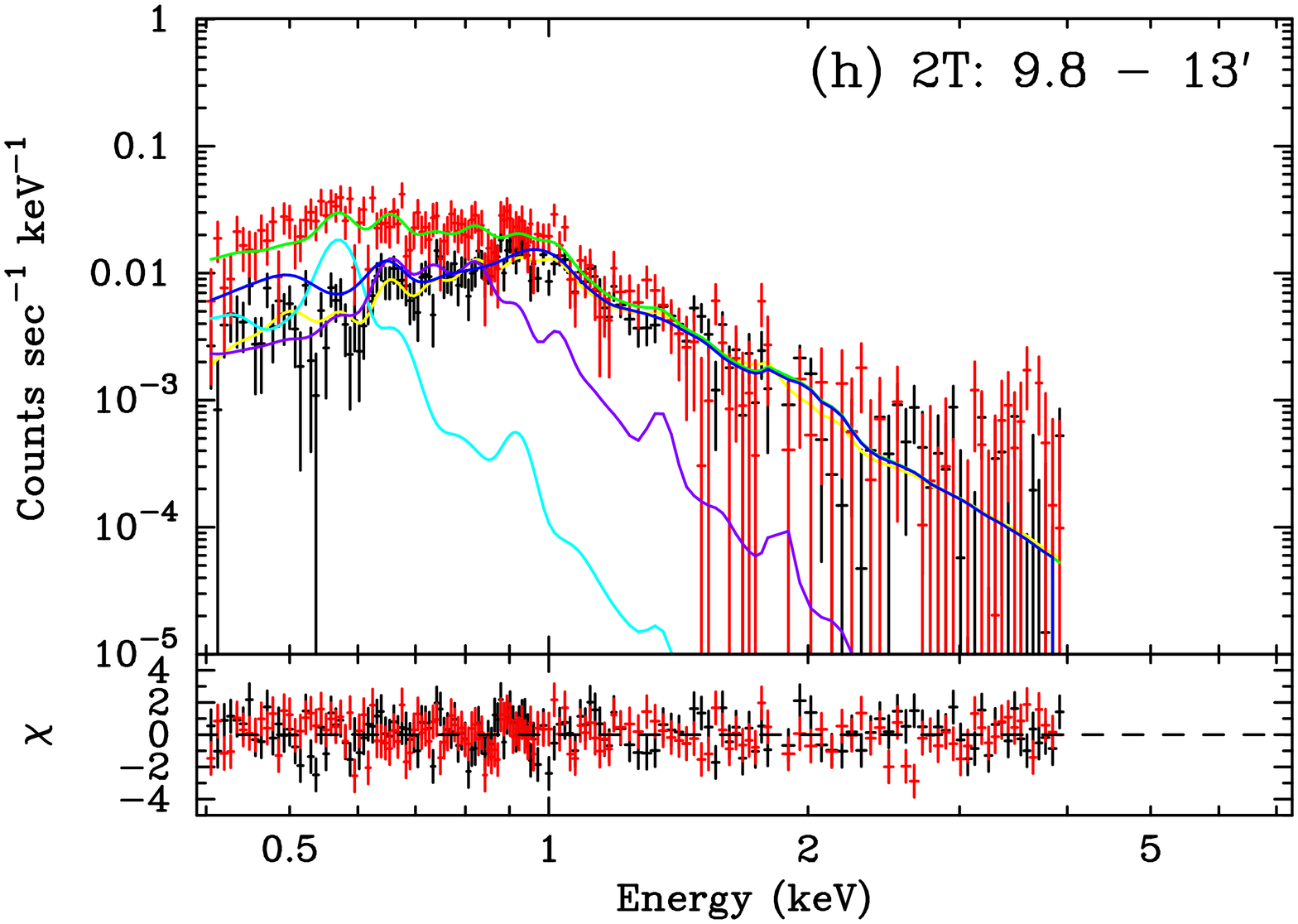}
\caption{
The observed spectra are plotted by red and black crosses
for BI and FI, respectively, and they are simultaneously fitted with
(a)--(d) Gal 1T: $constant\times [\,phabs\times (vapec + vapec) + apec\,]$, or
(e)--(h) Gal 2T: $constant\times [\,phabs\times (vapec + vapec) + apec+apec\,]$
model, drawn by green and yellow lines for the BI and FI spectra.
The hot and cool ICM components (two {\it vapec}) for the BI spectra
correspond to magenta and blue lines.
The {\it apec} component(s) for the BI spectra are indicated
by cyan (and purple) line(s).
The spectral fit was conducted in the energy range of 0.4--7.5~keV
for (a)--(c) and (e)--(g), and 0.4--4~keV for (d) and (h).
The energy range around the Si K-edge (1.825--1.840~keV) 
was ignored in the spectral fit.
The lower panels show the fit residuals in unit of $\sigma$.
}\label{fig:spectrum}
\end{figure*}

In figure~\ref{fig:nxb_cxb}, spectra of the back-illuminated sensor
(BI = XIS1) and sum of three front-illuminated sensors (FI = XIS0, XIS2, XIS3)
for four annular regions after the background subtraction are presented.
The estimated spectra of the CXB and NXB are overlaid.
Uncertainty in the CXB spectrum is $\sim 10$\% or more \citep{Kushino2002}
depending on the accumulation area of the spectrum,
and the reproducibility of the NXB is $\lesssim 5$\% for 100~ks
observation spanning 2--3 days \citep{Tawa2007}.
The ${}^{55}$Fe calibration source regions at two CCD corners
were included when we accumulated spectral data for the 9.8--13$'$ annulus,
however the energy range of the fit was limited to 0.4--4~keV\@.
The calibration source regions in the 6.5--9.8$'$ annulus were excluded,
and the fitting energy range was set at 0.4--7.5~keV for
0.0--3.3$'$, 3.3--6.5$'$, and 6.5--9.8$'$ annuli.
In addition, the energy range around the Si K-edge (1.825--1.840~keV) 
was ignored in the spectral fit.
Table~\ref{tab:region} summarizes area, coverage, energy range,
and the BI and FI counts for the observed spectra and
the estimated NXB and CXB spectra. The fraction of the background,
$f_{\rm BGD}\equiv\rm (NXB + CXB)/OBS$, was less than 50\%
even at the outermost annulus, although the Galactic component
is not considered here.

The response of the XRT and XIS were calculated by
``xissimarfgen'' version 2006-08-26 ancillary response file (ARF)
generator \citep{Ishisaki2007}
and ``xisrmfgen'' version 2006-10-26 response matrix file (RMF) generator.
Slight degradation of the energy resolution was considered in the RMF,
and decrease of the low energy transmission of the XIS optical blocking
filter (OBF) was included in the ARF\@.
The ARF response was calculated assuming a surface brightness profile, $S(r)$,
based on the combined analysis of Chandra and XMM-Newton data in 0.5--4~keV
within $r<14'$. The profile is described by a $3\beta$-model,
which is given in table~3 of \citet{Morita2006}.
Fraction of the assumed intensity from the corresponding sky region
for each annulus is presented in the {\scriptsize SOURCE\_RATIO\_REG} column
of table~\ref{tab:region}.
We also checked the fraction of events
coming from near-by regions other than the corresponding sky for each annulus
using the ``xissim'' simulator \citep{Ishisaki2007},
as shown in table~\ref{tab:stray}.
The stray fraction was not significant and less than 1/3 in the worst case.
The deprojection analysis was not conducted in this paper,
however, we have confirmed that this analysis would not change the best-fit 
parameters significantly.

\begin{table}
\caption{
Estimated fractions of the ICM photons accumulated
in the annular detector regions
coming from the corresponding sky by ``xissim'' simulation
for BI (XIS1) at 1~keV\@. These numbers are not much different
($\lesssim 1$\%) for other sensors and the examined energy bands.
}\label{tab:stray}
\centerline{
\begin{tabular}{lrrrrr}
\hline\hline
\raisebox{-0.25ex}{Detector} & \multicolumn{5}{c}{Sky Region} \\
\cline{2-6}\\[-2ex]
\multicolumn{1}{c}{\raisebox{0.25ex}{Region}} & \makebox[2.8em][c]{0.0--3.3$'$} & \makebox[2.8em][c]{3.3--6.5$'$} & \makebox[2.8em][c]{6.5--9.8$'$} & \makebox[2.8em][c]{9.8--13$'$} & \makebox[2.8em][c]{$r > 13'$} \\
\hline\\[-2ex]
0.0--3.3$'$ & 95.6\% &  4.4\% &  0.0\% &  0.0\% & 0.0\% \\
3.3--6.5$'$ & 17.2\% & 68.6\% & 14.1\% &  0.0\% & 0.0\% \\
6.5--9.8$'$ &  0.2\% &  8.2\% & 81.3\% & 10.2\% & 0.0\% \\
9.8--13$'$  &  0.0\% &  0.2\% &  8.8\% & 90.0\% & 1.0\% \\
\hline
\end{tabular}
}
\end{table}

\section{X-ray Image}\label{sec:image}

The XIS images in 0.5--4~keV and 6--10~keV
are shown in figure~\ref{fig:image}(a) and (b).
Estimated contributions of NXB and CXB were subtracted
and exposure was corrected, but no vignetting correction was applied.
Magnified image of the central region in 0.5--4~keV without
gaussian smoothing is also shown in figure~\ref{fig:arc}(a),
which will be discussed in section~\ref{sec:arc}.
There is a slight positional shift of
the X-ray peak by less than $\sim 0.3'$  from the region center,
due to the source position error of $19''$ corresponding to the 90\%
error circle radius of Suzaku telescopes after the astrometry correction
for the thermal distortion of the spacecraft \citep{Uchiyama2007}.
Though this astrometry correction was not incorporated in our analysis,
the shift of the average pointing
direction due to the thermal distortion is estimated to be 6.4$''$
which gives a minor effect.

The 0.5--4~keV image is dominated by the ICM emission,
however intensity gradient becomes weaker at the outer two annuli,
where the Galactic emission has comparable intensity.
In 6--10~keV, the thermal ICM emission can hardly be seen as shown in
figure~\ref{fig:image}(b), which is a linear-scale plot of the intensity.
However, some extended emission can be seen within $r<3.3'$.
It is notable that this emission seems elongated along the direction of
the two cavities (32$''$ north-east and 20$''$ south-west) which were observed
with Chandra \citep{Vrtilek2002,Morita2006}. There are
other extended hard emissions which do not match point sources
(white circles) as detected with XMM-Newton. The statistical significance
of the hard X-ray emission is discussed
in section~\ref{sec:hard-emission}.

Figure~\ref{fig:image}(c) shows a hardness ratio image based on the 1-2~keV
and 0.5--1~keV intensities. 
Although the vignetting correction was not performed,
the two energy bands show very similar vignetting features.
The transmission degradation due to the OBF contaminant is not corrected for either,
which is larger for the 0.5--1~keV band and at the central part of the CCD\@.
Therefore the hardness ratio shows a systematic drop from the center
to outer regions for a constant temperature distribution.
In the hardness image, there is a clear doughnut-like structure with
high hardness ratio around the annulus of 3.3--6.5$'$ .
This is due probably to a relative dominance of the high 
temperature component
in the ICM which is mostly occupied by the cool component
at the group core, as already reported by \citet{Morita2006}.
They also found a sharp temperature drop from $kT_{\rm Hot}\sim 1.5$~keV
at $r\sim 3'$ to $\sim 0.6$~keV at $r\sim 10'$,
which may well form the doughnut-like hot structure.
However, the Galactic component becomes comparable to the ICM
in the outer annuli, so the overall temperature structure needs to be 
examined in detail.

\begin{table*}[ht]
\caption{
Summary of the best-fit parameters for single (Gal 1T) or two (Gal 2T)
temperature Galactic component models.
}\label{tab:best-fit}
\begin{tabular}{lccccccccc}
\hline\hline\\[-2.2ex]
Gal 1T &$kT_{\rm Hot}$ &$kT_{\rm Cool}$ &O &Ne &Mg, Al &Si & \makebox[0in][c]{S, Ar, Ca} &Fe, Ni &$\chi^{2}$/dof  \\
    &(keV) & (keV) & (solar) & (solar) & (solar) & (solar) & (solar) & (solar)&\\
\hline\\[-2.2ex]
0.0--3.3$'$	&$1.65^{+0.03}_{-0.03}$  &$0.766^{+0.007}_{-0.007}$ 
&$0.32^{+0.08}_{-0.07}$  &$1.66^{+0.24}_{-0.22}$  &$1.24^{+0.14}_{-0.12}$     
&$0.97^{+0.07}_{-0.07}$  &$1.12^{+0.13}_{-0.12}$  &$0.82^{+0.06}_{-0.05}$ 
& 654/473 \\
3.3--6.5$'$	&$1.48^{+0.06}_{-0.06}$  &$0.753^{+0.027}_{-0.031}$ 
&$0.10^{+0.13}_{-0.10}$  &$0.55^{+0.18}_{-0.17}$  &$0.36^{+0.11}_{-0.10}$     
&$0.32^{+0.07}_{-0.06}$  &$0.38^{+0.10}_{-0.10}$  &$0.26^{+0.03}_{-0.03}$ 
& 512/421 \\
6.5--9.8$'$	&$1.23^{+0.06}_{-0.05}$  &$0.588^{+0.048}_{-0.062}$ 
&$0.00^{+0.10}_{-0.00}$  &$0.04^{+0.16}_{-0.04}$  &$0.22^{+0.10}_{-0.09}$     
&$0.20^{+0.07}_{-0.07}$  &$0.41^{+0.14}_{-0.13}$  &$0.17^{+0.02}_{-0.02}$ 
& 467/405 \\
9.8--13$'$	&$1.43^{+0.20}_{-0.17}$  &$0.301^{+0.060}_{-0.043}$ 
&$0.07^{+0.08}_{-0.06}$  &$0.72^{+0.34}_{-0.26}$  &$0.41^{+0.56}_{-0.39}$     
&$0.39^{+0.51}_{-0.39}$  &$0.00^{+0.54}_{-0.00}$  &$0.25^{+0.19}_{-0.12}$ 
& 301/285 \\
\multicolumn{1}{c}{All} &&&&&&&&& \makebox[0in][c]{1934/1584}\\
\hline\hline\\[-2.2ex]
Gal 2T &$kT_{\rm Hot}$  &$kT_{\rm Cool}$ &O &Ne & Mg, Al & Si & \makebox[0in][c]{S, Ar, Ca} &Fe, Ni &$\chi^{2}$ /dof\\
    &(keV) & (keV) & (solar) & (solar) & (solar) & (solar) & (solar) & (solar)&\\
\hline\\[-2.2ex]
0.0--3.3$'$	&$1.66^{+0.03}_{-0.03}$  &$0.773^{+0.007}_{-0.007}$ 
&$0.35^{+0.08}_{-0.08}$  &$1.71^{+0.25}_{-0.23}$  &$1.29^{+0.14}_{-0.13}$     
&$1.01^{+0.10}_{-0.09}$  &$1.16^{+0.13}_{-0.12}$  &$0.84^{+0.06}_{-0.06}$ 
& 657/473 \\
3.3--6.5$'$	&$1.63^{+0.17}_{-0.12}$  &$0.990^{+0.060}_{-0.114}$ 
&$0.23^{+0.14}_{-0.13}$  &$0.36^{+0.24}_{-0.25}$  &$0.39^{+0.13}_{-0.11}$     
&$0.35^{+0.08}_{-0.07}$  &$0.38^{+0.11}_{-0.10}$  &$0.29^{+0.04}_{-0.04}$ 
& 500/421 \\
6.5--9.8$'$	&$1.81^{+0.28}_{-0.31}$  &$1.047^{+0.036}_{-0.074}$ 
&$0.15^{+0.17}_{-0.15}$  &$0.00^{+0.15}_{-0.00}$  &$0.25^{+0.15}_{-0.13}$     
&$0.20^{+0.09}_{-0.08}$  &$0.36^{+0.16}_{-0.16}$  &$0.20^{+0.05}_{-0.05}$ 
& 451/405 \\
9.8--13$'$	&$1.02^{+0.06}_{-0.06}$  &---------
&$0.38^{+0.38}_{-0.30}$  &$0.00^{+0.26}_{-0.00}$  &$0.00^{+0.20}_{-0.00}$     
&$0.11^{+0.22}_{-0.11}$  &$0.00^{+0.45}_{-0.00}$  &$0.11^{+0.04}_{-0.03}$ 
& 311/287 \\
\multicolumn{1}{c}{All} &&&&&&&&& \makebox[0in][c]{1919/1586}\\
\end{tabular}


\begin{tabular}{lccrrccrrcccc}
\hline\hline\\[-2.2ex]
Gal 1T & $kT_{\rm Gal}$ && ${\it K}_{\rm Hot}{}^\ast$ & $K_{\rm Cool}{}^\ast$ & $K_{\rm Gal}{}^\dagger$ && \makebox[1.85em][r]{$S_{\rm Hot}$\makebox[0in][l]{$^\ddagger$}} & \makebox[1.85em][r]{$S_{\rm Cool}$\makebox[0in][l]{$^\ddagger$}} & \makebox[1.85em][c]{$S_{\rm Gal}$\makebox[0in][l]{$^\ddagger$}} && \makebox[1.85em][r]{$S_{\rm CXB}$\makebox[0in][l]{$^\ddagger$}} & \makebox[1.85em][c]{$S_{\rm LMXB}$\makebox[0in][l]{$^\ddagger$}} \\
 & (keV) & & & & & & \\
\hline\\[-2.2ex]
0.0--3.3$'$ & $\downarrow$              && $32.6_{-1.6}^{+1.6}$ & $30.3_{-1.9}^{+1.9}$ & $\downarrow$        && 23.3 & 43.5 & $\downarrow$ && $\downarrow$ & 0.9 \\
3.3--6.5$'$ & $\downarrow$              && $11.9_{-0.8}^{+0.8}$ & $ 4.1_{-0.4}^{+0.5}$ & $\downarrow$        &&  6.2 &  2.7 & $\downarrow$ && $\downarrow$ & --- \\
6.5--9.8$'$ & $\downarrow$              && $ 9.6_{-0.9}^{+0.8}$ & $ 3.1_{-0.5}^{+0.6}$ & $\downarrow$        &&  4.5 &  1.4 & $\downarrow$ && $\downarrow$ & --- \\
9.8--13$'$  & \makebox[0in][c]{$0.208_{-0.011}^{+0.007}$} && $ 4.4_{-0.9}^{+1.0}$ & $ 5.6_{-1.4}^{+1.5}$ & $1.2_{-0.1}^{+0.1}$ &&  2.3 &  2.2 & 1.8 && 1.6 & --- \\
\hline\hline\\[-2.2ex]
Gal 2T & $kT_{\rm G1}$ & $kT_{\rm G2}$ & $K_{\rm Hot}{}^\ast$ & $K_{\rm Cool}{}^\ast$ & $K_{\rm G1}{}^\dagger$ & $K_{\rm G2}{}^\dagger$ & \makebox[1.85em][r]{$S_{\rm Hot}$\makebox[0in][l]{$^\ddagger$}} & \makebox[1.85em][r]{$S_{\rm Cool}$\makebox[0in][l]{$^\ddagger$}} & \makebox[1.85em][c]{$S_{\rm G1}$\makebox[0in][l]{$^\ddagger$}} & \makebox[1.85em][c]{$S_{\rm G2}$\makebox[0in][l]{$^\ddagger$}} & \makebox[1.85em][r]{$S_{\rm CXB}$\makebox[0in][l]{$^\ddagger$}} & \makebox[1.85em][c]{$S_{\rm LMXB}$\makebox[0in][l]{$^\ddagger$}} \\
 & (keV) & (keV) & & & & & \\
\hline\\[-2.2ex]
0.0--3.3$'$ & $\downarrow$ & $\downarrow$ & $31.8_{-1.6}^{+1.6}$ & $29.2_{-1.9}^{+1.9}$ & $\downarrow$        & $\downarrow$        & 22.9 & 43.0 & $\downarrow$ & $\downarrow$ & $\downarrow$ & 0.9 \\
3.3--6.5$'$ & $\downarrow$ & $\downarrow$ & $ 9.1_{-1.7}^{+1.8}$ & $ 5.3_{-2.0}^{+2.1}$ & $\downarrow$        & $\downarrow$        &  4.8 &  3.3 & $\downarrow$ & $\downarrow$ & $\downarrow$ & --- \\
6.5--9.8$'$ & $\downarrow$ & $\downarrow$ & $ 4.0_{-0.8}^{+1.9}$ & $ 5.9_{-1.9}^{+1.9}$ & $\downarrow$        & $\downarrow$        &  1.9 &  3.1 & $\downarrow$ & $\downarrow$ & $\downarrow$ & --- \\
9.8--13$'$  & \makebox[4em][c]{0.138 (fix)} & \makebox[4em][c]{0.344 (fix)} & ---------            & $ 7.5_{-1.2}^{+1.3}$ & $1.3^{+0.1}_{-0.1}$ & $0.9_{-0.1}^{+0.1}$ &  --- &  3.5 & 1.2 & 1.5 & 1.6 & --- \\
\hline
\end{tabular}

\medskip
\parbox{\textwidth}{\footnotesize
\footnotemark[$*$] Normalization of the {\it vapec} component
scaled with a factor of {\scriptsize SOURCE\_RATIO\_REG} / {\scriptsize AREA}
in table~\ref{tab:region},\\
${\it K}=\frac{\makebox{\scriptsize\rm SOURCE\_RATIO\_REG}}
{\makebox{\scriptsize\rm AREA}}
\int n_{\rm e} n_{\rm H} dV \,/\, (4\pi\, (1+z)^2 D_{\rm A}^{\,2})$
$\times 10^{-20}$~cm$^{-5}$~arcmin$^{-2}$,
where $D_{\rm A}$ is the angular distance to the source.

\footnotemark[$\dagger$] Normalization of the {\it apec} component
divided by the solid angle,
$\Omega^{\makebox{\tiny U}} = \pi\times (20')^2$,
assumed in the uniform-sky ARF calculation
(20$'$ radius from the optical axis of each XIS sensor),
${\it K} = \int n_{\rm e} n_{\rm H} dV \,/\,
(4\pi\, (1+z)^2 D_{\rm A}^{\,2}) \,/\, \Omega^{\makebox{\tiny U}}$
$\times 10^{-20}$ cm$^{-5}$~arcmin$^{-2}$.

\footnotemark[$\ddagger$] Surface brightness in 0.4--4~keV
in unit of photons~cm$^{-2}$~Ms$^{-1}$~arcmin$^{-2}$,
calculated in the same way as the normalization,
and corrected for the Galactic absorption of
$N_{\rm H} = 3.03 \times 10^{20}$~cm$^{-2}$.
}
\end{table*}

\section{Spectral Analysis}\label{sec:spec}

We have carried out spectral fits to the observed spectra
in the four annular regions.
The estimated NXB and CXB were subtracted from the spectra,
but we estimate the parameter error ranges by adjusting
these background intensities by $\pm 10$\%. The fitting model consists
of the following components.

\smallskip\noindent
1. Galactic Hot Gas: we tried both single (Gal 1T; $kT_{\rm Gal}$) and
two temperature (Gal 2T; $kT_{\rm G1}$, $kT_{\rm G2}$) models
for this component.
In both models, surface brightness were constrained to be equal
among the four annuli, and {\it apec} models with solar
abundance at $z=0$ were utilized.
When single temperature was assumed,
the common temperature over the four annuli and
its surface brightness were varied as two free parameters.
In the two temperature case, we had to fix
both the temperatures because of strong coupling with the ICM temperatures.
Based on the previous reports (e.g., \cite{Lumb2002,Kuntz2000,Snowden1997}),
we assumed $kT_{\rm G1} = 0.138$~keV attributed to the Local Hot Bubble
and $kT_{\rm G2} = 0.344$~keV as the Milky-Way Halo, respectively. 
The intensities of these two components were varied as free parameters.
Though these temperatures are slightly higher than those obtained with
ROSAT \citep{Kuntz2000} or XMM-Newton \citep{Lumb2002},
we chose the values which gave the largest difference from
the Gal 1T results within the ROSAT error range. We also note that
the location of HCG~62, ($l$, $b$) = (303$^\circ$, 54$^\circ$) 
in the Galactic coordinate,
is within $\sim 15^\circ$ 
from the Galactic Bulge and the North Polar Spur -- Loop~I
\citep{Snowden1995}. Influence of the assumed temperatures
on the ICM abundance is examined in section~\ref{sec:abun}.

\smallskip\noindent
2. Low-Mass X-ray Binaries (LMXBs): the discrete source contribution in
member galaxies of HCG~62 was included only for 0.0--3.3$'$ annulus,
assuming thermal bremsstrahlung emission, {\it zbrems}, at $z=0.0145$
with a fixed temperature of $kT=7$ keV and fixed intensity.
The intensity, $F_{\rm X} = 9.3\times 10^{-14}$ erg~cm$^{-2}$~s$^{-1}$
(0.3--8 keV), employed here is based on the Chandra results
by \citet{Kim2004}, in which the integrated LMXB luminosity,
$L_{\rm X}$, was related to the $B$ band optical luminosity, $L_B$, as
\begin{equation}\label{eq:lmxb}
L_{\rm X} / L_B = (0.9\pm 0.5)\times 10^{30}~\makebox{erg~s$^{-1}$} 
 / L_{B\odot}.
\end{equation}

\smallskip\noindent
3. ICM in HCG~62: we assumed two temperature
($kT_{\rm Hot}$, $kT_{\rm Cool}$)
optically-thin thermal plasma model, {\it vapec}, at $z=0.0145$ for ICM 
based on the previous Chandra and XMM-Newton studies \citep{Morita2006}.
As an exception, only single temperature was considered
in the Gal 2T model for the outermost annulus (9.8--13$'$),
because we could not constrain $kT_{\rm Hot}$.
As for the metal abundance, we grouped relevant metals into six groups,
namely, Mg and Al; S, Ar and Ca; Fe and Ni
were combined, and  O, Ne and Si were varied individually.

\smallskip
Consequently,
Gal 1T: $constant\times [\,phabs\times (vapec + vapec) + apec\,]$, or
Gal 2T: $constant\times [\,phabs\times (vapec + vapec) + apec + apec\,]$
model was adopted for the spectral fit.
The {\it phabs}\/ factor denotes the Galactic absorption due to
neutral atoms, and the hydrogen column density was fixed to
$N_{\rm H} = 3.03\times 10^{20}$~cm$^{-2}$ \citep{Dickey1990}.
Note that this absorption model is also affected by the assumed solar
abundance table as discussed in Appendix~2 of \citet{Sato2007}.
In order to constrain the surface brightness of the Galactic component,
a total of 8 spectra, with BI and FI sensors for the four annuli,
were simultaneously fitted. The {\it constant}\/ factor was introduced
to compensate the possible flux discrepancy between the BI and FI sensors,
however, it turned out to be unity within $\pm 5$\%.

The fitted spectra are presented in figure~\ref{fig:spectrum}, and
the best-fit parameters are summarized in table~\ref{tab:best-fit}.
Flux and normalization in the table are for the FI spectra.
The reduced $\chi^2$ values are $\simeq 1.2$ for the two models,
and slightly better for the Gal 2T model with $\Delta\chi^2 = 15$.
Both models are statistically rejected if we consider only the 
statistical errors,
however, they become acceptable when systematic error of 5\%
is added in each bin. The reduced $\chi^2$ is particularly
large in the innermost region ($r < 3.3'$),
where temperature gradient was observed with Chandra \citep{Morita2006}.
The residuals of the fit show large scatter around 1--1.5~keV in
figure~\ref{fig:spectrum}(a) and (e). This is presumably because
the Fe-L line complex of multi-temperature ICM is
not well-fitted by the simple two temperature {\it vapec} model.
We will further describe the results on temperature and
metal abundance separately.

\begin{figure*}[tbg]
\FigureFile(0.32\textwidth,1cm){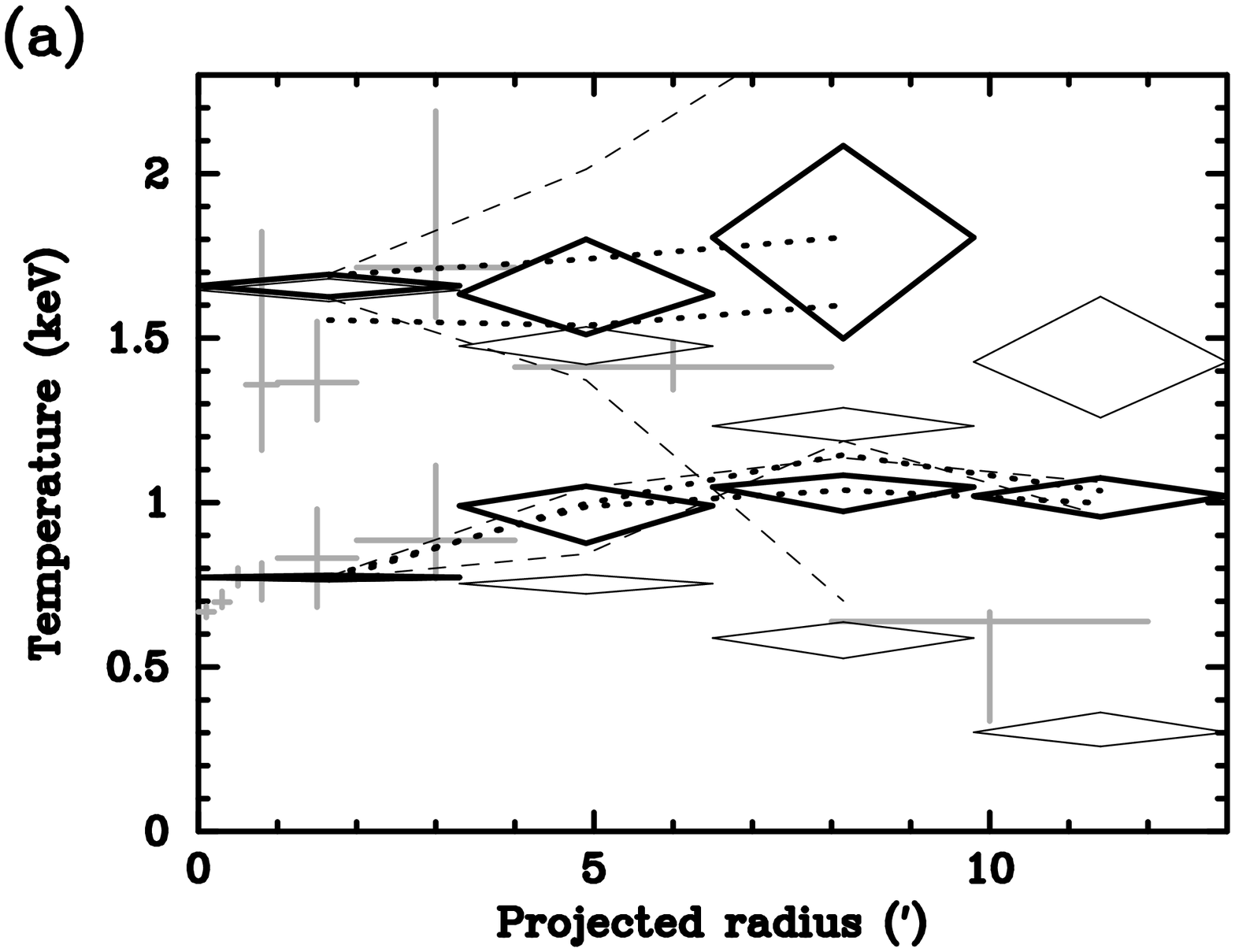}\hfill
\FigureFile(0.32\textwidth,1cm){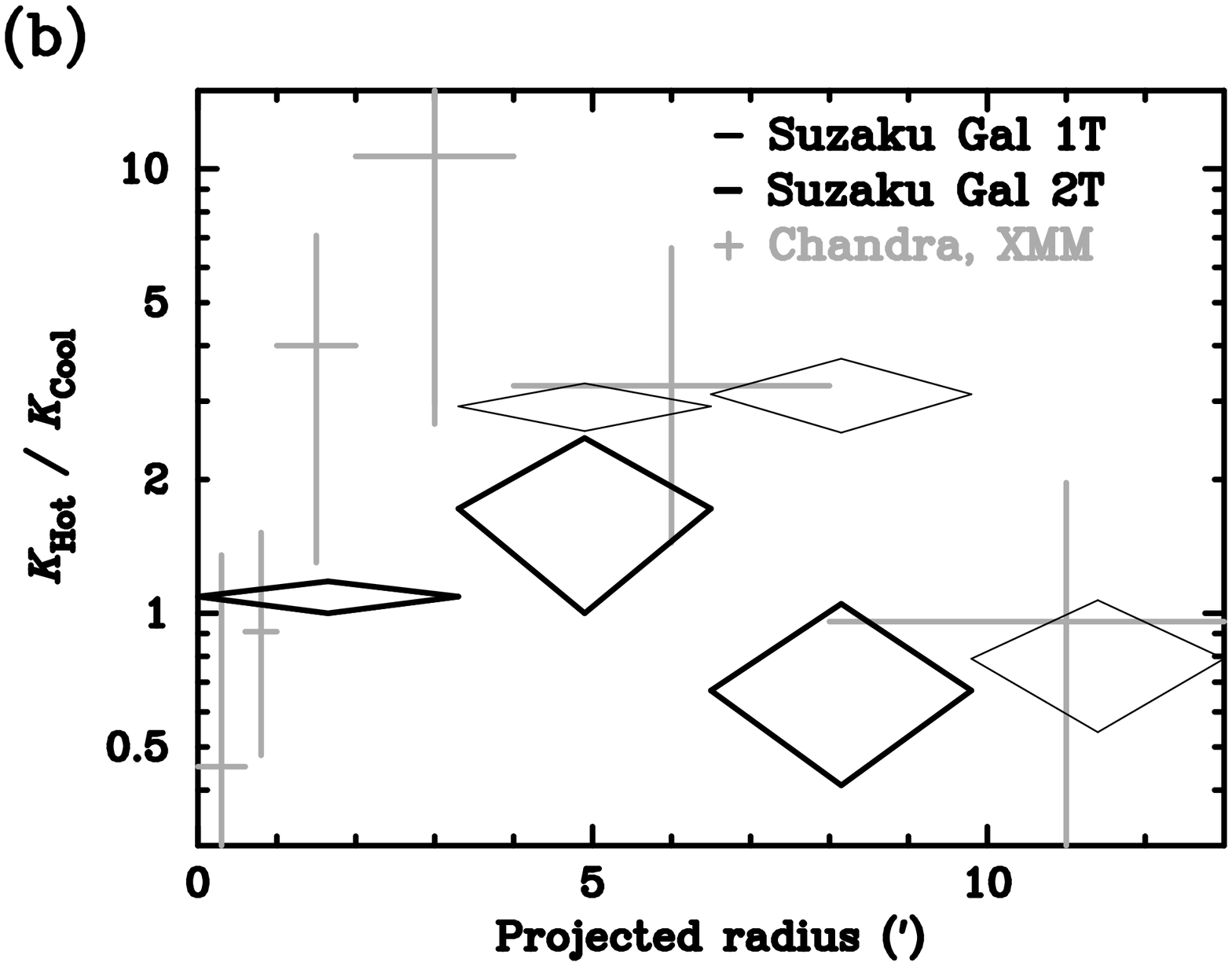}\hfill
\FigureFile(0.32\textwidth,1cm){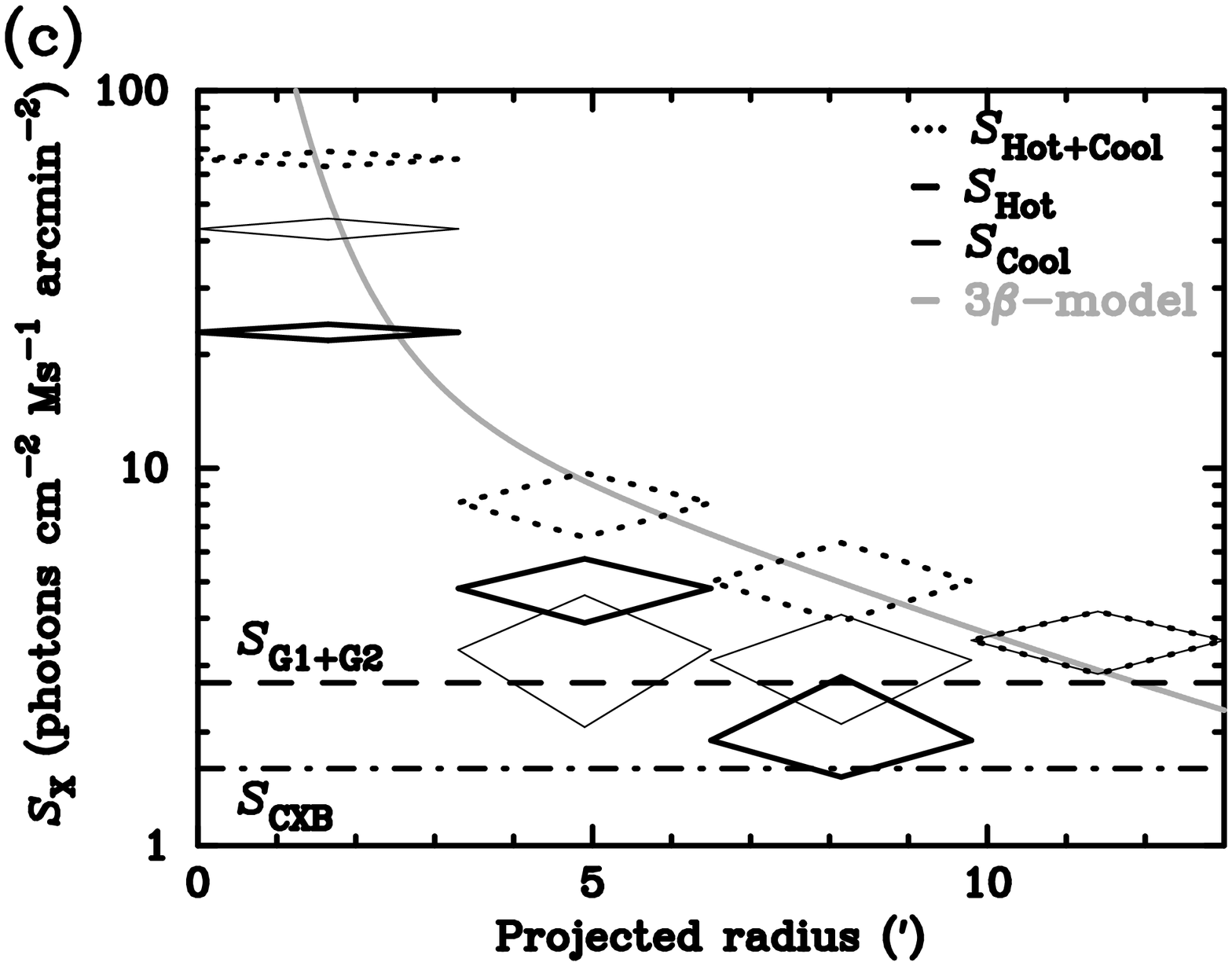}
\caption{
(a) Temperatures of the ICM components, $kT_{\rm Hot}$ and $kT_{\rm Cool}$,
(b) ratio of the hot ICM component normalization to the cool ICM,
$K_{\rm Hot}\, /\, K_{\rm Cool}$, and
(c) surface brightness profiles corrected for the Galactic absorption,
are plotted against radii from the group center.
Diamonds in thin lines represent the Suzaku best-fit results
with the Gal 1T model and diamonds in thick lines represents
the results with the Gal 2T model in (a) and (b).
Only the result with the Gal 2T model is presented in (c).
Dashed lines in (a) denote the movable range of the best-fit temperature
when the NXB + CXB components are changed by $\pm 10$\%,
and dotted lines denote the range when the absorption column of
the XIS contaminant is changed by $\pm 20$\%.
Gray crosses in (a), (b) and the gray line ($3\beta$-model) in (c) are adopted
from the previous results with Chandra and XMM-Newton by \citet{Morita2006}.
}\label{fig:kT}
\FigureFile(0.32\textwidth,1cm){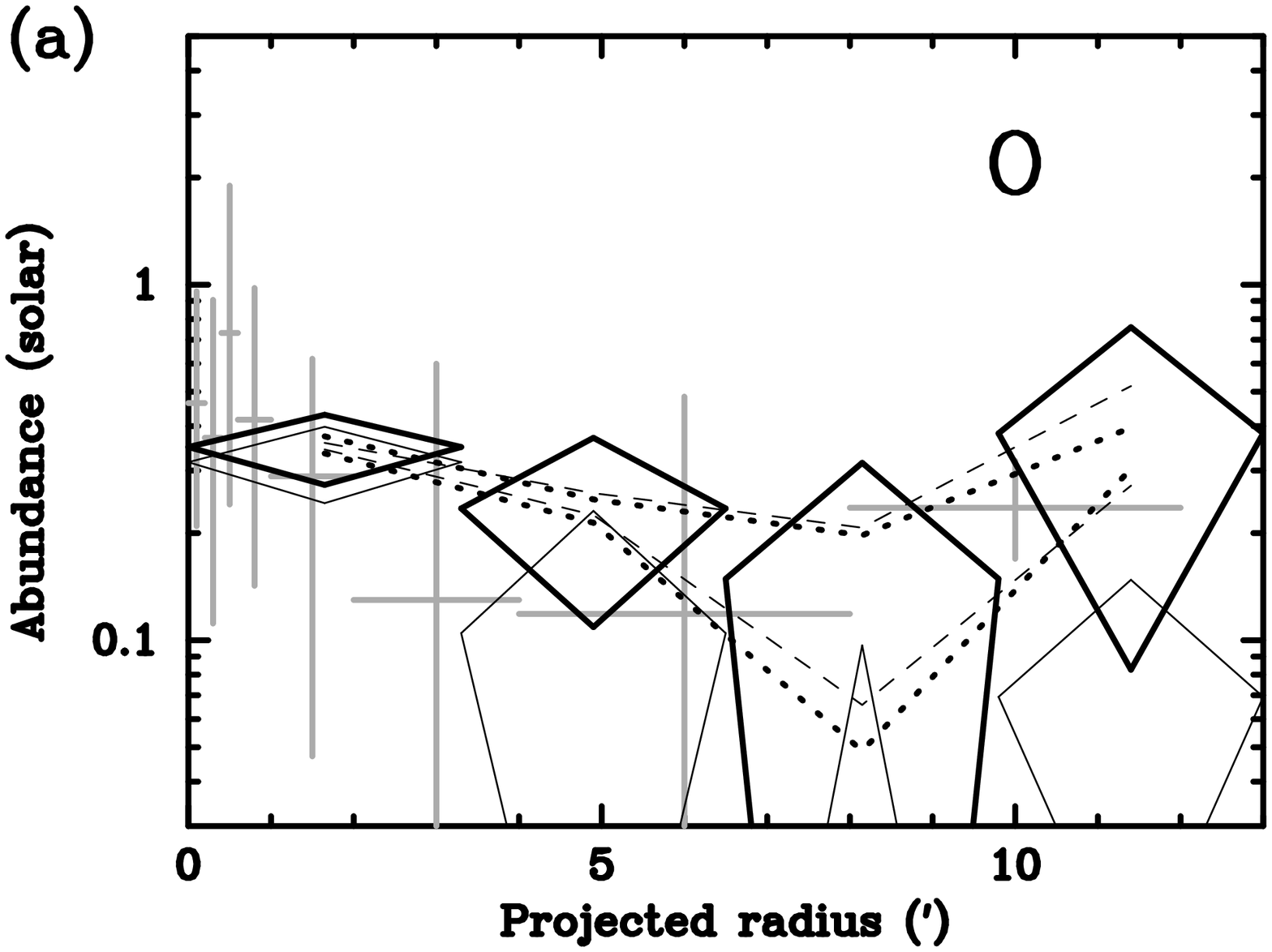}\hfill
\FigureFile(0.32\textwidth,1cm){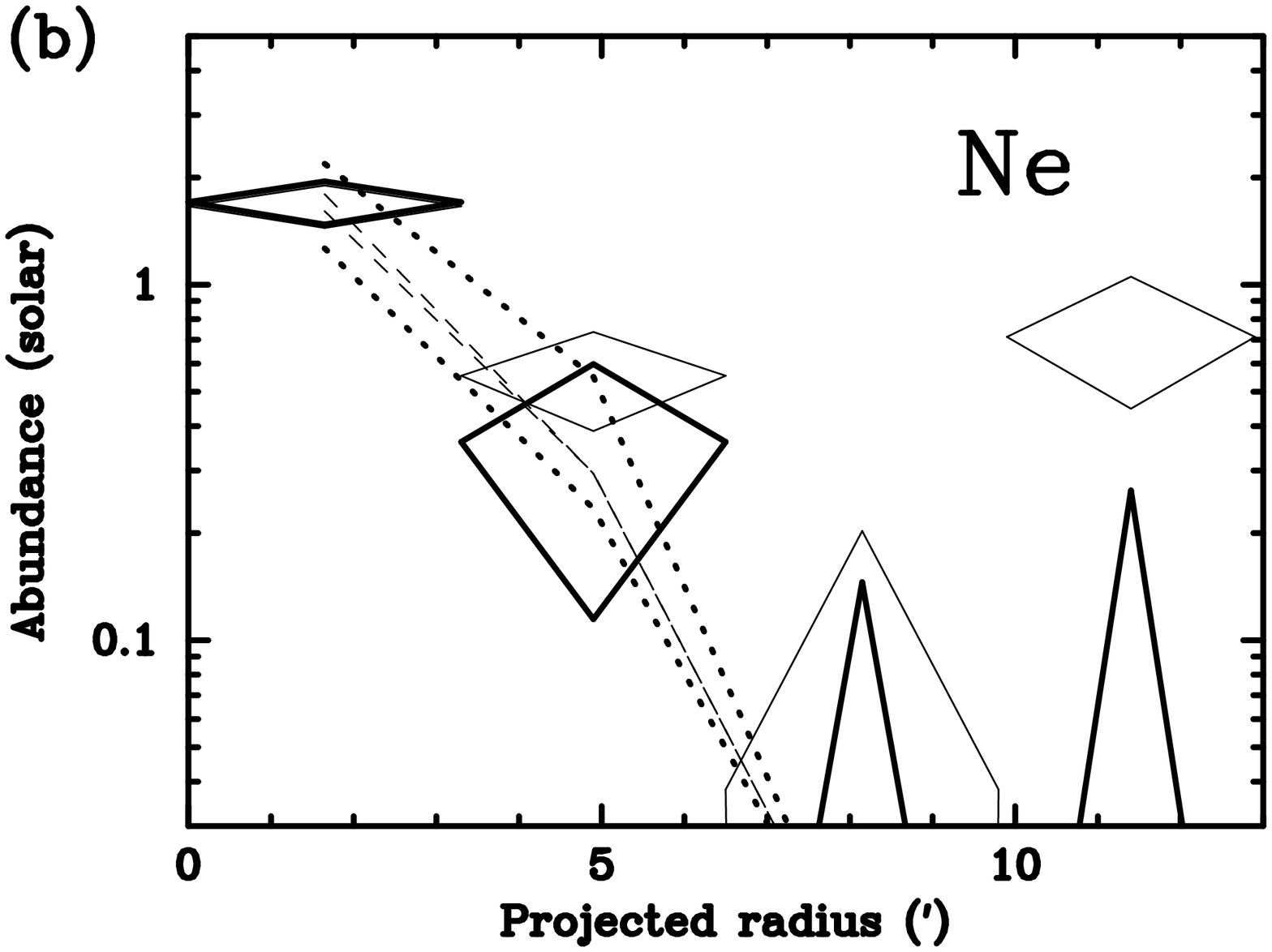}\hfill
\FigureFile(0.32\textwidth,1cm){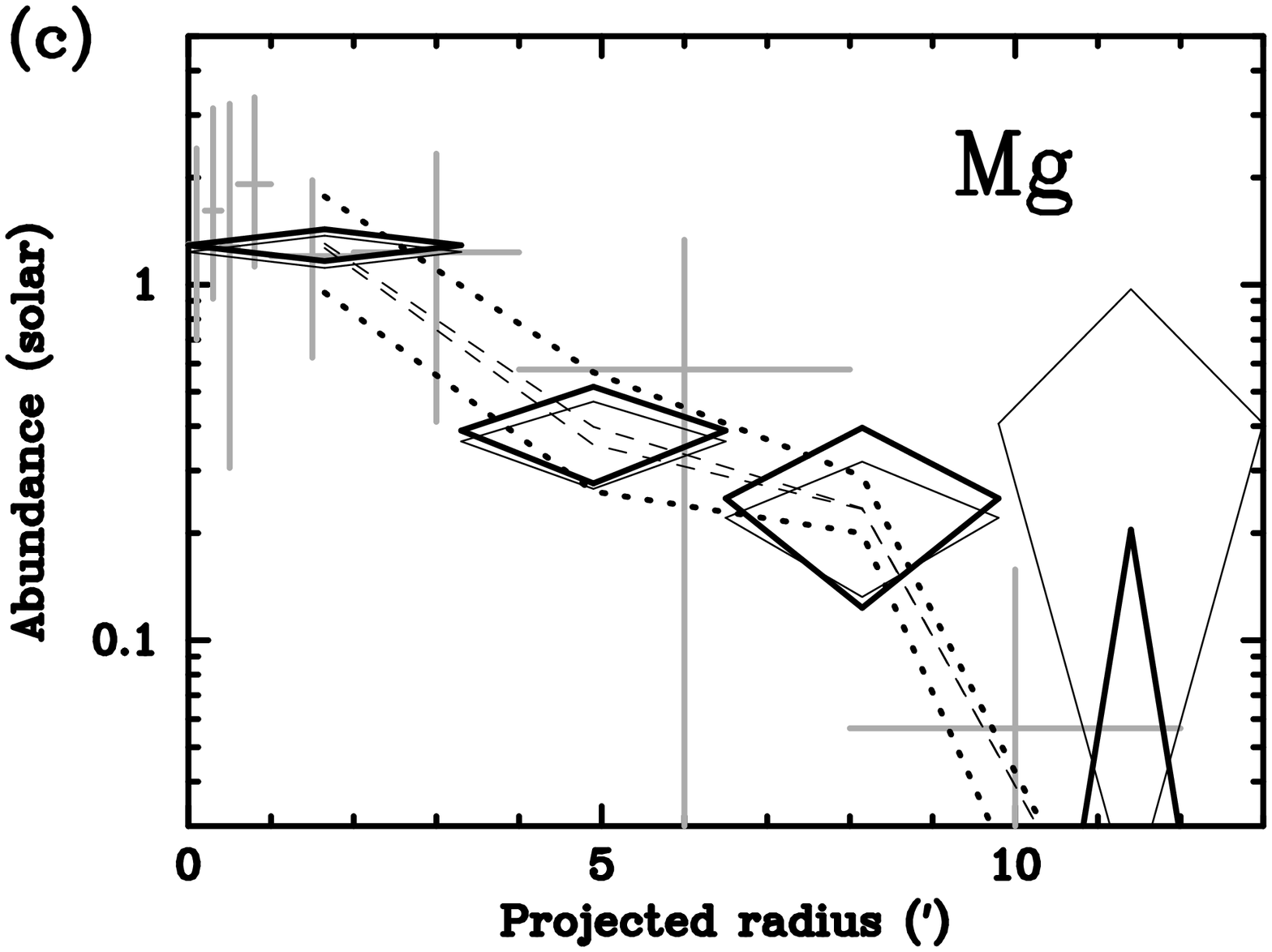}
\FigureFile(0.32\textwidth,1cm){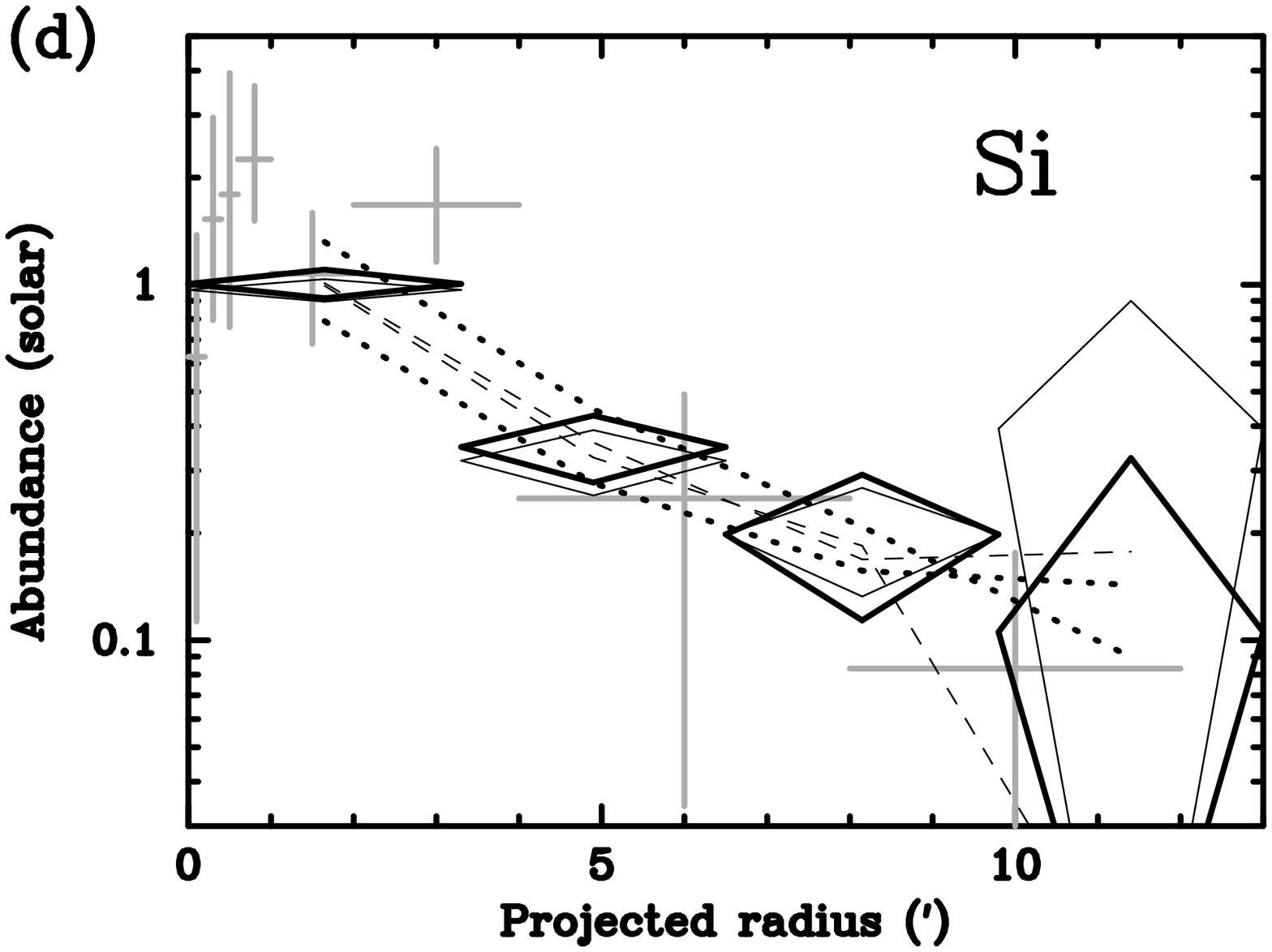}\hfill
\FigureFile(0.32\textwidth,1cm){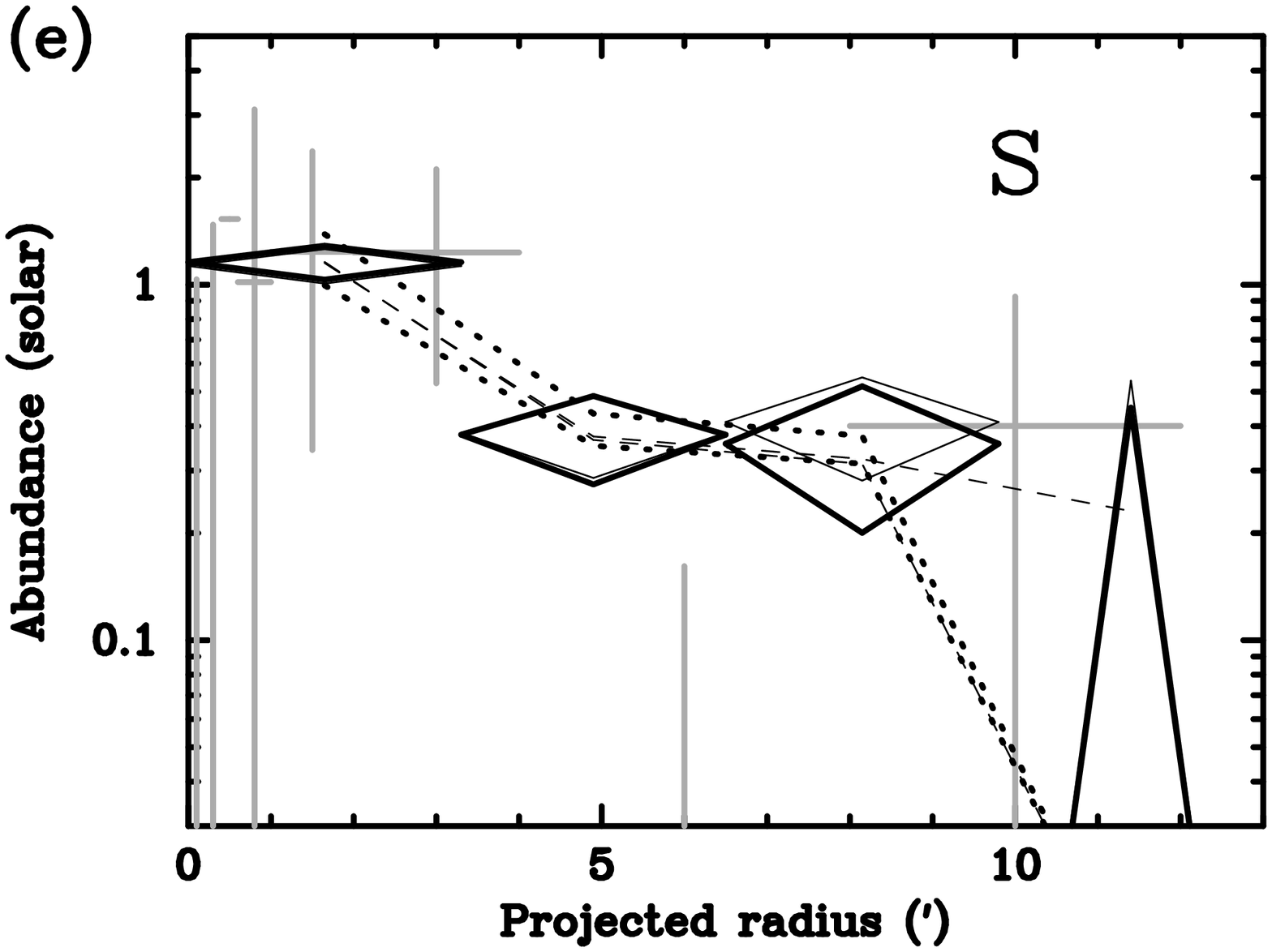}\hfill
\FigureFile(0.32\textwidth,1cm){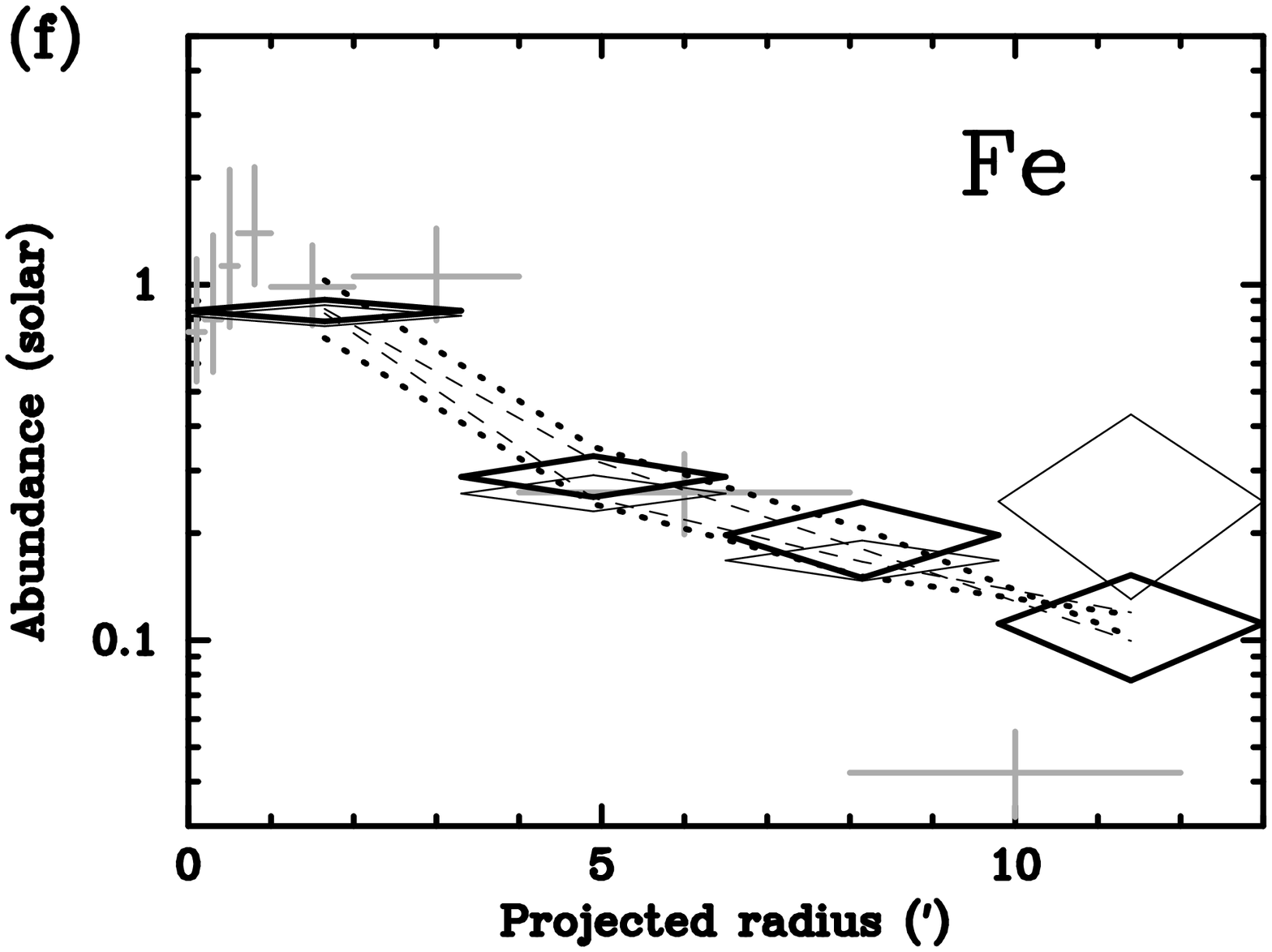}
\caption{
Elemental abundances are plotted against radii from the group center.
Diamonds in thin lines represents the Suzaku best-fit results
with the Gal 1T model and diamonds in thick lines represents
the results with the Gal 2T model.
Dashed lines denote the movable range of the best-fit temperature
when the NXB + CXB components are changed by $\pm 10$\%,
and dotted lines denote the range when the absorption column of
the XIS contaminant is changed by $\pm 20$\%.
Gray crosses are adopted from the previous results
with Chandra and XMM-Newton by \citet{Morita2006},
in which Ne was fixed to have the same value with the O abundance.
}\label{fig:abundance}
\end{figure*}

\section{Temperature Profile}\label{sec:temperature}

Radial temperature profile and the ratio of the {\it vapec}
normalizations between hot and cool ICM components are shown
in figure~\ref{fig:kT}(a) and (b).
The different models of the Galactic emission (1T or 2T) gave
a difference in ICM temperature by about 0.5~keV for the hot and cool
ICM components, except for the innermost region.
Though the previous Chandra and XMM-Newton result
seems to be consistent with the Gal 1T case,
the Galactic component had been neglected in that analysis.
Very crudely, $kT_{\rm Hot}\sim 1.5$~keV, and $kT_{\rm Cool}\sim 0.8$~keV\@.
The cool component is strongest in the innermost region, and
even though it seems to decline in the outer regions, the possible
coupling with the Galactic emission makes the precise estimation difficult.
The radius of $10'\sim 180$~kpc corresponds to
$\sim 0.16\; r_{\rm 180}$, and the temperature decline,
observed in several other clusters, is not clearly recognized in this system
due partly to the multi-phase nature of ICM\@.
However, hardening of the spectra at the 3.3--6.5$'$ annulus
and softening in the outer two annuli suggested by the hardness ratio
image in figure~\ref{fig:image}(c) is evident in figure~\ref{fig:kT}(a) and (b)
for both models of the Galactic emission.

Figure~\ref{fig:kT}(c) show the surface brightness profile with Suzaku
for the Gal 2T model, compared with the $3\beta$-model obtained with
Chandra and XMM-Newton \citep{Morita2006}.
The observed profiles look almost consistent with this model,
while the Suzaku intensity for the Gal 1T model is slightly higher 
at the outermost annulus. We note that, although the surface brightness 
of the hot ICM component decreases steeply, 
the cool component stays almost constant at $r>3.3'$.
The intensity is comparable to the Galactic component, $S_{\rm G1+G2}$,
and it is very difficult to separate these two components clearly.
Further offset observations with Suzaku and/or high resolution
spectroscopy with, e.g., microcalorimeters, are anticipated.

We have estimated the systematic errors caused by the
background subtraction and by the estimation of the XIS filter contamination.
We varied the sum of the NXB and CXB by $\pm 10\%$,
and the contamination thickness by $\pm 20\%$, respectively.
The results are shown by dotted lines for the Gal 2T model
in figure~\ref{fig:kT}(a).
The impact on the temperature is generally less than the systematic
temperature shift, particularly on $kT_{\rm Cool}$, with the 
different models of the Galactic emission.
On the other hand,
$kT_{\rm Hot}$ is much affected by the background uncertainty
most notably in the 6.5--9.8$'$ annulus, where the hot-component intensity is
comparable to the Galactic emission and CXB,
as seen in figure~\ref{fig:kT}(c).

\section{Abundance Profile}\label{sec:abun}

\begin{figure}[tbg]
\centerline{
\FigureFile(0.16\textwidth,1cm){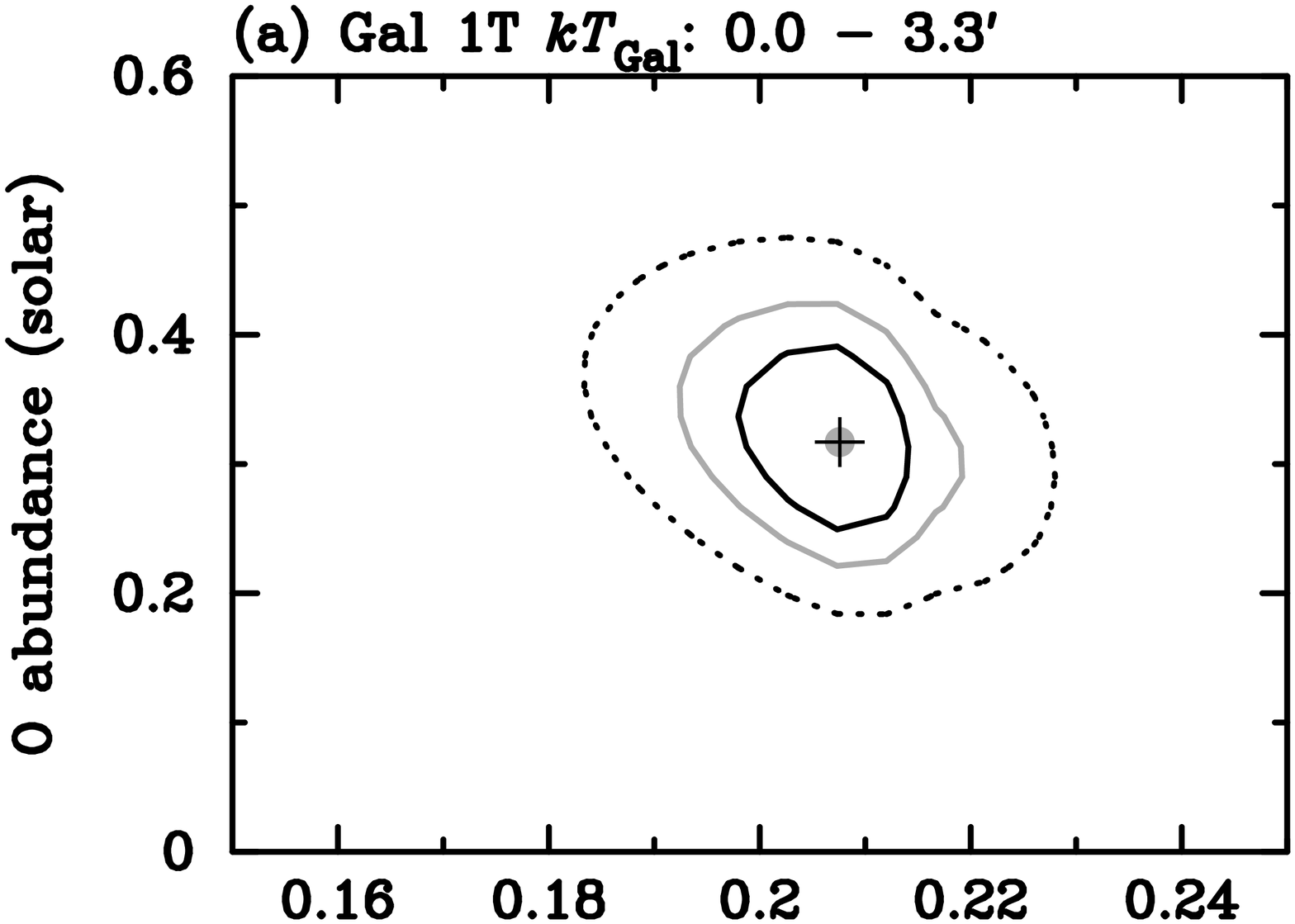}
\FigureFile(0.16\textwidth,1cm){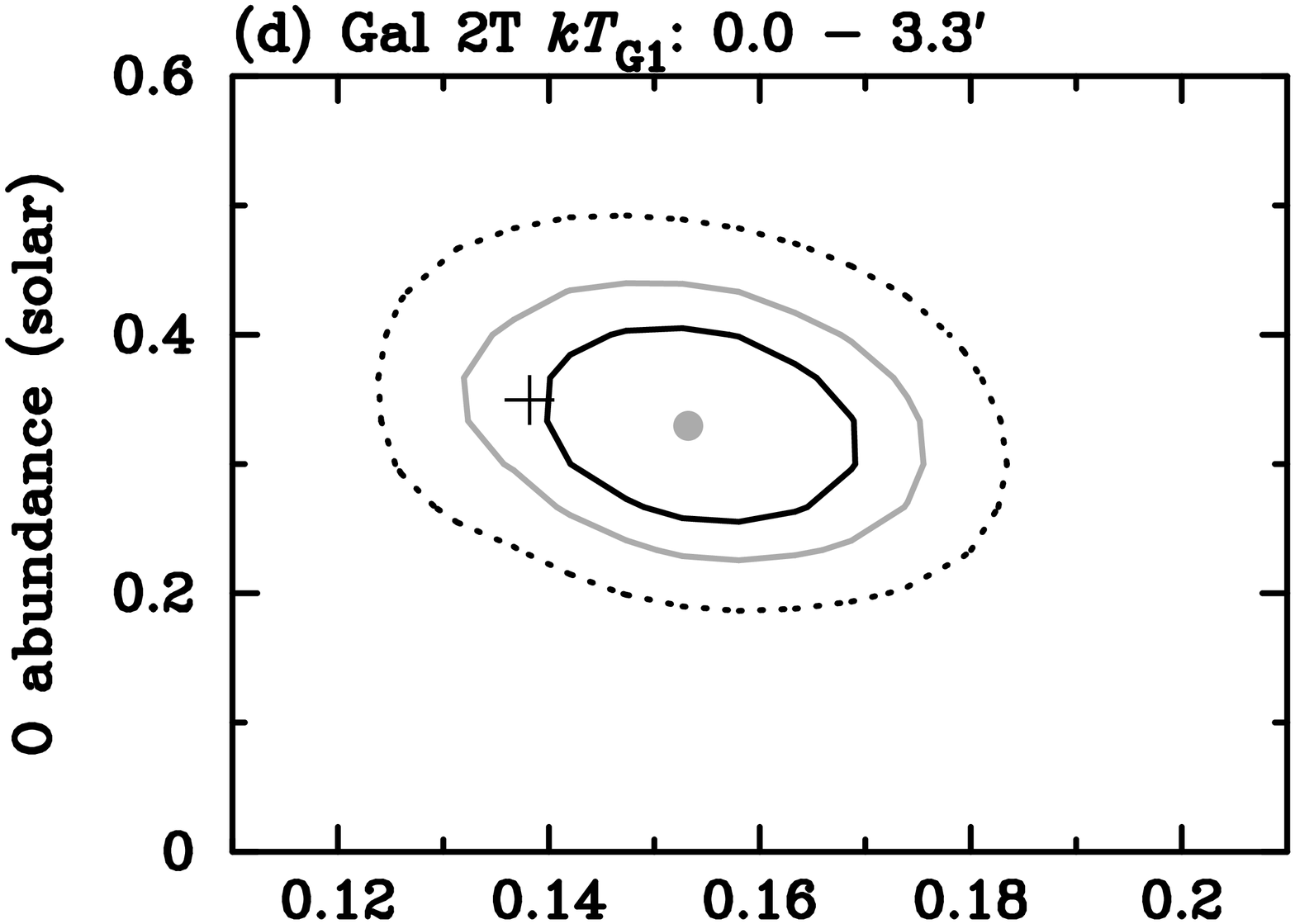}
\FigureFile(0.16\textwidth,1cm){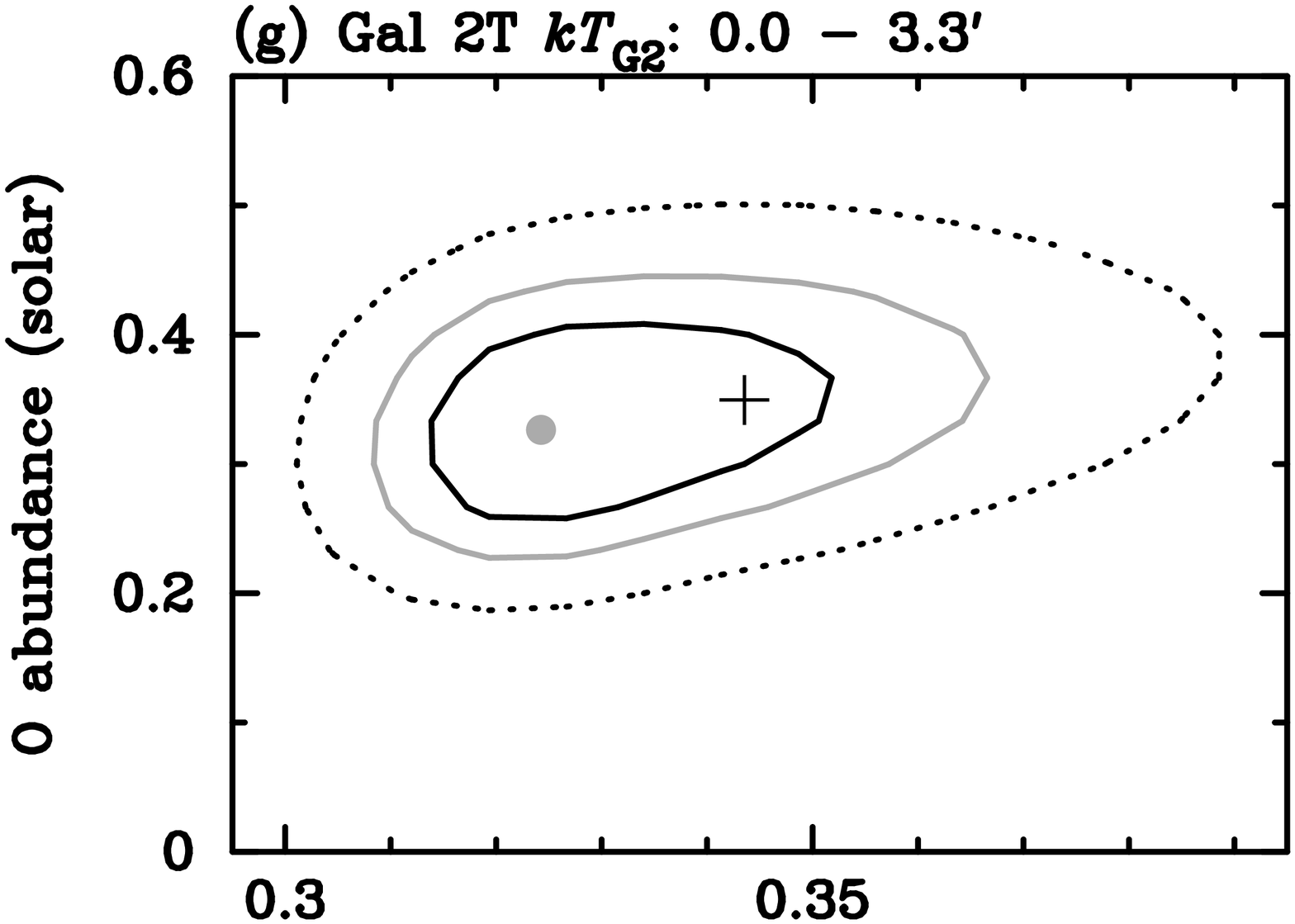}
}
\centerline{
\FigureFile(0.16\textwidth,1cm){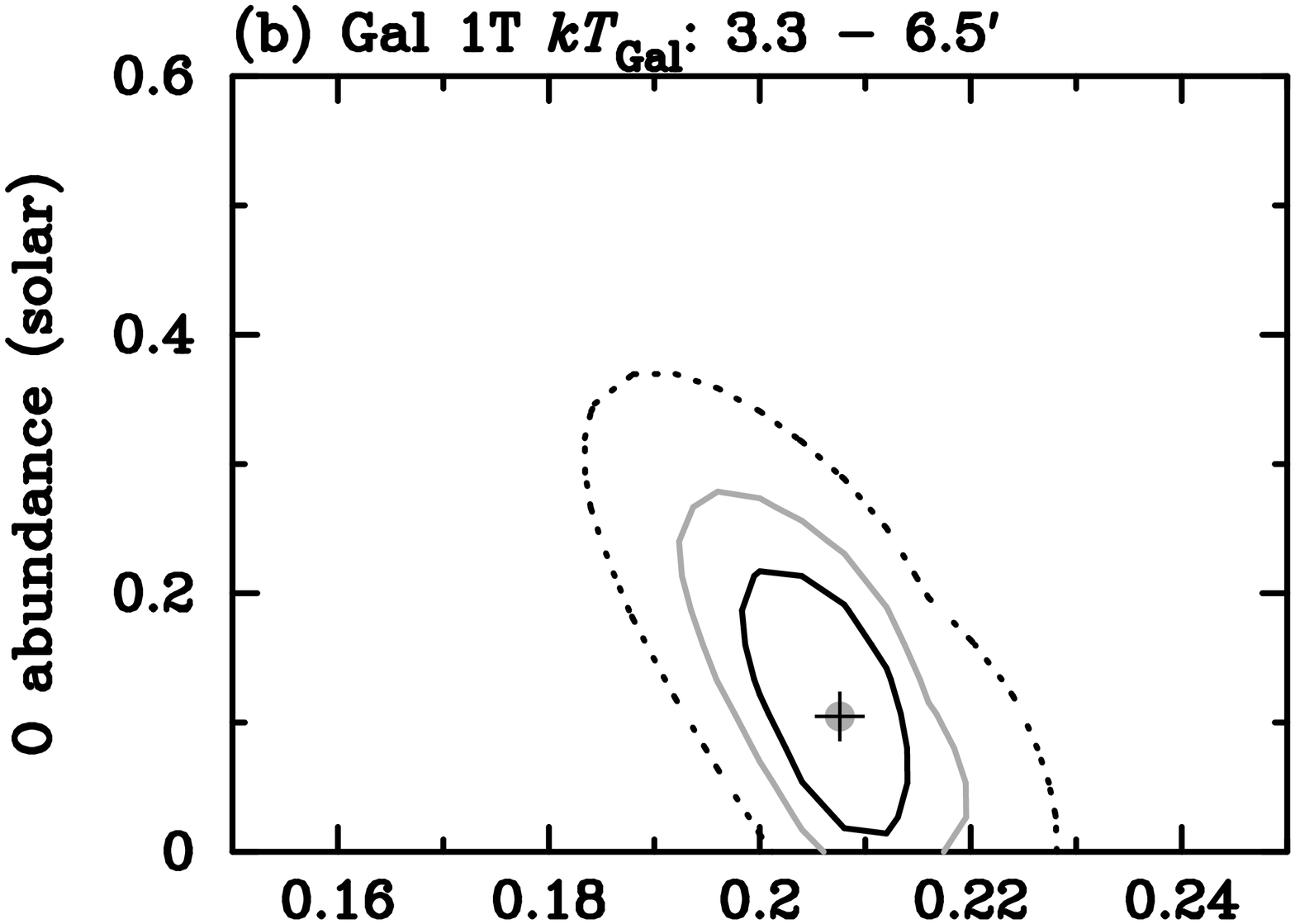}
\FigureFile(0.16\textwidth,1cm){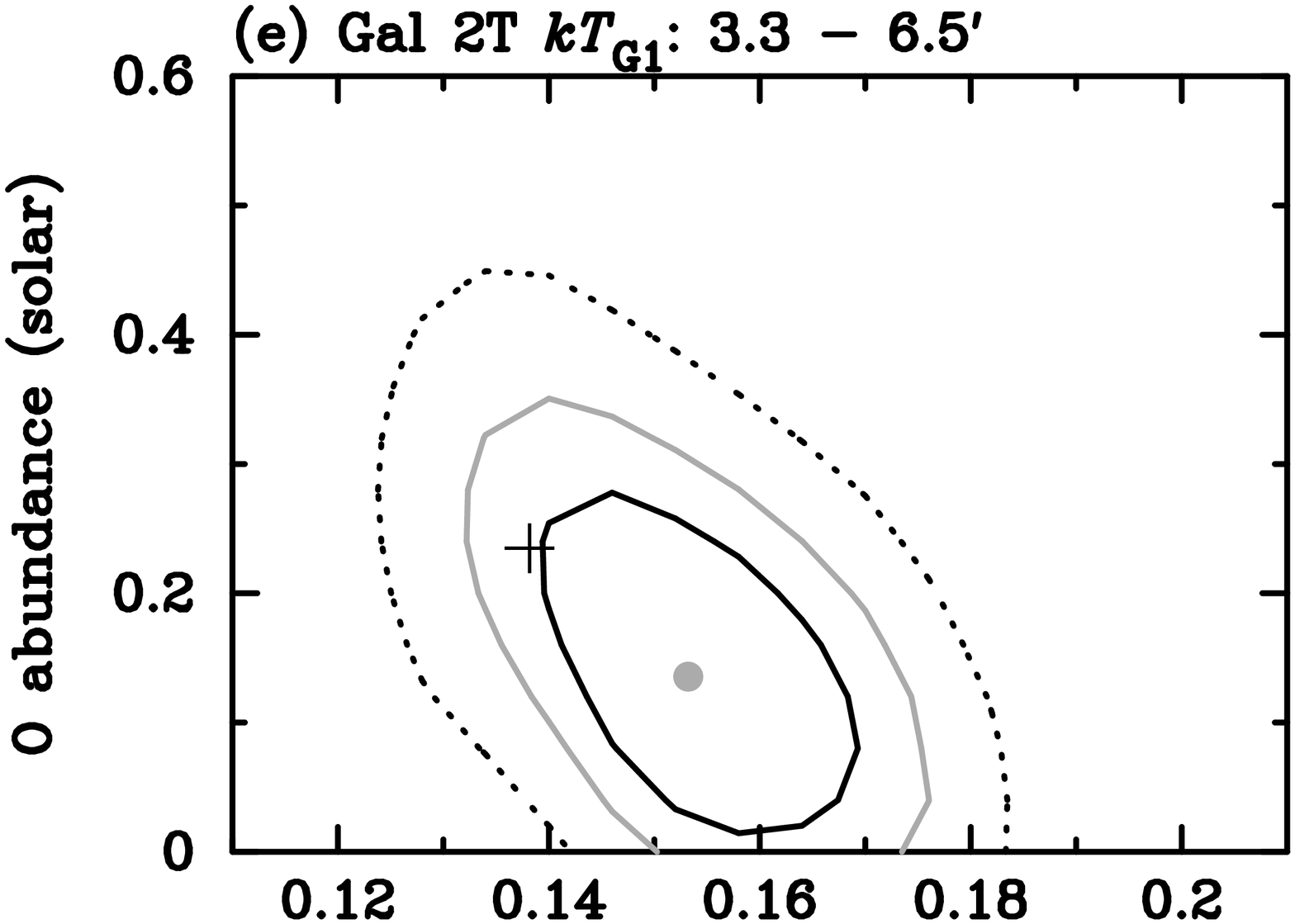}
\FigureFile(0.16\textwidth,1cm){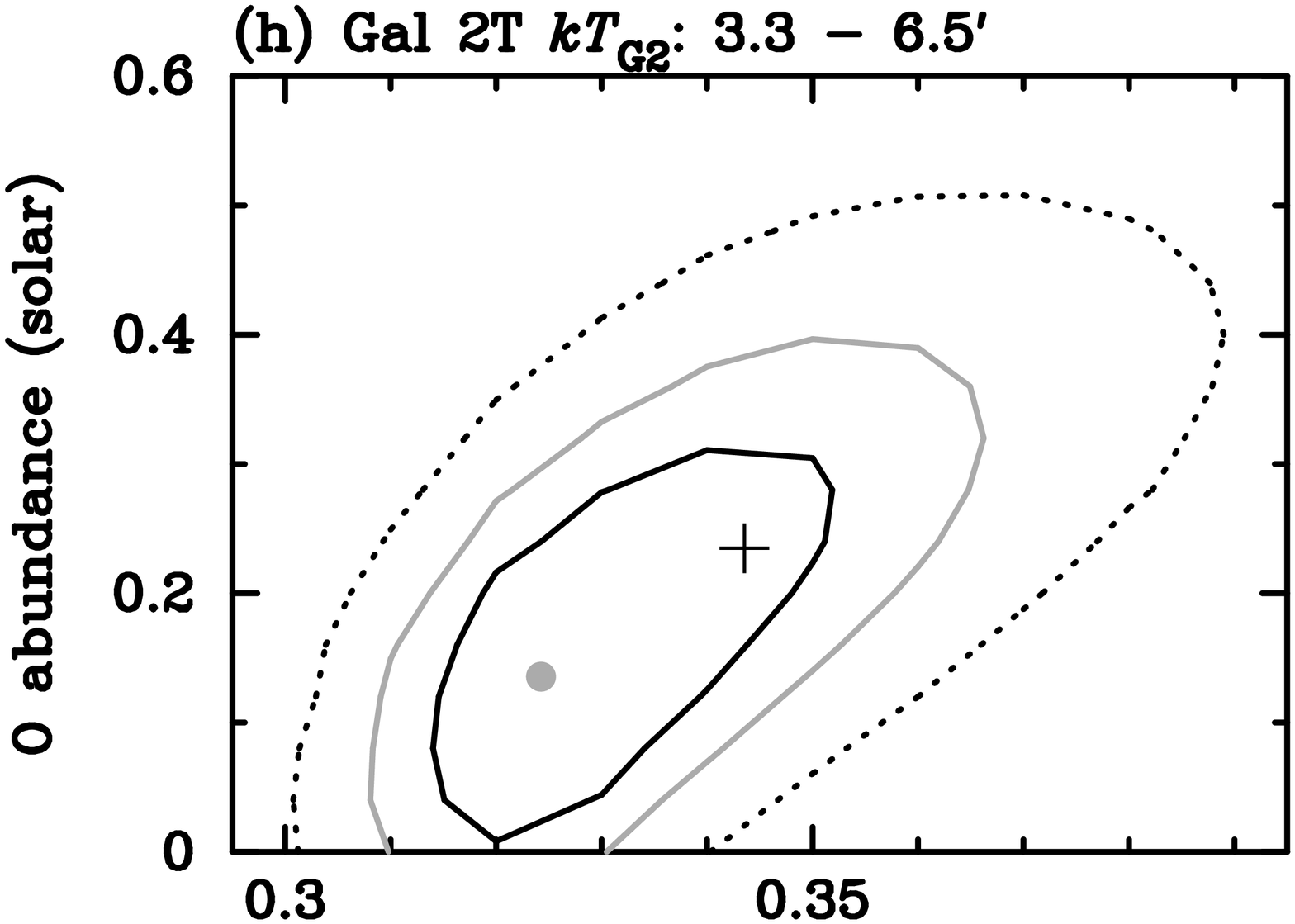}
}
\centerline{
\FigureFile(0.16\textwidth,1cm){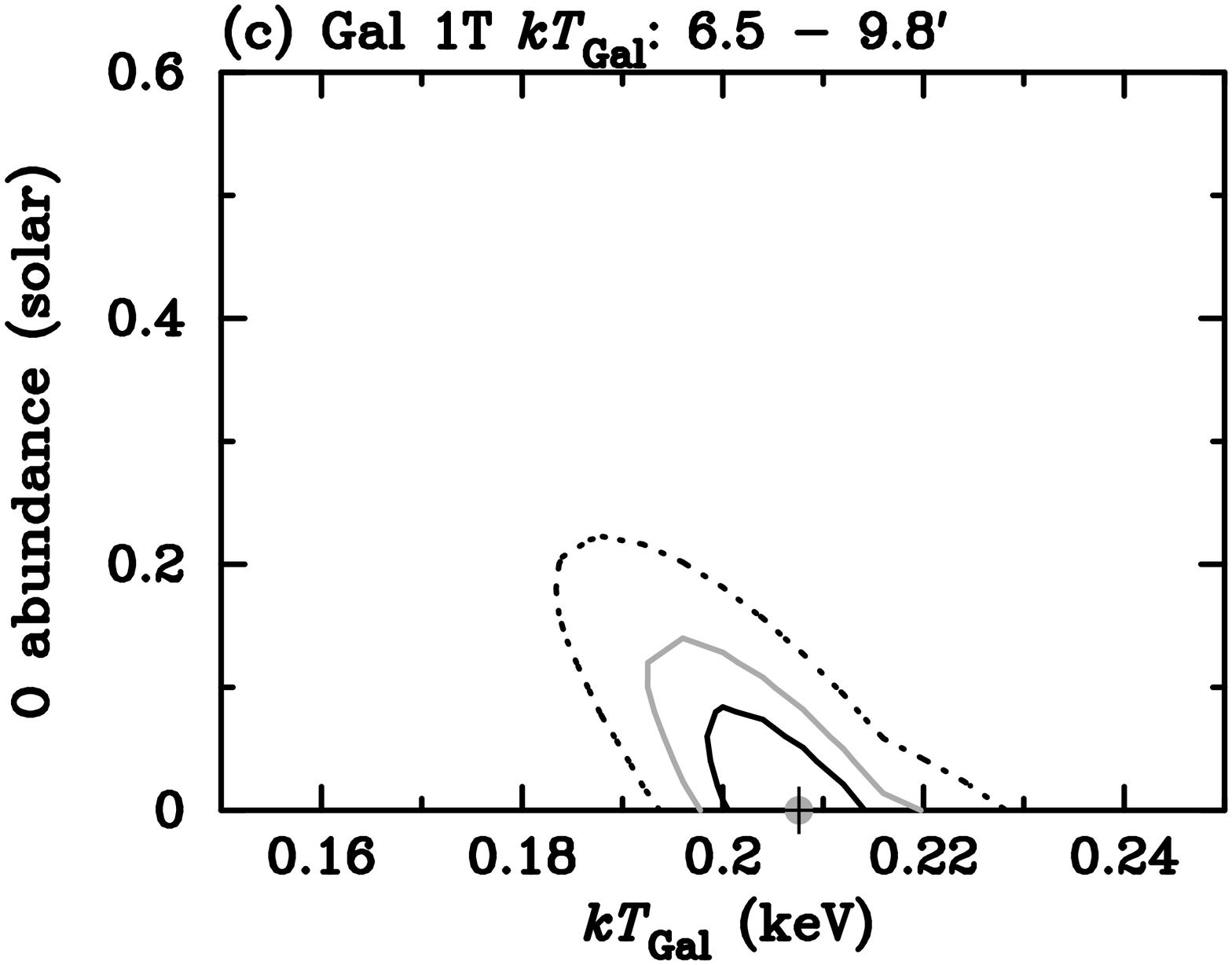}
\FigureFile(0.16\textwidth,1cm){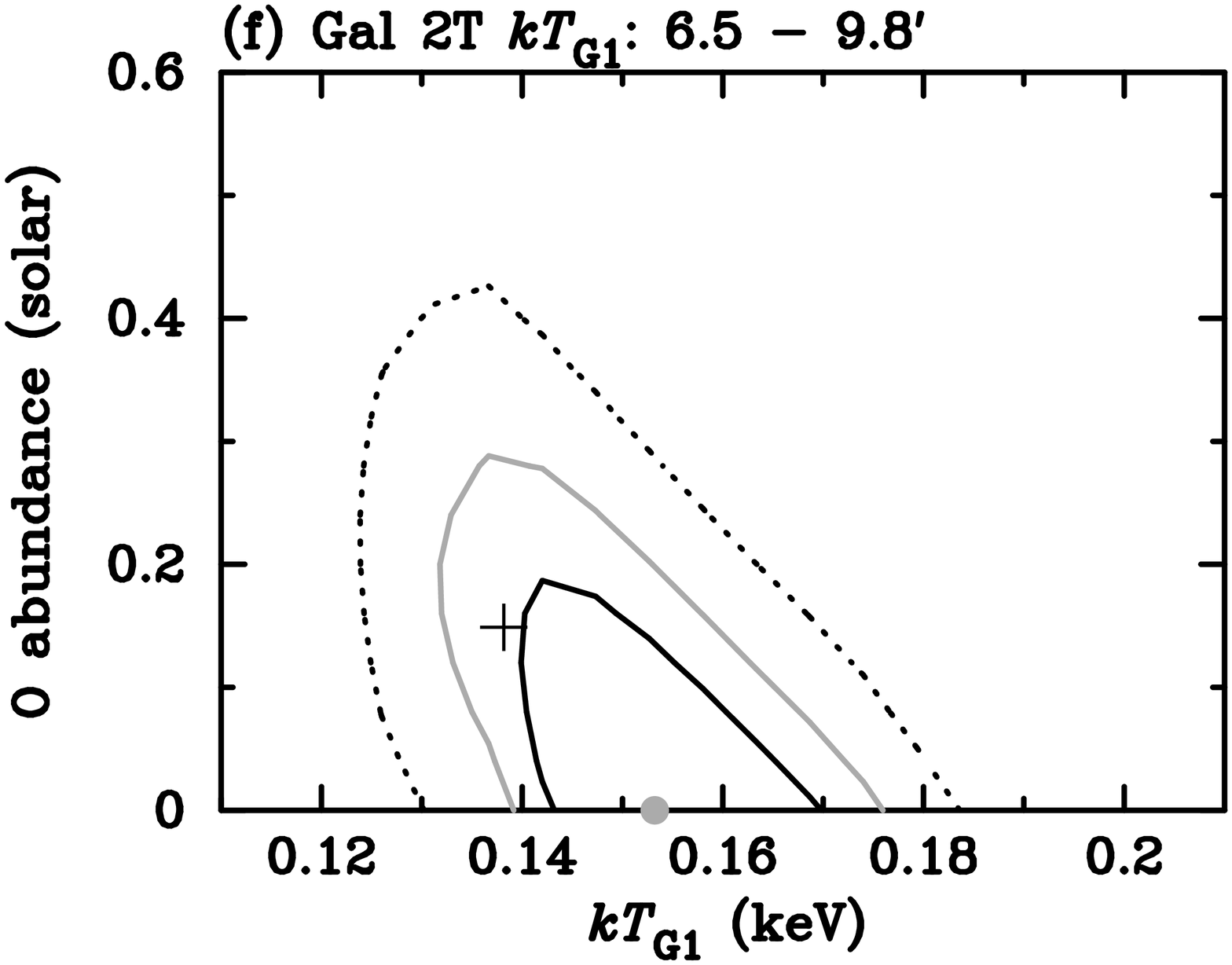}
\FigureFile(0.16\textwidth,1cm){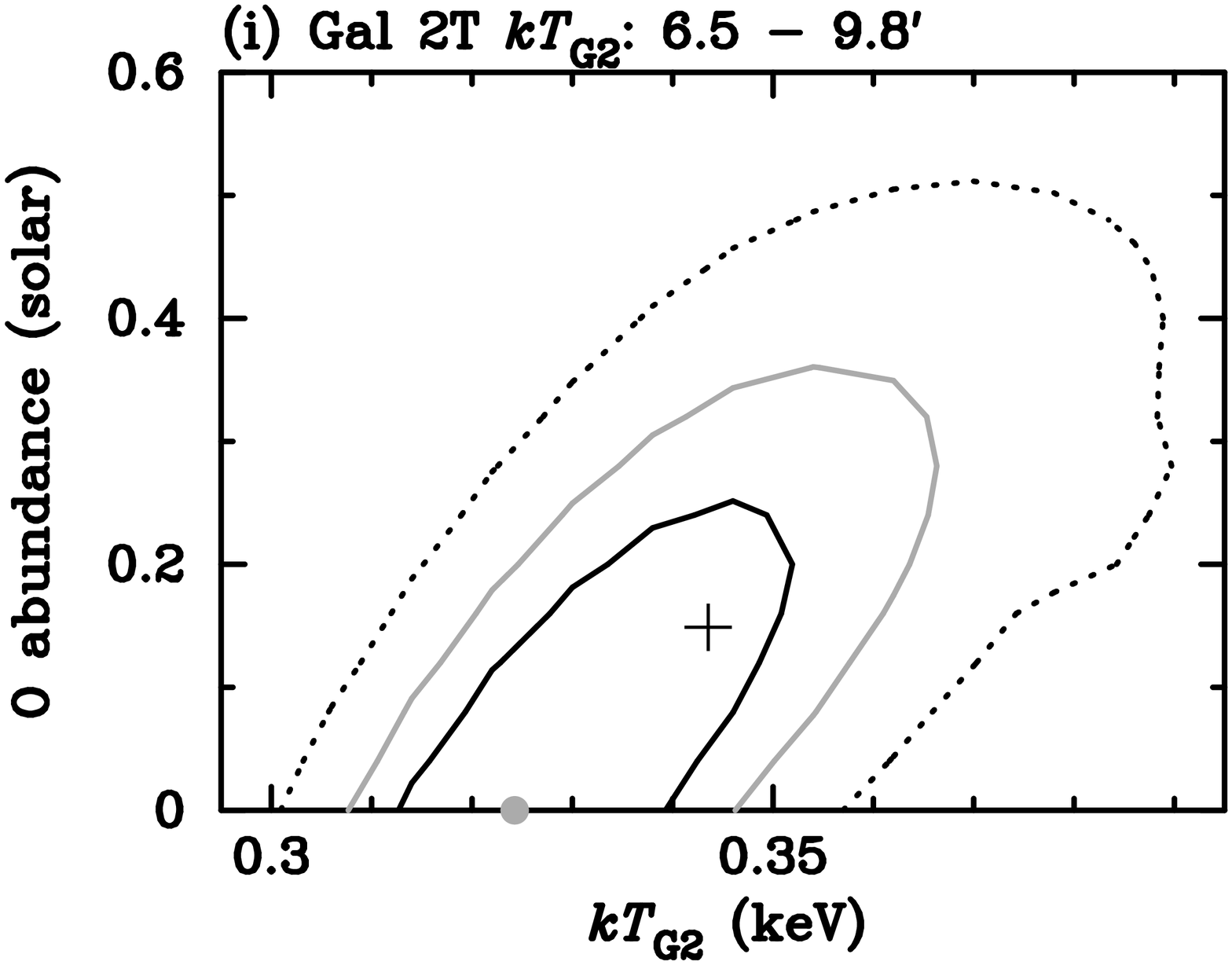}
}
\caption{
Confidence contours between O abundance of the ICM component
and temperature of the Galactic component with the Gal 1T model
for (a)--(c), and with the Gal 2T model for (d)--(i).
The contours represents 1$\sigma$, 90\%, and 99\% confidence regions
from inner to outer, corresponding $\Delta\chi^2 = +2.3$, +4.6, and +9.21
from the $\chi^2$-minimum.
The cross markers represents the best-fit locations
in table~\ref{tab:best-fit}, while the filled gray circles represents the
$\chi^2$-minimum, which are different from the best-fit for the Gal 2T model
because the temperatures of the {\it apec} models were fixed
in the fitting whereas $kT_{\rm G1}$ or $kT_{\rm G2}$ was
allowed to change in these plots.
}\label{fig:contour}
\end{figure}

Metal abundances are determined for the six element groups individually
as shown in figures~\ref{fig:abundance}(a)--(f).
O abundance is strongly affected by how we model the Galactic emission.
The two cases give largely different abundance values
up to 0.3 solar in the regions of $r > 3.3'$,
although they overlap within 90\% statistical errors.
Figures~\ref{fig:contour}(a)--(i) show confidence contours
between O abundance for the inner three annuli and the Galactic temperatures.
In the Gal 2T model, either of the two temperatures was 
fixed to the assumed value when we calculated the contours.
Only for the group center region, we obtained a tight and consistent value of
$0.35\pm 0.08$ solar (Gal 2T case).
The assumed temperatures are slightly lower
or higher than the temperatures at the $\chi^2$-minimum for the $kT_{\rm G1}$
or $kT_{\rm G2}$ case, respectively, and give higher O abundance,
though it is only at 1$\sigma$ level.
Gal 2T model gives higher O abundance than Gal 1T case 
in the outer three annuli, 
however O abundance is less than 0.32 solar at the 90\% confidence
excluding the outer most annulus which has large statistical errors.
Thus spatial profile of O abundance is likely to be
almost flat or declining with the radius.

We note that Ne abundance has a  problem in the spectral fit due
to the coupling with Fe-L feature. Therefore, regarding the spatial
structure, we deal with the remaining four elemental groups: Mg, Si, S and Fe.
The four abundance results look quite similar to each other,
including the central value and the radial gradient.
The central abundances are between 0.8--1.2 solar, and the abundances
decline to
about 1/5 of the central value in the 6.5--9.8$'$ annulus.
Note also that the abundance results for the four elements in $r<9.8'$
all agree within 15\% between the two choices of the Galactic component,
even though the resultant temperatures exhibit large discrepancy
as seen in figure~\ref{fig:kT}(a).
This indicates that these abundance results are fairly robust.

Again, we looked into the effect of the errors in the NXB and CXB
intensities and the OBF contamination thickness. As shown by dotted lines
in figures~\ref{fig:abundance}(a)--(f), the systematic effect is less
than the statistical error for all regions.
Our Suzaku results are also consistent with the previous results
with Chandra and XMM-Newton plotted with gray crosses.
However, errors are much smaller in the outer three annuli,
owing to much deeper exposure and good sensitivity of Suzaku XIS\@.

\begin{table*}[pbt]
\caption{
Best-fit parameters for spectra in $r<1.1'$, NE arc, and SW arc with Gal 2T model.
}\label{tab:arc}
\begin{tabular}{lccccccccc}
\hline\hline\\[-2.2ex]
Gal 2T &$kT_{\rm Hot}$ &$kT_{\rm Cool}$ &O &Ne &Mg, Al &Si & \makebox[0in][c]{S, Ar, Ca} &Fe, Ni &$\chi^{2}$/dof  \\
    &(keV) & (keV) & (solar) & (solar) & (solar) & (solar) & (solar) & (solar)&\\
\hline\\[-2.2ex]
$r<1.1'$                 &$1.63^{+0.06}_{-0.06}$  &$0.755^{+0.009}_{-0.009}$ 
&$0.45^{+0.17}_{-0.14}$  &$1.99^{+0.52}_{-0.45}$  &$1.80^{+0.19}_{-0.29}$     
&$1.42^{+0.15}_{-0.14}$  &$1.66^{+0.29}_{-0.26}$  &$1.22^{+0.18}_{-0.15}$ 
&549/473\\
NE arc                   &$1.74^{+0.08}_{-0.09}$  &$0.778^{+0.012}_{-0.015}$ 
&$0.31^{+0.15}_{-0.13}$  &$1.76^{+0.40}_{-0.44}$  &$1.23^{+0.25}_{-0.24}$    
&$0.96^{+0.18}_{-0.17}$  &$0.84^{+0.22}_{-0.20}$  &$0.71^{+0.10}_{-0.10}$ 
&545/473\\
SW arc                   &$1.55^{+0.05}_{-0.05}$  &$0.785^{+0.013}_{-0.013}$ 
&$0.37^{+0.16}_{-0.14}$  &$1.65^{+0.45}_{-0.41}$  &$1.34^{+0.25}_{-0.23}$     
&$1.00^{+0.17}_{-0.15}$  &$1.26^{+0.23}_{-0.21}$  &$0.86^{+0.11}_{-0.10}$ 
&520/473\\
\end{tabular}


\begin{tabular}{lccrrccrrcccc}
\hline\hline\\[-2.2ex]
Gal 2T & $kT_{\rm G1}$ & $kT_{\rm G2}$ & $K_{\rm Hot}$ & $K_{\rm Cool}$ & $K_{\rm G1}$ & $K_{\rm G2}$ & \makebox[1.85em][r]{$S_{\rm Hot}$} & \makebox[1.85em][r]{$S_{\rm Cool}$} & \makebox[1.85em][c]{$S_{\rm G1}$} & \makebox[1.85em][c]{$S_{\rm G2}$} & \makebox[1.85em][r]{$S_{\rm CXB}$} & \makebox[1.85em][c]{$S_{\rm LMXB}$} \\
 & (keV) & (keV) & & & & & \\
\hline\\[-2.2ex]
$r<1.1'$ & \makebox[4em][c]{0.138 (fix)} & \makebox[4em][c]{0.344 (fix)} & $91.2_{-9.5}^{+9.8}$ & $\makebox[0in][r]{1}38.0_{-17.}^{+18.}$& \makebox[2.7em][c]{1.3 (fix)} & \makebox[2.7em][c]{0.9 (fix)} & 77.8 & 274.2 & 1.2 & 1.5 & 1.6 & 0.9 \\
NE arc & $\uparrow$ & $\uparrow$ & $14.3_{-1.1}^{+1.4}$ & $12.1_{-1.4}^{+1.6}$ & $\uparrow$        & $\uparrow$        &  9.4 & 15.8 & $\uparrow$ & $\uparrow$ & $\uparrow$ & $\uparrow$ \\
SW arc & $\uparrow$ & $\uparrow$ & $17.1_{-1.4}^{+1.5}$ & $10.1_{-1.0}^{+1.1}$ & $\uparrow$        & $\uparrow$        & 13.1 & 14.9 & $\uparrow$ & $\uparrow$ & $\uparrow$ & $\uparrow$ \\
\hline
\end{tabular}
\end{table*}

\begin{figure*}[ptbg]
\begin{minipage}{0.43\textwidth}
\FigureFile(\textwidth,1cm){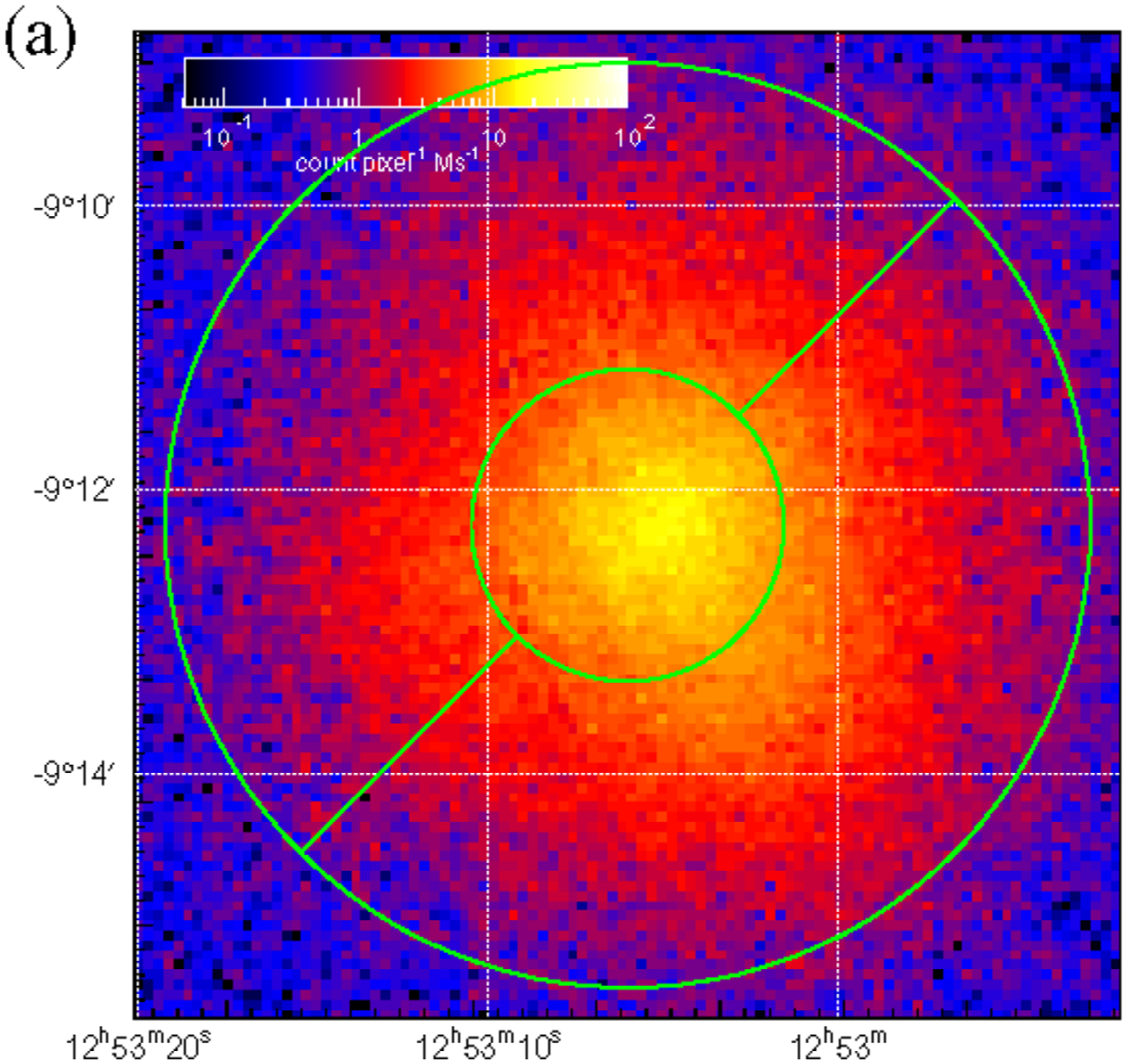}
\end{minipage}%
\begin{minipage}{0.28\textwidth}
\FigureFile(\textwidth,1cm){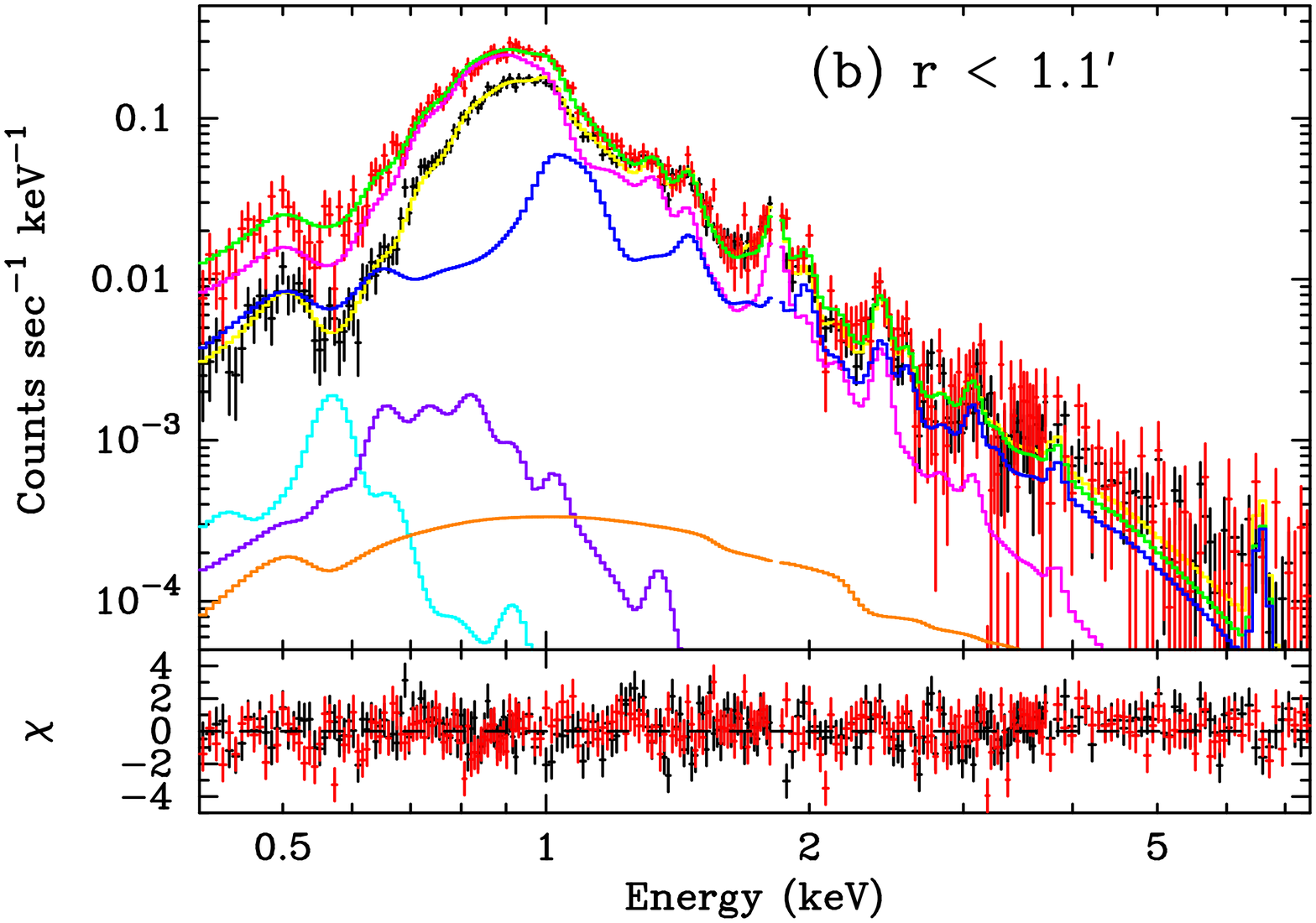}
\end{minipage}\hfill
\begin{minipage}{0.28\textwidth}
\FigureFile(\textwidth,1cm){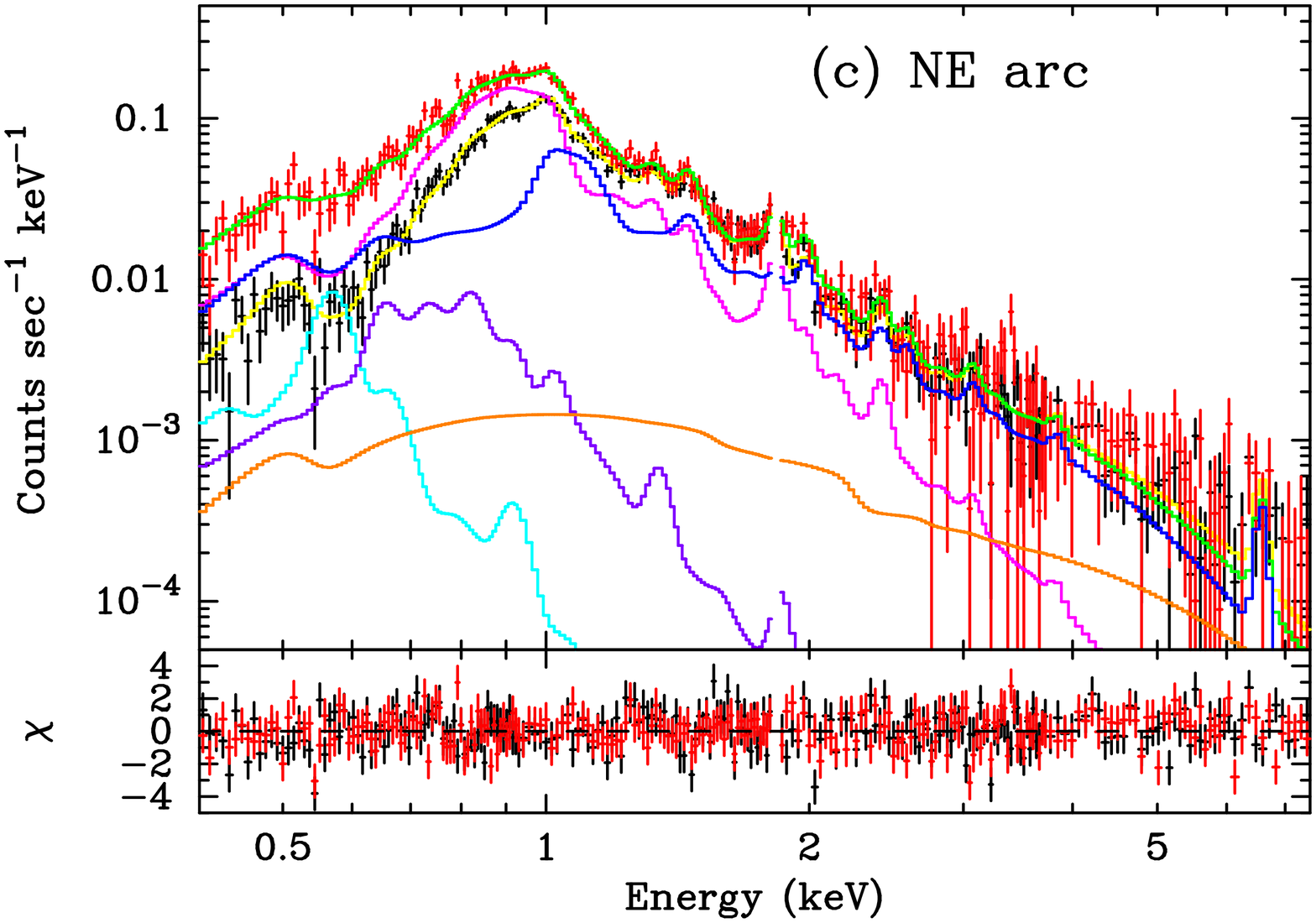}
\FigureFile(\textwidth,1cm){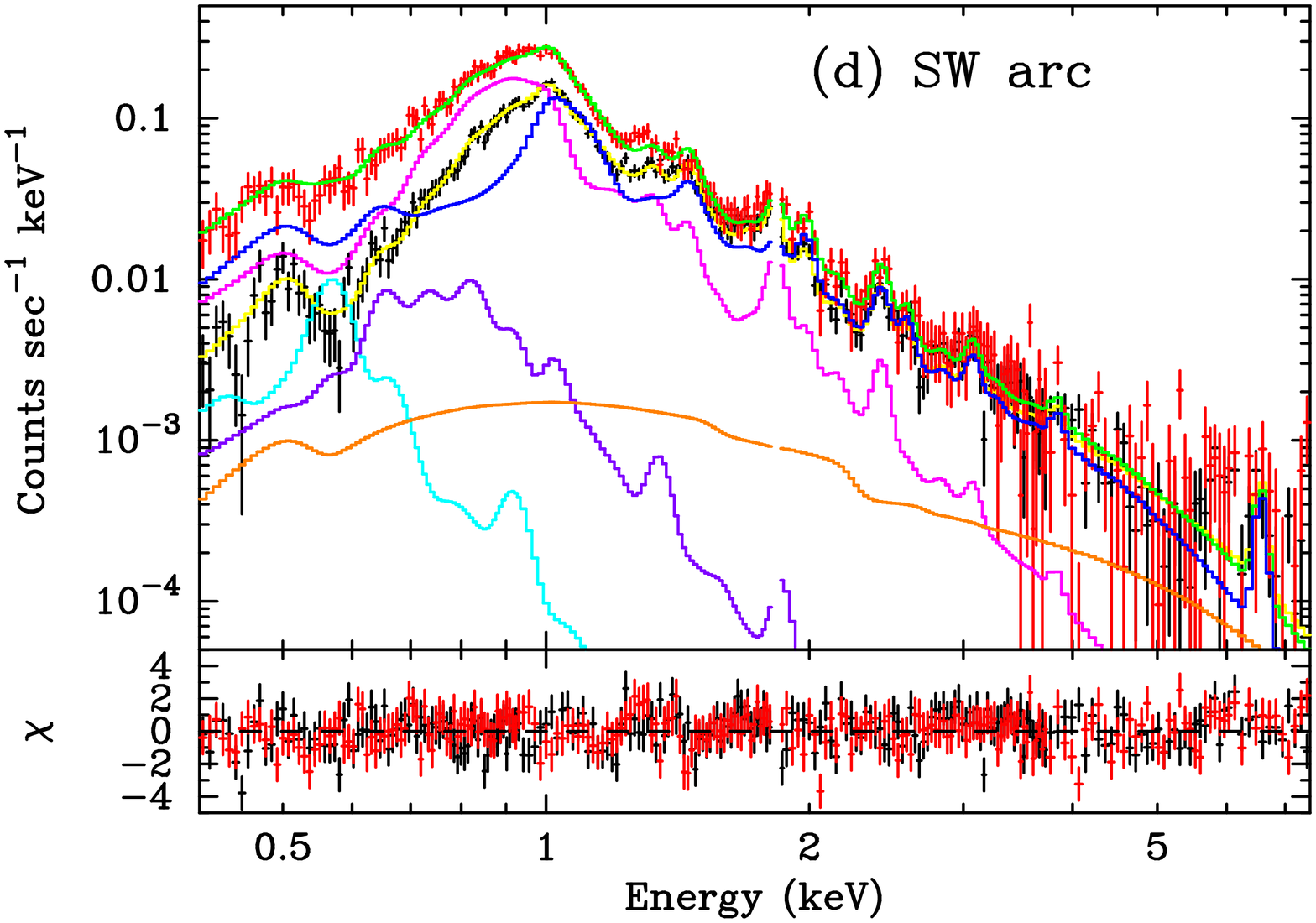}
\end{minipage}%
\caption{
(a) Magnification of the 0.0--3.3$'$ annulus in figure~\ref{fig:image}(a),
which is divided into three regions of $r<1.1'$, NE arc, and SW arc.
The image is binned to $4.2''\times 4.2''$,
and the smoothing with gaussian is not conducted.
(b)--(d) Energy spectra of the three regions.
}\label{fig:arc}

\smallskip

\begin{minipage}{0.5\textwidth}
\centerline{
\FigureFile(0.9\textwidth,1cm){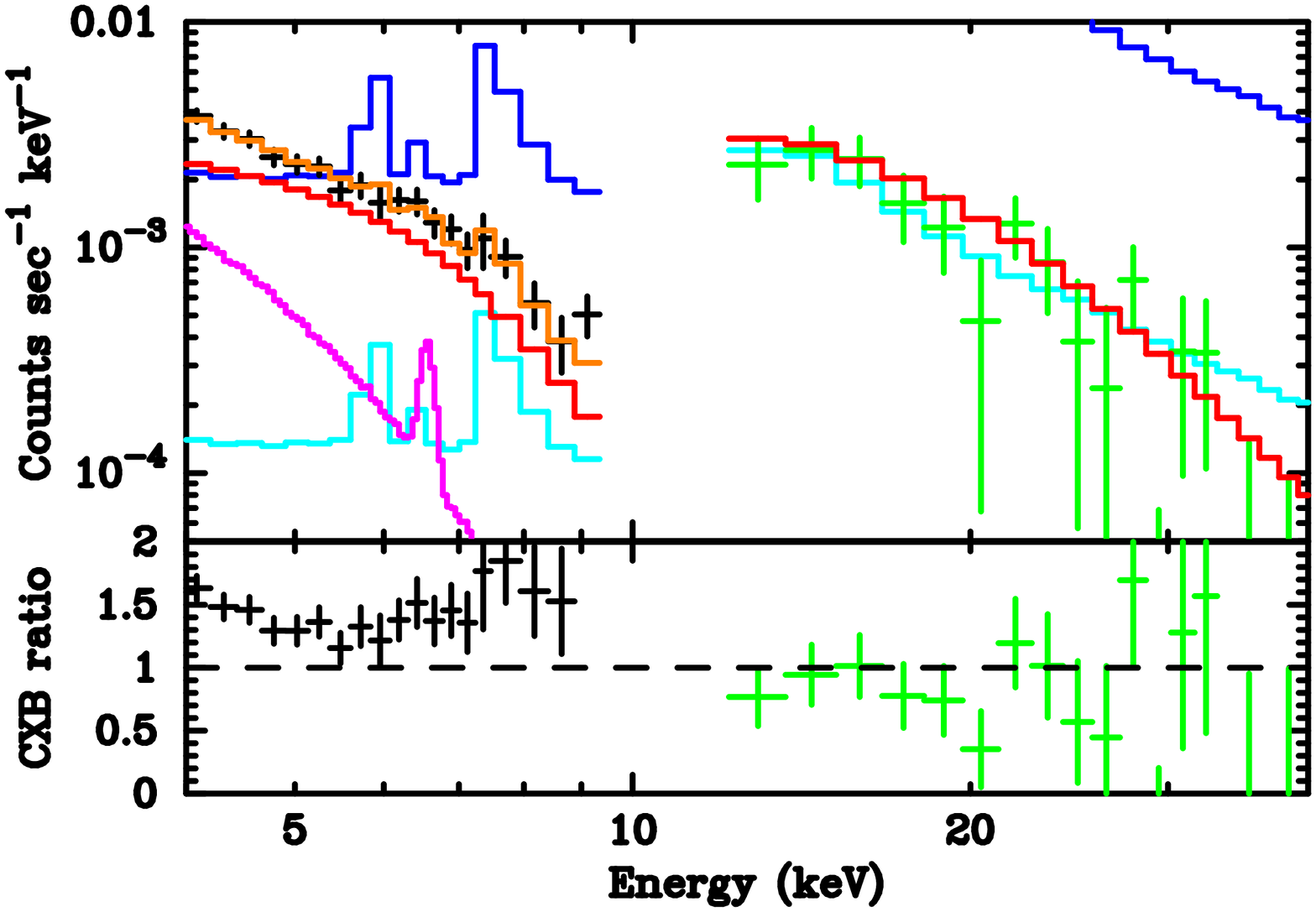}
}
\caption{
The 4--10~keV band XIS spectrum of FI sensors (black)
at 3.3--9.8$'$ annulus and the 12--40~keV band spectrum of HXD-PIN (green).
The estimated NXB spectra (blue) are subtracted from
both spectra, but the CXB spectra (red) are not subtracted.
The ICM component for XIS is shown by magenta line,
and the 90\% confidence levels of the NXB reproducibility
for XIS ($\pm 6.5$\%) and HXD-PIN ($\pm 5.6$\%) are
indicated by cyan histograms.
The orange histogram represents the ICM + CXB + 6.5\% NXB spectrum for XIS\@.
Bottom panel show the ratio of each spectrum to CXB\@.
}\label{fig:xis-pin-spec}
\end{minipage}%
\begin{minipage}{0.5\textwidth}
\centerline{
\FigureFile(0.9\textwidth,1cm){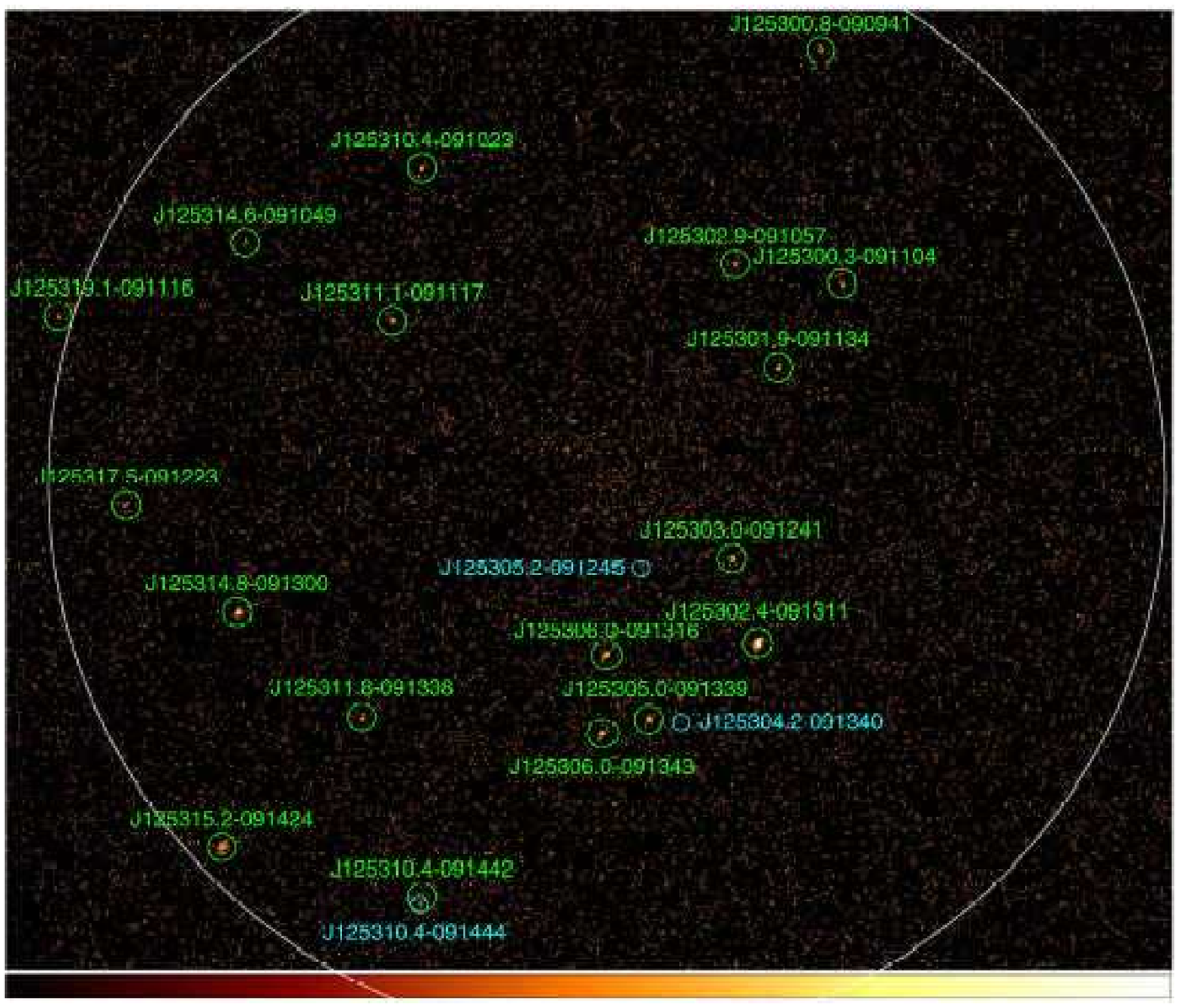}
}
\caption{
The 2--10~keV band image of the central region of HCG~62 obtained with 
Chandra ACIS-S3 with 49.15~ks exposure.
A large white circle represents 3.3$'$ radius
from the group center. Point sources detected in 2.1--7~keV are
indicated by small green circles, and cyan circles correspond
to sources only detected in 0.3--2.1~keV \citep{Harrison2003}.
}\label{fig:Chandra-image}
\end{minipage}

\end{figure*}

\section{Does ``High-Abundance Arc'' Exist?}\label{sec:arc}

\citet{Gu2007} reported that there was a high-abundance arc region
at about 2$'$ from the X-ray peak spanning from south to northwest, and
a part of it roughly coincides with the outer edge of the southwest
X-ray cavity. The reported abundance result with Chandra ACIS-S3
as $0.84^{+0.19}_{-0.15}$ solar in the radial range 1.77--2.21$'$ of
southwest half arc (region B+C in table~1 of \cite{Gu2007}).
The abundance was even higher than the level at the group center ($r<0.44'$).
The projected 0.7--7~keV spectrum was fitted with a single
absorbed {\it apec} model with the absorption fixed to the Galactic value.
They used the solar abundance table of \citet{Grevesse1998},
where the iron number abundance relative to hydrogen is $3.16\times 10^{-5}$,
which is less than our value of $4.68\times 10^{-5}$ by
\citet{Anders1989}.

In order to confirm the ``high-abundance arc'' with Suzaku,
we split the central 0.0--3.3$'$ region into three parts;
$r<1.1'$, northeast (NE) arc, and southwest (SW) arc, respectively,
as shown in figure~\ref{fig:arc}(a).
The extracted spectrum for each region and the best-fit parameters
with the Gal 2T model are presented in figures~\ref{fig:arc}(b)--(d)
and table~\ref{tab:arc}. Area, coverage, and observed/estimated
counts are given in table~\ref{tab:region}.
In the spectral fit, the Galactic and LMXB components were
fixed to the same shape as in the previous fit for the 0.0--3.3$'$ region
with the same surface brightness. We note that the modeling of
the Galactic component does not affect the best-fit results
in this region, as shown in figures~\ref{fig:kT}~and~\ref{fig:abundance}.

Though we confirmed the abundance gradient in $r<3.3'$
as suggested by \citet{Morita2006}, no excess abundance
larger than the central $r<1.1'$ region
was observed in SW arc which contains the ``high-abundance arc''.
Comparing NW and SW arcs, SW one tend to show slightly higher
value, although their 90\% error ranges
of the abundances overlap.
One outstanding feature in SW arc is that the intensity of
the hot ICM component is much higher than the cool one among
the three regions examined, as seen in figure~\ref{fig:arc}(d)
with blue and magenta lines.

Since the angular resolution of Suzaku
($\sim 2'$ in half power diameter; \cite{Serlemitsos2007})
is much worse than that of Chandra ($\lesssim 1''$),
X-ray emission from the ``high-abundance arc'' may be substantially
diluted by the surrounding emission.
However, spectral sensitivity and photon statistics
with Suzaku are quite high, which are important
to determine the elemental abundance.
One possible explanation is that the excess intensity of
the hot ICM component in SW arc could have been 
mis-identified as the ``high-abundance arc''.

We also note that the abundances were 0.35, 0.20, and 0.28 solar
for $r<1.1'$, NW arc, and SW arc, respectively,
when the Suzaku spectra were fitted with the single absorbed {\it apec} model
in the 0.7--7~keV using the \citet{Grevesse1998} abundance just
as performed by \citet{Gu2007}. Though somewhat high abundance
in SW arc was obtained, the abundance value
was much lower than our results in table~\ref{tab:arc}.
Moreover, the model was rejected
with high statistical significance of $\chi^2/{\rm dof}\sim 2$.
The multi-phase nature of ICM is an important factor
as discussed in section~\ref{sec:temperature},
and treating the data with a single temperature model 
can result in quite a different abundance value.

\section{Extended Hard X-ray Emission}\label{sec:hard-emission}

\subsection{Observed Counts and NXB Reproducibility}

The XIS spectra in figures~\ref{fig:spectrum}(a)--(h) suggest
hard X-ray emission in excess of the group thermal and LMXB emission 
in the energy range above $\sim 5$ keV\@.
As already mentioned in section \ref{sec:image}, 
the 6--10~keV hard-band image shows elongated feature along the
NE and SW directions around the group center.
It is partly because
of a clump of point sources $\sim 1.5'$ aside the group center,
but the feature looks to direct to the two cavities.
Point source contribution at the group center is
discussed later in this section.
On the other hand, in the region at $r > 3.3'$ there seems
no clear correlation with point sources.
Since diffuse hard excess emission in HCG~62 was reported based on an ASCA
observation
\citep{Fukazawa2001,Nakazawa2007}, we investigate the properties of
the hard emission with the Suzaku data.

Table~\ref{tab:counts-cor8} summarizes observed total counts compared
 with the expected contributions
from background (NXB and CXB) and the group emission (ICM and LMXB)\@.
With statistics, the excess counts are significant by more than 3$\sigma$
level in all the three annuli.
We compared these results with the previous ASCA flux
\citep{Fukazawa2001,Nakazawa2007}.
\citet{Nakazawa2007} reported that the hard excess flux
within 3--15$'$ region was $0.92_{-0.17}^{+0.18}\times10^{-12}$
erg~cm$^{-2}$~s$^{-1}$ (2--10 keV),
which is roughly $\sim 1/3$ of the CXB intensity.
We calculated the expected counts with
this flux assuming a spatially flat distribution.
The results are shown in  ``ASCA'' raw
in table~\ref{tab:counts-cor8}.
The present Suzaku counts look generally consistent
with the ASCA results.
However, systematics associated with the reproducibility of the NXB
must be examined carefully.

\citet{Tawa2007} report that the intrinsic variability of NXB
is 4.46/5.60/3.20\% for XIS0/2/3 in 5--12~keV
after the standard screening procedure. In this study, the NXB counts
over 5~ks, which corresponds to 2--3 days of observation, was compared
with an exposure weighted sum of the {\it COR}\/ sorted data for
night Earth observations.
Therefore, the 90\% confidence range of the NXB fluctuation for FI sensors
is $1.6\times\!\sqrt{(4.46^2 + 5.60^2 + 3.20^2)/3} = 7.2\pm 1.1$\%.
They also claim that this level can be reduced to
$6.5\pm 1.0$\% 
when an updated cut-off rigidity calculation of ${\it COR2}$
is used as the NXB indicator and the following additional screening
criteria are applied;
\begin{eqnarray}\label{eq:screening}\nonumber
{\it T\_SAA\_HXD} > 436\quad & \makebox{\scriptsize AND} \\\nonumber
({\it SAT\_LAT}>-23\quad \makebox{\scriptsize OR}\quad {\it SAT\_ALT}<576.5)
	\quad & \makebox{\scriptsize AND} \\\nonumber
({\it SAT\_LAT}<+29\quad \makebox{\scriptsize OR}\quad {\it SAT\_ALT}<577.5)
	\quad & \makebox{\scriptsize AND} \\
{\it TIME} \ge 181470000\makebox[0in][l]{,}\quad &
\end{eqnarray}
where {\it SAT\_LAT}\/ (deg) denotes the orbital location of the satellite
in the geographic latitude, and {\it SAT\_ALT}\/ (km)
is the altitude of the satellite.
Though we have selected ${\it COR} > 8$~GV, which is different from the
\citet{Tawa2007} study, we also tested the ${\it COR2}$\/ sorted 
NXB estimation
after the data screening with equation~(\ref{eq:screening}).
The result was almost identical to the values in table~\ref{tab:counts-cor8},
so that we adopt $\pm 6.5$\% as the 90\% confidence range
of NXB systematic error for XIS-FI in 5--12~keV\@.

\subsection{Outer Region and Study with HXD-PIN }

Taking this NXB error into account,
the excess hard counts in 3.3--6.5$'$ and 6.5--9.8$'$ annuli as shown
in table~\ref{tab:counts-cor8}(a) are no more significant.
In figure~\ref{fig:xis-pin-spec}, we plot the (OBS$-$NXB) spectra
of the XIS-FI and HXD-PIN in comparison with the CXB, NXB,
and ICM spectra. The XIS-FI spectrum in the 3.3--9.8$'$ annulus
exhibits an apparent excess over the nominal CXB spectrum
by \citet{Kushino2002}, however, it becomes quite consistent
if the ICM and 6.5\% NXB spectra are added (orange histogram).
The observed counts above 12~keV shown in
table~\ref{tab:counts-cor8}(b) also suggests that the
NXB counts are a few percent higher
than our estimation. The effective area of the XRT above 12~keV
is so small that almost all the observed counts
should be due to the NXB, although the reproducibility
in this energy band has not been studied.

In addition, the HXD-PIN data in 12--40~keV is fairly
consistent with the nominal CXB spectrum,
given by equation~(1) of \citet{Gruber1999}
in the energy range of 3--60~keV as;
\begin{equation}
S(E) = 7.877\; E^{-0.29} \exp\left(-\frac{E}{41.13}\right)\;\;
\frac{\rm keV}{\rm keV~cm^2~s~sr}.
\end{equation}
The residual counts after subtracting
the official NXB events amounts to
$(1.3\pm 0.2\pm 1.0)\times 10^{-2}$ cts~s$^{-1}$ (15--40~keV)\@.
The former error stands for statistical 1$\sigma$ error,
while the latter is a systematic error of $\sim 3.5$\% ($1\sigma$),
derived from a document in the Suzaku web page.\footnote{
http://www.astro.isas.jaxa.jp/suzaku/doc/suzakumemo/suzakumemo-2006-42.pdf}
We used the released response of the HXD-PIN for the HXD nominal position,
{\tt ae\_hxd\_pinhxnom\_20060814.rsp}, and scaled by the opening angle of
0.3~deg$^{2}$ of ``fine collimators'' \citep{Takahashi2007}.
The CXB contribution corrected for the normalization difference of
13\% between XIS and PIN is derived as $1.9\times 10^{-2}$ cts~s$^{-1}$,
which agrees with the observed intensity if one takes into account
the background systematics.
Therefore, HXD-PIN shows no signature of strong excess hard X-rays in 
the 15--40~keV spectrum.
If we take the PIN background systematics to be 5.6\%
at the 90\% confidence, the upper limit on the flux is
$6.2\times 10^{-12}$ erg~cm$^{-2}$~s$^{-1}$ (15--40~keV)\@.

\subsection{Excess in Central Region}

On the other hand, the 0.0--3.3$'$ region shows significant
excess counts with XIS-FI,
even considering the uncertainty in the NXB flux.
This region contains the cavities and possibly an X-ray weak AGN,
together with four bright member galaxies \citep{Zabludoff2000}.
As shown in table~\ref{tab:counts-cor8}(a),
the estimated LMXB flux from these galaxies
can account for only 30\% of the observed excess
after subtraction of the NXB, CXB, and ICM components.
Even though the X-ray to optical luminosity ratio
scatters by about 60\% as indicated by equation~(\ref{eq:lmxb}),
this scatter is unable to explain the whole excess flux.

\begin{figure}[tbg]
\centerline{
\FigureFile(0.45\textwidth,1cm){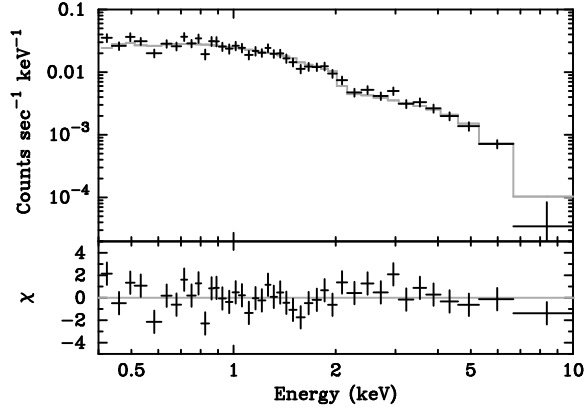}
}
\caption{
Summed spectra of 17 point sources detected with Chandra ACIS-S
within $r < 3.3'$ from the group center.
The gray histogram represents the best-fit power-law model
multiplied by the Galactic absorption of
$N_{\rm H} = 3.03\times 10^{20}$~cm$^{-2}$.
The best-fit parameters are a photon index, $\Gamma=1.38\pm 0.06$,
and 2--10~keV flux, $F_{\rm X} = (2.81\pm 0.24)\times 10^{-13}$
erg~cm$^{-2}$~s$^{-1}$, with $\chi^2/{\rm dof} = 44.6/40$.
The power-law model is acceptable with null hypothesis probability of 0.283.
The bottom panel shows the fit residuals in unit of $\sigma$.
}\label{fig:spec-sum}
\end{figure}

The residual excess can be explained if there are more hard sources
in excess of the nominal LMXBs in a similar order.
To examine this, we looked into the high resolution Chandra image to
directly measure the point source contribution.
There are 17 sources cataloged by \citet{Harrison2003} in this region,
and their locations are presented in figure~\ref{fig:Chandra-image}
overlaid on the Chandra ACIS-S3 image with an exposure of 49.15~ks.
The 2--10 keV fluxes of these sources distribute around
(0.7--8.2)$\times 10^{-14}$ erg~cm$^{-2}$~s$^{-1}$,
which can be converted to a luminosity of
(3.3--$39)\times 10^{39}$ erg~s$^{-1}$,
at the redshift of HCG~62.
This luminosity range is higher than the level of typical LMXBs.
We extracted the source spectra within $r < 3''$ 
and took the background from the 3--6$''$ annulus for each source.
The combined spectrum, shown in figure~\ref{fig:spec-sum},
can be fitted well with a power-low model with $\Gamma = 1.38\pm 0.06$,
pretty consistent with the nominal CXB slope.
However, the total flux amounts to
$(2.81\pm 0.24)\times 10^{-13}$ erg~cm$^{-2}$~s$^{-1}$,
which is higher than the expected CXB intensity,
$\pi\cdot (3.3')^2\cdot 5.97\times 10^{-8}$ erg~cm$^{-2}$~s$^{-1}$~sr$^{-1}$
$= 1.73\times 10^{-13}$ erg~cm$^{-2}$~s$^{-1}$,
by a factor of $1.62\pm 0.14$.
This almost completely explains the relative excess flux of $70\pm 19$\%
over the CXB as shown in table~\ref{tab:counts-cor8}(a).

The point source contribution is much higher than the expected level,
however it is still consistent with the CXB fluctuation.
\citet{Kushino2002} reported a field-to-field intensity fluctuation of 
$\sigma_{\rm CXB} = 6.5$\%, with a detector beam size of
$\Omega\sim 0.4$ deg$^2$ in the 2--10 keV band with the ASCA GIS\@.
Assuming a simple Euclidean log$N$-log$S$ relation as $N(>S)\propto S^{-2.5}$,
the CXB fluctuation scales as $\sigma_{\rm CXB}\propto\Omega^{-0.5}$.
This gives relative $1\sigma$ fluctuations of 42\%, 25\% and 19\%
for the 0.0--3.3$'$, 3.3--6.5$'$, and 6.5--9.8$'$ annuli, respectively,
so that the observed excess of $\sim 70$\% at $r<3.3'$ is
within the 90\% confidence range.

To summarize, there is a suggestion of extended hard X-ray emission
reported by the ASCA GIS in the Suzaku XIS data. However,
the uncertainties in
the NXB and CXB prevent us from claiming significant detection.
Some of the point sources detected with
the Chandra ACIS-S3 may well be associated with HCG~62\@.
Lastly, we warn that the hard excess flux increases by a large factor
if we include all the XIS data, in particular with ${\it COR}\le 8$~GV, 
in the spectral analysis, which is due certainly to the artificial effect
of the NXB reproducibility.
This indicates that suppression of the background counts 
with better screening criteria
is essential in the analysis of such extended hard X-ray emission.

\begin{table}[tbp]
\caption{
Summary of counts in (a) 5--12~keV and (b) above 12~keV bands
with the XIS FI sensors (XIS0 + XIS2 + XIS3) under the condition
of ${\it COR} > 8$~GV with 85.4~ks exposure.
}\label{tab:counts-cor8}
\centerline{
\begin{tabular}{lrrr}
\hline\hline\\[-1.5ex]
& \multicolumn{3}{l}{(a) Counts in 5--12~keV}\\[1ex]
\hline\\[-2ex]
Region & 0.0--3.3\makebox[0in][l]{$'$} & 3.3--6.5\makebox[0in][l]{$'$} & 6.5--9.8\makebox[0in][l]{$'$} \\
\hline
OBS\,\footnotemark[$*$] &		2619 & 5759 & 7396 \\
NXB\,\footnotemark[$\dagger$] &		1451 & 4298 & 5944 \\
CXB\,\footnotemark[$\dagger$] &		 420 &  989 &  905 \\
ICM\,\footnotemark[$\ddagger$] &	 333 &  180 &   90 \\
LMXB\,\footnotemark[$\ddagger$] &	 121 &  --- &  --- \\
\hline
Excess\,\footnotemark[\S] &	$294\pm 82 $ & $292\pm 121$& $420\pm 138$  \\
NXB fraction\makebox[0in][l]{\,\footnotemark[$\|$]} &
	$21\pm 6$\%  & $6.8\pm 2.8$\% & $7.1\pm 2.3$\% \\
CXB fraction\makebox[0in][l]{\,\footnotemark[$\|$]} &
	$70\pm 19$\% & $30\pm 12$\% & $46\pm 15$\% \\
\hline
ASCA\,\footnotemark[$\#$] &	 140 &  330 &  302 \\
\hline\hline\\[-1.5ex]
& \multicolumn{3}{l}{(b) Counts above 12~keV}\\[1ex]
\hline\\[-2ex]
Region & 0.0--3.3\makebox[0in][l]{$'$} & 3.3--6.5\makebox[0in][l]{$'$} & 6.5--9.8\makebox[0in][l]{$'$} \\
\hline
OBS\,\footnotemark[$*$] &		 598 & 1787 & 2318 \\
NXB\,\footnotemark[$\dagger$] &		 546 & 1650 & 2182 \\
CXB\,\footnotemark[$\dagger$] &		   4 &    5 &    4 \\
\hline
Excess\,\footnotemark[\S]  &	$48\pm 39 $ & $132\pm 68$& $132\pm 77$  \\
NXB fraction\makebox[0in][l]{\,\footnotemark[$\|$]} &
	$8.8\pm 7.1$\%  & $8.0\pm 4.1$\% & $6.0\pm 3.5$\% \\
\hline
\\[-1ex]
\end{tabular}
}
\footnotesize
\noindent
\footnotemark[$*$]
Observed counts with FI sensors including the NXB and CXB\@. \\
\footnotemark[$\dagger$]
Estimated NXB and CXB counts presented in figure~\ref{fig:nxb_cxb}.\\
\footnotemark[$\ddagger$]
ICM and LMXB counts for the Gal-2T model in table~\ref{tab:best-fit}. \\
\footnotemark[\S]
$\rm Excess\equiv OBS - NXB - CXB - ICM - LMXB$, and the 90\% confidence
Poisson error calculated as $1.6\times\sqrt{\rm OBS}$.\\
\footnotemark[$\|$]
NXB or CXB fraction calculated as
$\frac{\rm Excess}{\rm NXB}$ or $\frac{\rm Excess}{\rm CXB}$\@.\\
\footnotemark[$\#$]
Predicted excess hard counts by the ASCA results, as $\frac{\rm CXB}{3}$\@.
\normalsize
\end{table}

\section{Discussion}\label{sec:discussion}

\subsection{High-Temperature Ring}

We found the doughnut-like high-temperature ring at 3.3--6.5$'$
surrounding the group center, as shown in the hardness image of figure~\ref{fig:image}(b).
Spectral fits, shown in figure~\ref{fig:kT}(b), indicated 
an excess in the intensity ratio of
hot to cool ICM components in this annulus. 
We note that the co-existence of the
hot and cool components was already observed with
Chandra and XMM-Newton \citep{Morita2006} based on the deprojection analysis.
However, as shown in figure~11(a)
of \citet{Morita2006}, the hot component should occupy 
more than $\gtrsim 90$\% of the volume if two components are 
under a pressure balance.
In other words, the total thermal energy given by
a product of pressure and volume, $PV$,
is dominated by the hot component in this annulus.

We therefore assume that the 3.3--6.5$'$ (60--120~kpc) annulus is
filled with the hot ICM with temperature
$kT_{\rm Hot}\simeq 1.6$~keV and the electron density
$n_{\rm e}\simeq 7\times 10^{-4}$ cm$^{-3}$ for simplicity.
Then the gas pressure and the total energy are
$P = 1.92\; n_{\rm e} kT_{\rm Hot} = 2.2$ eV~cm$^{-3}$ and
$E = PV = 5.8\times 10^{59}$ erg~s$^{-1}$, respectively.
The bolometric luminosity is
$L_{\rm bol} = 1.1\times 10^{42}$ erg~s$^{-2}$
using the {\it vapec} model with the elemental abundances
given in table~\ref{tab:best-fit} (Gal 2T) for this annulus.
This gives the radiation cooling time as
\begin{equation}
\tau_{\rm cool} = E/L_{\rm bol} = 17~\makebox{Gyr},
\end{equation}
which is longer than the Hubble time.
The thermal conduction time is calculated as
\begin{eqnarray}
\tau_{\rm cond} &=& r^2/\kappa_{\rm S} = 0.33~\rm Gyr \\\nonumber
&\times&
  \Big(\frac{r}{100~\rm kpc}\Big)^2
  \Big(\frac{kT_{\rm Hot}}{1.6~\rm keV}\Big)^{-{\frac{5}{2}}}
  \Big(\frac{n_{\rm e}}{7\makebox{$\cdot$} 10^{-4}~\rm cm^{-3}}\Big),
\end{eqnarray}
assuming the Spitzer thermal conductivity \citep{Spitzer1962}.
Though this time scale is much shorter than the Hubble time,
the thermal conduction may be suppressed by a large factor
if there are turbulent magnetic fields,
and the time scale may become comparable to or longer than 
the group age of a few Gyr.

\citet{Morita2006} suggest that the hydrostatic equilibrium is
broken at $r\sim 5'$, and that the outflow of the hot ICM may occur,
based on the result of steep temperature drop at $r\sim 5$--10$'$.
In this case, the hot ICM may be expanding with nearly the sound velocity,
$v_{\rm s} = \sqrt{\gamma kT_{\rm Hot}/(\mu m_{\rm p})} = 640$~km~s$^{-1}$,
where $\gamma=5/3$ and $\mu = 0.62$. The expanding time scale is
\begin{equation}
\tau_{\rm s} = r/v_{\rm s} = 0.15\;
  \Big(\frac{r}{100~\rm kpc}\Big)
  \Big(\frac{kT_{\rm Hot}}{1.6~\rm keV}\Big)^{-0.5}
  ~\rm Gyr.
\end{equation}
Thus, the total power input required to form the high-temperature ring
can be roughly estimated as,
\begin{equation}
W = E/\tau_{\rm s} = 1.2\times 10^{44}~\makebox{erg~s$^{-1}$}.
\end{equation}
This power is by about two orders of magnitude higher than
that required to generate the two cavities
in the central region of HCG~62\@. The large implied power and the absence of strong AGN activity
in this group suggest that the hot ring may be caused by some
large-scale dynamical process in the central region of the group.

\subsection{Metallicity Distribution in ICM}

\begin{figure}
\centerline{
\FigureFile(0.45\textwidth,1cm){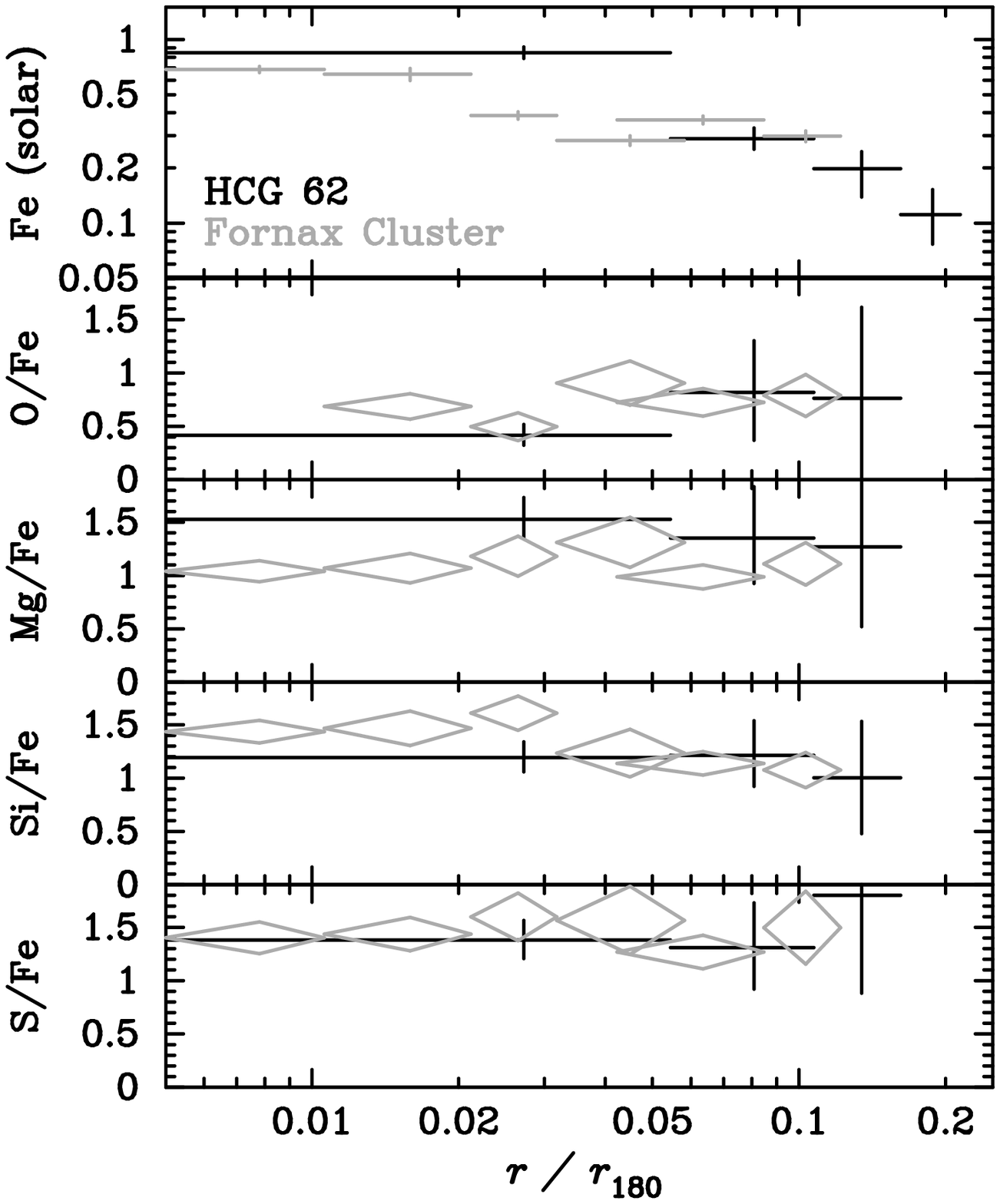}
}
\caption{
Comparison of the Gal 2T results for HCG~62 (black) with
results for the Fornax cluster (gray; \cite{Matsushita2007b}).
Fe abundance and the O/Fe, Mg/Fe, Si/Fe, and S/Fe ratios
in solar unit \citep{Anders1989} are plotted against
the projected radius scaled by the virial radius, $r_{180}$,
from upper to lower panels.
Results in $r>9.8'$ for HCG~62 are not plotted
for the Fe ratios due to the large uncertainty.
The innermost point of the O/Fe ratio is not plotted
because of the unsatisfactory fit.
}\label{fig:fe_ratio}
\end{figure}

The present Suzaku observation of HCG~62 showed abundance distribution
of O, Ne, Mg, Si, S, and Fe out to a radius of
$10'\simeq 180$~kpc. Ne abundance has large ambiguity
due to a strong coupling with Fe-L lines, as mentioned 
in section~\ref{sec:abun}.
Distributions of Mg, Si, S, and Fe are quite
similar to each other, while O profile in the outer region 
has a large uncertainty.
We plotted abundance ratios of
O, Mg, Si, and S against Fe as a function of the projected radius
in figure~\ref{fig:fe_ratio}. Here, the values in the outermost
region ($r>9.8'$) were excluded
because of large uncertainties.
The ratios Mg/Fe, Si/Fe and S/Fe are consistent to be a constant value
around 1.5--2, while O/Fe ratio for the innermost
region ($r<3.3'$) is significantly lower around 0.5.
In the Gal 2T fit, the O/Fe ratio seems to increase with radius.
These features have been seen with Chandra and XMM-Newton
by \citet{Morita2006}, and the present result gave a good
confirmation in the outer region.

Recent Suzaku observations have presented abundance profiles
in several other systems: an elliptical galaxy 
NGC~720 \citep{Tawara2007},
the Fornax cluster and NGC~1404 \citep{Matsushita2007b},
and a cluster of galaxies Abell~1060 \citep{Sato2007}.
While Si/Fe ratio is almost the same among all the systems,
Mg/Fe ratio is slightly higher in HCG~62 and Abell ~1060 than in
NGC~720, Fornax cluster, and NGC~1404\@.
We compare the abundances in HCG~62 with those in the Fornax cluster
as shown in figure~\ref{fig:fe_ratio}.
Regarding the Fornax cluster, we use
$r_{180}= 1.00 \sqrt{k\langle T\rangle / 1.3~\rm keV}$~Mpc,
and $z=0.00429$ corrected to the reference frame defined by
the 3~K microwave background radiation
(NASA/IPAC Extragalactic Database; NED)\@.
The solar abundance by \citet{Feldman1992}
with $\rm [Fe/H] = 3.24\times 10^{-5}$ employed by \citet{Matsushita2007b}
was scaled by a factor of 0.7 to match 
 the \citet{Anders1989} value of
$\rm [Fe/H] = 4.68\times 10^{-5}$.
HCG~62 shows lower Fe abundance than the Fornax cluster
in the central region at $r\lesssim 0.05\; r_{180}$, but
abundances become similar at $r\sim 0.1\; r_{180}$.
Abundance ratios of O/Fe, Mg/Fe, Si/Fe, and S/Fe are quite similar
between HCG~62 and the Fornax cluster at $r\sim 0.1\; r_{180}$.

\citet{Tamura2004} reported abundance ratios for 19 clusters
studied with XMM-Newton, and the mean Si/Fe ratio in cool and
medium temperature clusters with $kT < 6$~keV was $\sim 1.4$,
consistent with our HCG~62 result.
Their O/Fe ratio, $\sim0.6$, in the cluster core 
also agrees with our result.
\citet{Matsushita2003,Matsushita2007a} also reported abundance ratio
for M87 and the Centaurus cluster, respectively.
M87 showed the Mg/O ratio to be $\sim 1.3$ in the central region, and
the Centaurus cluster indicated the O/Fe and Si/Fe ratios within $8'$ 
consistent with our results.
The Mg/O ratio for HCG~62 with XMM-Newton \citep{Morita2006} is
$\sim 3.3$ within $1'$, which is almost the same as our result of $\sim 3.6$
within 3.3$'$. Mg/O ratios in other groups are
$\sim2.5$ for NGC~5044 \citep{Tamura2003},
and $\sim1.3$ for NGC~4636 \citep{Xu2002} both measured with XMM-Newton RGS\@.

\citet{Sato2007c} studied contributions of type Ia and II supernovae 
to the metal enrichment, based on Suzaku results of HCG~62,
Abell~1060, AWM~7, NGC~570 \citep{Sato2007,Sato2007b}.

\begin{figure*}
\centerline{
\FigureFile(0.45\textwidth,1cm){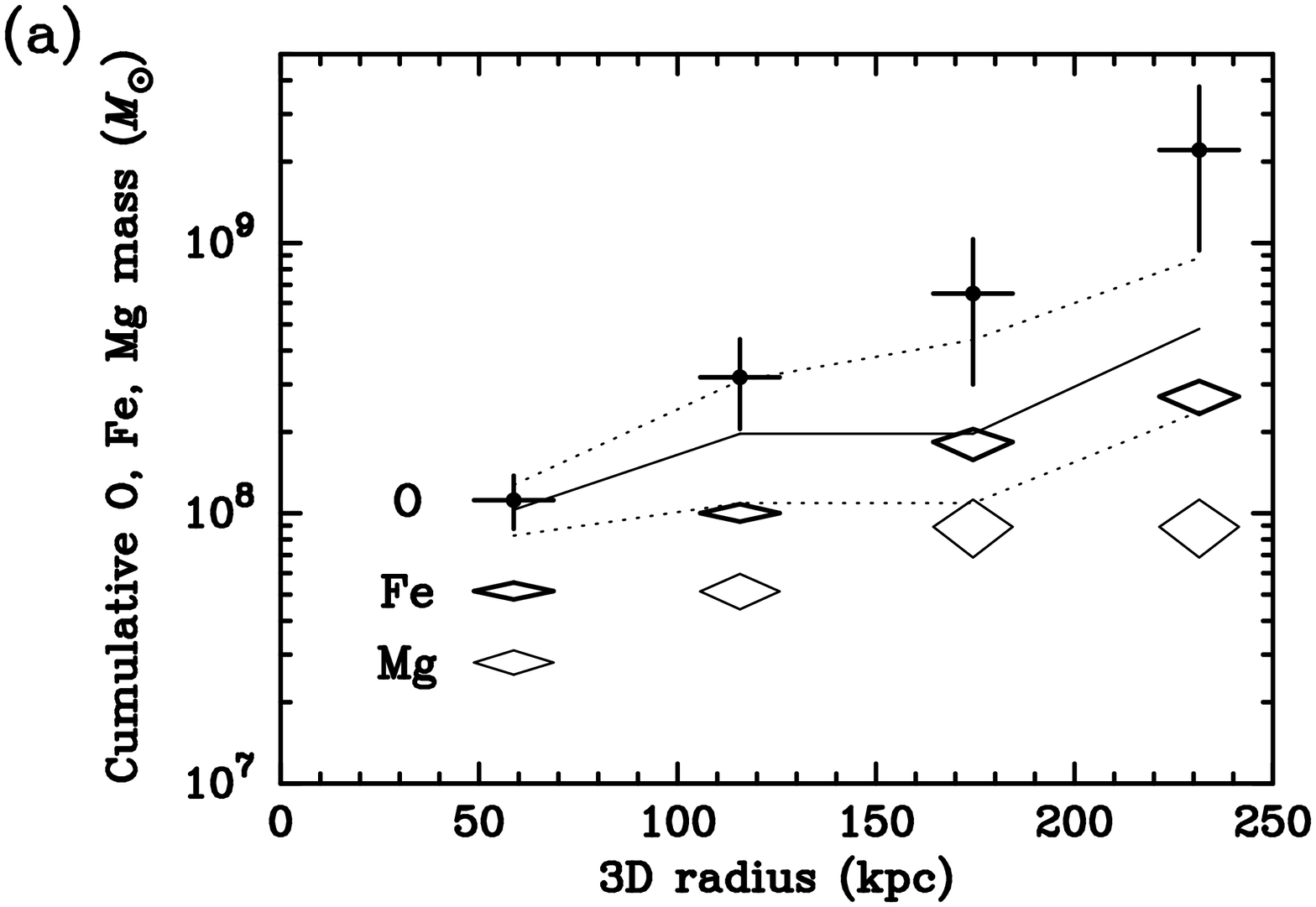}%
\hspace*{0.05\textwidth}
\FigureFile(0.45\textwidth,1cm){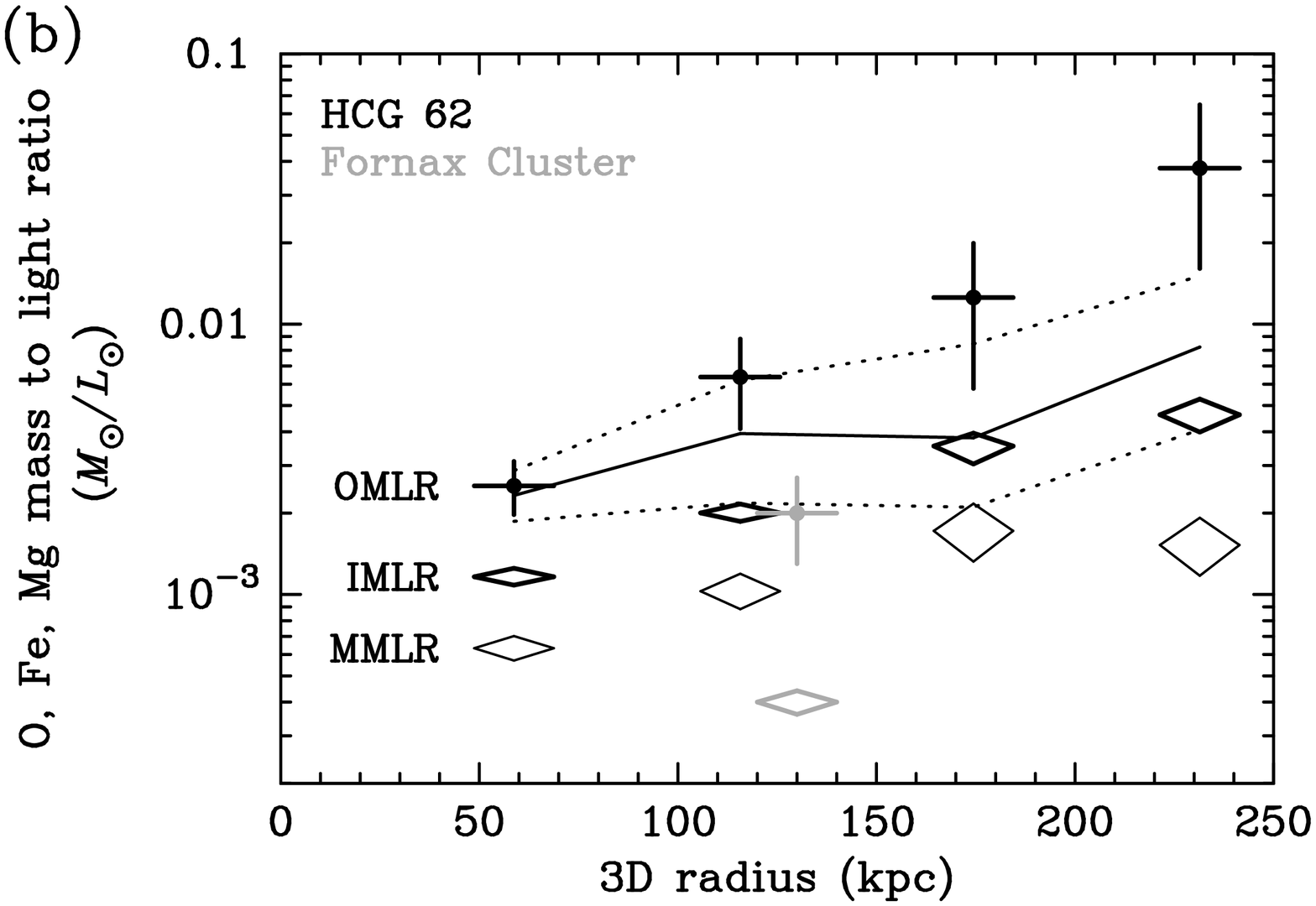}%
}
\caption{
(a) Cumulative mass, $M(<R)$, within the 3-dimensional radius, $R$,
for O, Fe, and Mg are plotted for HCG~62,
combining the abundance determination with Suzaku
and the gas mass profile with Chandra and XMM-Newton.
The cross and diamond markers show the mass profiles
with the Gal 2T model, and the O mass profile and its error region
with the Gal 1T model are shown in solid and dotted lines.
(b) Ratios of the oxygen, iron, and magnesium mass in unit of $M_\odot$
to the $B$ band optical luminosity in unit of $L_\odot$
(OMLR, IMLR, and MMLR, respectively) are plotted against
the 3-dimensional radius. Black markers represent our results
for HCG~62, and gray markers of OMLR and IMLR show
the results for the Fornax cluster by \citet{Matsushita2007b}.
}\label{fig:mass_mlr}
\end{figure*}

\subsection{Metal Mass-to-Light Ratio}

Metal mass-to-light ratios for oxygen, iron,
and magnesium (OMLR, IMLR, and MMLR, respectively),
were examined.
First, we show the metal mass profiles in figure~\ref{fig:mass_mlr}(a),
based on the 3-dimensional gas mass profile
by \citet{Morita2006}
and the abundance profile measured with Suzaku.
The derived iron mass within the 3-dimensional radius of $R< 200$~kpc is
(1.5--$3)\times 10^8$ $M_\odot$, which is quite consistent with
the previous measurement in figure~11(e) of \citet{Morita2006}.
We obtained the magnesium mass
to be (0.7--$1.1)\times 10^{8}$ $M_\odot$.
Though the oxygen mass has large errors as seen
in figure~\ref{fig:mass_mlr}(a), it is unlikely that
oxygen has smaller mass than iron.
Therefore, the Gal 2T model (crosses) is preferred
to the Gal 1T one (solid and dotted lines).
The estimated oxygen mass is $\sim 10^9$ $M_\odot$,
within $R< 200$~kpc.

Secondly, we adopted \citet{Zabludoff2000} results 
as the member galaxy catalog
of HCG~62 (12 galaxies in $r < 13'\sim 230$~kpc),
and their redshifts were used to estimated the 3-dimensional
distribution.
Since only the $R$-band optical magnitudes are provided
in this catalog, we converted them into the $B$-band magnitudes
using the color of $B-R = 2.0$~mag for the HCG~62a galaxy at the center
\citep{Hickson1989}, and $A_B = 0.224$ and $A_R = 0.139$ after
NASA/IPAC Extragalactic Database (NED) in the direction of HCG~62\@.

We thus calculated the integrated values of OMLR, IMLR, and MMLR within $r\lesssim 230$~kpc
as shown in figure~\ref{fig:mass_mlr}(b),
and they turned out to be
$\sim 4\times 10^{-2}$, 
$\sim 4.6\times 10^{-3}$,
and $\sim 1.5\times 10^{-3}$ $M_{\odot}/L_{\odot}$,
respectively.
The errors are only based on the statistical errors of
metal abundances in the spectral fit, and
the uncertainties of the gas mass profiles and the luminosities
of the member galaxies are not considered.
The IMLR values are consistent with the collective
results with ASCA by \citet{Makishima2001}.
The MMLR and IMLR show similar steep increase with radius up to
$r \sim 100$~kpc and seem to reach almost
a plateau at 100--200~kpc.
This feature is not apparent for the OMLR,
due partly to the large uncertainty.
The behavior of MLR curves would be related to different
enrichment processes, as discussed in \citet{Morita2006}.

We also point out that the derived OMLR and IMLR for HCG~62
are much larger than those of the Fornax cluster \citep{Matsushita2007b}
as shown in figure~\ref{fig:mass_mlr}(b).
Particularly the difference in IMLR is significant,
although that in OMLR is marginal due to the uncertainty
in the Galactic emission.
These two systems have a similar potential depth
with $k\langle T\rangle \sim 1.3$~keV\@.
This feature may indicate that the Fornax cluster is a younger system
with less number of SN~Ia for the Fe production since the cluster formation,
and/or that the metal distribution in the Fornax may be much more extended.
Possible difference of the initial mass function (IMF) between the two
may account for some fraction of the discrepancy in the MLRs.
\citet{Ikebe1996} discovered that there are two distinct length scales
of dark matter concentration in the Fornax cluster, and the cD galaxy
NGC~1399 is off-centered by $\sim 50$~kpc with respect to 
the cluster hot gas.
There is also an X-ray luminous elliptical galaxy NGC~1404 at 10$'$
south-west of NGC~1399\@. These features suggest that the Fornax system
may be dynamically young and that galaxy interactions
in ICM may have caused extended metal distribution.

\ifnum1=0
\begin{table}
\caption{O, Mg, and Fe Masses for the observed clusters}
\label{tab:mass}
\begin{center}
\begin{tabular}{lccl} \hline
& $M_{\rm O}\ (M_\odot)^\ast $ & $M_{\rm Mg}\ (M_\odot)$ & $M_{\rm Fe}\ (M_\odot)$\\
& $\times10^{9}$ & $\times10^{7}$ &  $\times10^{8}$ \\
\hline
HCG~62 & 2.2/0.5 & 8.9 & 2.7\\
\hline
\multicolumn{4}{l}{\parbox{0.45\textwidth}{\footnotesize
\footnotemark[$\ast$]
These values correspond to the case with 2 or 1 {\it apec} models
as the Galactic emission}} \\
\end{tabular}
\end{center}
\end{table}
\fi

\subsection{AGN versus Mergers}

HCG~62 exhibits several interesting activities in both the central and the outer regions:
namely, two cavities \citep{Vrtilek2002, Morita2006},
extended hard X-ray emission \citep{Fukazawa2001,Nakazawa2007},
multi-phase ICM (\cite{Morita2006}; this work),
possible ``high-abundance arc'' \citep{Gu2007},
and the doughnut-like high-temperature ring (this work).
One might relate the central features with an AGN activity, however,
the central galaxy (HCG~62a = NGC~4761) currently shows
little evidence of AGN activities in both optical
\citep{Coziol1998,Shimada2000,Coziol2004} and radio bands,
as summarized in section~8.1 of \citet{Morita2006}
and section~3.1 of \citet{Gu2007}.
\citet{Morita2006} also have placed an upper limit on
the X-ray luminosity of the AGN to be
$L_{\rm X}\lesssim 10^{39}$~erg~s$^{-1}$ (0.5--4~keV)\@.
Since the Suzaku HXD-PIN shows no excess over the CXB flux 
(figure~\ref{fig:xis-pin-spec}), heavily absorbed AGN
with an intrinsic luminosity larger than
$\sim 3\times 10^{42}$ erg~s$^{-1}$ (15--40~keV) is also ruled out.
Another possibility related with the HCG~62 activity can be recent mergers,
as suggested in the above mentioned previous works.
Energy inputs from magneto-hydrodynamic interactions of the
member galaxies with the ICM \citep{Makishima2001} can be another
possible mechanism.

We found  a concentration of hard sources
in $r<3.3'$ with average photon indices $\Gamma\sim 1.4$.
As shown in section~8, this concentration itself is
within the range of CXB fluctuation, however,
the spatial coincidence with the group center is remarkable
as seen in figure~\ref{fig:image}(b).
Since the source luminosities are (3.3--$39)\times 10^{39}$ erg~s$^{-1}$
(2--10 keV) at HCG~62,
we suggest that some of the sources may be remnants of minor mergers
which were previously central black holes of the merged galaxies.
The overdensity of X-ray sources with
$L_{\rm X} = (4.0$--$250)\times 10^{39}$ erg~s$^{-1}$ (2--10~keV)
in the fields of A~194 and A~1060 was also reported by \citet{Hudaverdi2006}.
Optical identifications of these objects will be of much interest.

\section{Summary}\label{sec:summary}

\begin{itemize}\setlength{\itemsep}{1ex}

\item Suzaku confirmed the multi-phase nature of
the ICM out to a radius $\sim 10'$.
We found a doughnut-like high-temperature ring at 3.3--6.5$'$
in the hardness image,
which is caused by higher hot component intensity over the 
cool component as shown by the spectral fit.
Possible ICM heating by the mass accretion is suggested.

\item We could not confirm the ``high-abundance arc''
in SW arc in the 1.1--3.3$'$ annulus reported by \citet{Gu2007}, and 
possible misidentification of an excess hot component as the line
was suggested.

\item Mg to Fe ratio showed enhancement at the center
confirming the previous Chandra and XMM-Newton result.
Temperature, surface brightness, and O abundance at $r > 3.3'$
were subject to modeling of the Galactic component,
while Mg, Si, S, and Fe abundances were fairly robust.
The Gal 2T model was preferred in terms of the
surface brightness profile and the integrated O mass.

\item O abundance was $\sim 0.4$ solar at the center and
less than 0.5 at $r>3.3'$.
Abundance ratios, O/Fe, Mg/Fe, Si/Fe, and S/Fe, showed
similar values with those in the Fornax cluster
in $r\sim 0.1\; r_{180}$. Comparison of 19 clusters with HCG~62
showed consistent levels of $\rm O/Fe\sim 0.6$ and $\rm Si/Fe\sim 1.4$. 
On the other hand, HCG~62 showed $\rm Mg/O\sim 3.6$ at $r<3.3'$, 
significantly higher than in other groups and giant ellipticals.

\item The OMLR and IMLR values in HCG~62 are by about order of magnitude
higher than the Fornax cluster results at $r\sim 130$~kpc,
while our IMLR agrees with collection of the ASCA results.

\item Thermal fit and the 6--10~keV image
indicated an excess above $\sim 5$~keV in most of the radii,
however it was not significant considering uncertainties
in the NXB and CXB fluctuations.
ASCA detection of the hard excess is still consistent with our results.

\item We found an excess X-ray emission of
$70\pm 19$\% times the nominal CXB intensity (5--12~keV) within $r<3.3'$,
and most of it could be explained by a concentration
of hard X-ray sources detected with Chandra.
We suggested some of the sources could be remnants of minor mergers.

\end{itemize}

\bigskip
Part of this work was financially supported by the Ministry of
Education, Culture, Sports, Science and Technology of Japan,
Grant-in-Aid for Scientific Research
No.\ 14079103, 15340088, 15001002, 16340077, 18740011.

\end{document}